\documentclass[11pt, reqno,notitlepage]{article}
\usepackage{eurosym}
\usepackage{graphicx,amsmath,amssymb}
\usepackage{natbib,amsfonts}
\usepackage{amsthm}
\usepackage{color,url}
\usepackage{bm, multirow}
\usepackage{hyperref}
\usepackage{algpseudocode}
\usepackage{algorithm}
\usepackage{subcaption}
\usepackage{comment}
\usepackage{enumitem}
\usepackage{setspace}
\usepackage[normalem]{ulem}

\hypersetup{
    colorlinks=true,
    linkcolor=blue,
    filecolor=magenta,
    urlcolor=cyan,
}
\providecommand{\abs}[1]{\lvert#1\rvert}
\providecommand{\norm}[1]{\lVert#1\rVert}
\providecommand{\erps}{\varepsilon}
\newtheorem{theorem}{Theorem}
\theoremstyle{plain}

\newtheorem{myalgorithm}{Algorithm}

\newtheorem{corollary}{Corollary}

\newtheorem{definition}{Definition}
\newtheorem{example}{Example}[section]

\newtheorem{lemma}{Lemma}

\newtheorem{proposition}{Proposition}

\newtheorem{assumption}{Assumption}
\newenvironment{varproof}[1][Proof.]{\begin{trivlist}
\item[\hskip \labelsep{\bfseries #1}]}{\end{trivlist}}
\numberwithin{equation}{section}
\def\bkR{{\rm I\kern-.17em R}}
\theoremstyle{definition}

\def\uniset{{\rm 1\kern-.40em 1}}
\graphicspath{{../plots/}}
\input{tcilatex}
\begin{document}

\title{\textbf{Cupid's Invisible Hand: }\\
{\Large \textbf{Social Surplus and Identification in Matching Models}}}
\date{June 1, 2021\thanks{
{\tiny This paper builds on and very significantly extends our earlier
discussion paper Galichon and Salani\'{e} (2010), which is now obsolete. The
authors are grateful to Pierre-Andr\'{e} Chiappori, Eugene Choo, Chris
Conlon, Facundo Danza, Gautam Gowrisankaran, Leon Guzman lizardo, Jim Heckman, Sonia Jaffe, Robert McCann,
Jean-Marc Robin, Aloysius Siow, the editor and referees and many seminar
participants for very useful comments and discussions. Part of the research
underlying this paper was done when Galichon was visiting the University of
Chicago Booth School of Business and Columbia University, and when Salani%
\'{e} was visiting the Toulouse School of Economics. Galichon thanks the
Alliance program for its support, and Salani\'{e} thanks the Georges Meyer
endowment. Galichon's research has received funding from NSF DMS-1716489,
and ERC grants FP7--313699 and CoG--866274.}}}
\author{\textbf{Alfred Galichon}\thanks{{\tiny {\ Economics and Mathematics
Departments, New York University, and Economics department, Sciences Po;
e-mail: ag133@nyu.edu.}}} \and \textbf{Bernard Salani\'e}\thanks{{\tiny {\
Department of Economics, Columbia University; e-mail: bsalanie@columbia.edu.}%
}}}

\maketitle

\begin{abstract}
{\small We investigate a model of one-to-one matching with transferable
utility and general unobserved heterogeneity. Under a separability assumption that
generalizes \citet{choo-siow:06}, we first show that the equilibrium
matching maximizes a social gain function that trades off exploiting
complementarities in observable characteristics and matching on unobserved
characteristics. We use this result to derive simple closed-form formul\ae\ %
that identify the joint matching surplus and the equilibrium utilities of
all participants, given any known distribution of unobserved heterogeneity.
 We  provide
efficient algorithms to compute the stable matching and to estimate parametric versions of the model. Finally, we revisit Choo and Siow's empirical
application to illustrate the potential of our more general approach. }
\end{abstract}

\section*{Introduction}

Since the seminal contribution of \citet{Becker:73}, many economists have
modeled the marriage market as a matching problem. When utility is perfectly
transferable, each potential match generates a marital surplus. The
distributions of tastes and of desirable characteristics determine
equilibrium shadow prices, which in turn explain how partners share the
marital surplus in any realized match. This insight is not specific to the
marriage market: it characterizes the ``assignment game'' of~\citet{SS:72},
i.e.\ models of matching with transferable utilities. Family economics makes
extensive use of this class of models; we refer the reader to the recent
book by~\citet{chiapporilove}. Matching with transferable utilities has also
been applied to competitive equilibrium in good markets with hedonic pricing %
\citep[][]{ChiapporiMcCannNesheim:10}, to trade %
\citep[e.g.,][]{Costinot-Vogel:15} to the labour market (\citet{Tervio} and %
\citet{gabland:08}) and to industrial organization (\citet{bajarifox}, %
\citet{fox:qe18}, \citet{foxyanghsu}) among other fields. Our results  apply in all of these contexts; however for concreteness, we will stick to the marriage metaphor
in our exposition of the main results.

While Becker presented the general theory, he focused on the special case in
which the types of the partners are one-dimensional and are complementary in
producing surplus. As is well-known, the social optimum then exhibits \emph{%
positive assortative matching}: higher types pair up with higher types.
Moreover, the resulting configuration is stable, and it is in the core of
the corresponding matching game. This sorting result is both simple and
powerful; but its implications are also at variance with the data, in which
matches are observed between partners with quite different characteristics.
To account for a wider variety of matching patterns, one solution consists
of allowing the matching surplus to incorporate latent
characteristics---heterogeneity that is unobserved by the analyst. %
\citet{choo-siow:06} have shown how it can be done in a way that yields a
highly tractable model in large populations, provided that the unobserved
heterogeneities enter the marital surplus quasi-additively, and that they
are independent and identically distributed as standard type I extreme value
terms. \citet{choo-siow:06} used their model to evaluate the effect of the
legalization of abortion on gains to marriage; and they applied it in~\cite{siow-choo:06} to
Canadian data to measure the impact of demographic changes. It has also been used to study increasing returns in
marriage markets (\citet{Botticini-Siow:11}), to compare the preference for
marriage versus cohabitation \citep[][]{MourifieSiow} and to estimate the
changes in the returns to education on the US marriage market %
\citep[][]{csw:17}. A continuous version of Choo and Siow's logit framework
has been developed by \citet{dg:14} to understand the affinities between
continuous characteristics personality traits on the marriage market, using
Dagsvik's theory of extreme value processes. \cite{cgg:19}\label{descr-DG}
used this approach to compare same-sex and different-sex couples.

We revisit here the theory of matching with transferable utility in the
light of Choo and Siow's insights. Three assumptions underlie their
contribution: the unobserved heterogeneities on the two sides of a match do
not interact in producing matching surplus; they are distributed as iid type
I extreme values; and populations are large. We maintain the first
``separability'' assumption, and the last one which is innocuous in many
applications. Choo and Siow's distributional assumption, on the other hand,
is very special; it generates a multinomial logit model that has quite
specific restrictions on cross-elasticities. We first show that this
distributional assumption can be completely dispensed with, and that the
Choo-Siow framework can be extended to encompass much less restrictive
assumptions on the unobserved heterogeneity. Our second contribution is to
spell out a complete empirical approach to identification, parametric
estimation, and computation in this class of models. Our third contribution
is to revisit the original \citet{choo-siow:06} dataset on marriage patterns
by age, making use of the new possibilities allowed by our extended
framework. We shall defer to Section~\ref{par:roadmap} the precise
description of each step of our paper.

\medskip

There are other approaches to estimating matching models with unobserved
heterogeneity; see the handbook chapter by~\cite{Graham:hdbk,
Graham:hdbkerrata} and the surveys by~\cite{csjel:16} and~\cite{chiappori:20}%
. For markets with transferable utility, %
\citet{Fox:identmatchinggames,fox:qe18} has proposed pooling data across
many similar markets and relying on a ``rank-order property'' that is valid
when unobserved heterogeneity is separable and exchangeable---which excludes
the nested logit, mixed logit, and other models considered in our paper. %
\citet{bajarifox} applied this approach to spectrum auctions. %
\citet{foxyanghsu} focus on identifying the complementarity between
unobservable characteristics. \citet{GualdaniSinha:19} study partial
identification issues in nonparametric matching models.

The literature on markets with non-transferable utility has evolved
separately, with some interesting similarities---in particular with %
\citet{Menzel:15}'s investigation of large NTU markets, building on a model of~\citet{dagsvik:00}. Many papers have
modeled school assignment, where preferences on one side of the market are
highly constrained by regulation (see \citet{agarwalsomainiare:20} for a
recent review.) \citet{Agarwal:15} estimates matching in the US medical
resident program; his work relies on the assumption that all hospitals agree
on how they rank candidates.

\medskip

\textbf{Notation and terminology} In the following, $X\sim P$ will denote
that random variable $X$ has probability distribution $P$. We use \textbf{%
bold} type to denote vectors and matrices.  Under perfectly transferable
utility, the stable matching  maximizes the social surplus over the set of
feasible matchings \citep{SS:72}; we sometimes use the terms ``social
optimum'' or ``equilibrium'' to denote the stable matching. For
simplicity, we also use ``joint surplus'' and ``joint utility''
interchangeably.  We hope that this creates no confusion.

\medskip

\section{Framework and Roadmap\label{sec:framework}}

We study in this paper a bipartite, one-to-one matching market with
transferable utility. We maintain throughout some of the basic assumptions
of \citet{choo-siow:06}: utility transfers between partners are
unconstrained, matching is frictionless, and there is no asymmetric
information among potential partners. We call the partners ``men'' and
``women'', as we have in mind an application to the heterosexual marriage
market; our results are not restricted to a marriage context, however.

\subsection{The setting\label{par:thePopulations}}

Following Choo and Siow, we assume that the analyst can only observe which
\emph{group} each individual belongs to. Each man $i\in \mathcal{I}$ belongs
to one group $x_{i}\in \mathcal{X}$; and, similarly, each woman $j\in
\mathcal{J}$ belongs to one group $y_{j}\in \mathcal{Y}$. We will say that
``man $i$ is in group $x$'' and ``woman $j$ is in group $y$.'' There is a
finite number of groups; they are defined by the intersection of the
characteristics which are observed by all men and women, and also by the
analyst. On the other hand, men and women of a given group differ along some
dimensions that they all observe, but which do not figure in the analyst's
dataset.

%
Like Choo and Siow, we assume that there is an (uncountably) infinite number
of men in any group $x$, and of women in any group $y$. We denote $n_{x}$
the mass of men in group $x$, and $m_{y}$ the mass of women in group $y$.
Since the problem is homogenous, we can assume that the total mass of
individuals is normalized to one, that is $\sum_{x}n_{x}+\sum_{y}m_{y}=1$.
Hence, $n_x$ and $m_y$ are not to be thought as numbers of individual of
each types, but as masses. We will often use the notation $\bm{r}=(\bm{n},%
\bm{m})$ for the vector that collects the ``margins'' of the problem.

A \emph{matching} is the specification of who matches with whom. It is
\emph{feasible} if each individual is matched to~0 or 1~partner. It is
\emph{stable} if no individual who has a partner would prefer to be
single, and if no two individuals would prefer forming a couple to their
current situation.

Our data can only describe matchings at the group level---that is, the mass
distribution of matched pairs across groups. Let $\mu _{xy}$ be the mass of
the couples where the man belongs to group $x$, and where the woman belongs
to group $y$. The (group-level) feasibility constraints state that the mass
of married individuals in each group cannot be greater than the mass of
individuals in that group, which is denoted $\bm{\mu}\in \mathcal{M}\left(%
\bm{r}\right)$, where $\mathcal{M}\left(\bm{r}\right) $ (or $\mathcal{M}$ in
the absence of ambiguity) is defined by:%
\begin{equation}
\mathcal{M}\left( \bm{n},\bm{m}\right) =\left\{ \bm{\mu}\geq 0:\forall x\in
\mathcal{X},~\sum_{y\in \mathcal{Y}}\mu _{xy}\leq n_{x}~;~\forall y\in
\mathcal{Y},~\sum_{x\in \mathcal{X}}\mu _{xy}\leq m_{y}\right\}
\label{feasible}
\end{equation}%
With mild abuse, we will call each element of $\mathcal{M}$ a \emph{feasible
matching}. For notational convenience, we shall denote $\mu
_{x0}=n_{x}-\sum_{y\in \mathcal{Y}}\mu _{xy}$ the corresponding mass of
single men of group $x$ and $\mu _{0y}=m_{y}-\sum_{x\in \mathcal{X}}\mu
_{xy} $ the mass of single women of group $y$. We also define the sets of
marital choices that are available to male and female agents, including
singlehood:
\begin{equation*}
\mathcal{X}_{0}=\mathcal{X}\cup \left\{ 0\right\} \text{, }\mathcal{Y}_{0}=%
\mathcal{Y}\cup \left\{ 0\right\} ,
\end{equation*}%
and we denote%
\begin{equation*}
\mathcal{A}=\left( \mathcal{X}\times \mathcal{Y}\right) \cup \left( \mathcal{%
X}\times \left\{ 0\right\} \right) \cup \left( \left\{ 0\right\} \times
\mathcal{Y}\right)
\end{equation*}
the set of marital arrangements.

%

\subsection{Separability\label{par:separ}}

Every match between a man $i$ and a woman $j$ generates a \emph{joint surplus%
}, which is the excess of the sum of their utilities when married over the
sum of their utilities when single. As shown in \citet{csw:17}, an important
assumption made implicitly in Choo and Siow is that the joint surplus
created when a man $i$ of group $x$ marries a woman $j$ of group $y$ does
not allow for interactions between their unobserved characteristics,
conditional on $(x,y)$. This leads us to assume:

\begin{assumption}[Separability]\label{ass:separ}
  There exist a matrix $\bm{\Phi}$ and random terms
$\bm{\varepsilon}$ and $\bm{\eta}$ such that

\begin{itemize}
\item[(i)] the joint utility from a match between a man $i$ in group $x\in
\mathcal{X}$  and a woman $j$ in group $y\in \mathcal{Y}$ is
\begin{equation}  \label{eq:sepPhi}
\tilde{\Phi}_{ij}=\Phi _{xy}+\varepsilon _{iy}+\eta _{xj},
\end{equation}

\item[(ii)] the utility of a single man $i$ is $\tilde{\Phi}%
_{i0}=\varepsilon_{i0}$

\item[(iii)] the utility of a single woman $j$ is $\tilde{\Phi}%
_{0j}=\eta_{0j}$
\end{itemize}

where, conditional on $x_{i}=x$, the random vector $\bm{\varepsilon}%
_{i}=(\varepsilon _{iy})_{y\in\mathcal{Y}_{0}}$ has probability distribution
$\bm{P}_{x}$, and, conditional on $y_{j}=y$, the random vector $\bm{\eta}%
_{j}=(\eta _{xj})_{x\in\mathcal{X}_{0}}$ has probability distribution $\bm{Q}%
_{y}$. The variables
\begin{equation*}
\max_{y\in \mathcal{Y}_0} \; \abs{\varepsilon_{iy}} \; \mbox{ and } \;
\max_{x\in \mathcal{X}_0} \; \abs{\eta_{xj}}
\end{equation*}
have finite expectations under $\bm{P}_x$ and $\bm{Q}_y$ respectively.
\end{assumption}

While separability is a restrictive assumption, it allows for ``matching on
unobservables'': a match between a man of group $x$ and a woman of group $y$
may occur because this woman has unobserved characteristics that make her
attractive to men of group $x$, and/or because this man has a strong
unobserved preference for women of group $y$. What separability does rule
out, however, is sorting on unobserved characteristics on both sides of the
market, e.g.\ some unobserved preference of man $i$ for some unobserved
characteristics of woman $j$.

Note that we did not constrain the distributions $\bm{P}_{x}$ and $\bm{Q}%
_{y} $ to belong to the extreme value class. We extend the logit framework
of~\cite{choo-siow:06} in several important ways: we allow for different
families of distributions, with any form of heteroskedasticity, and with any
pattern of correlation across partner groups. We will demonstrate the use of
these extensions on an application in Section~\ref{sec:appli}.

To summarize, a man $i$ in this economy is characterized by his full type $%
\left( x_{i},\bm{\varepsilon} _{i}\right) $, where $x_{i}\in \mathcal{X}$
and $\bm{\varepsilon} _{i}\in \mathbb{R}^{\mathcal{Y}_{0}}$; the
distribution of $\bm{\varepsilon}_{i}$ conditional on $x_{i}=x$ is $\bm{P}%
_{x}$. Similarly, a woman $j$ is characterized by her full type $\left(
y_{j},\bm{\eta} _{j}\right) $ where $y_{j}\in \mathcal{Y}$ and $\bm{\eta}
_{j}\in \mathbb{R}^{\mathcal{X}_{0}}$, and the distribution of $\bm{\eta}
_{j}$ conditional on $y_{j}=y$ is $\bm{Q}_{y}$.

\subsection{Objectives and a roadmap\label{par:roadmap}}

While the paper's final goal is to develop inference tools for matching
problems with transferable utility and separable unobserved heterogeneity,
this will require several intermediate steps.

\textbf{First, we show how given separability, the two-sided matching
problem resolves into a collection of one-sided problems of lower complexity.%
}

\textbf{Second, we provide new results on discrete choice (one-sided) models.%
} One-sided discrete choice problems will play a key role in our analysis.
Section~\ref{sec:one-sided} provides new results on this class of problems.
We introduce a convex function which we call the \emph{generalized entropy
of choice}. Theorem~\ref{thm:entropy-is-ot} shows that this function is the
value of an optimal transport problem, for which numerous computational
methods have been developed. Theorem~\ref{thm:onesidedident} then proves
that  given the choice probabilities and the distribution of errors, the
underlying mean utilities are identified by the gradient  of the generalized
entropy of choice. These results should be useful beyond the setting of this
paper.

\textbf{Third, we show how the stable matching solves a convex optimization
problem.} This is done in Section~\ref{par:ssid},  and formally  stated in
Theorem~\ref{MainThmSurplus}.

\textbf{Fourth, we use convex duality to identify the matching surplus.}
Identification consists of recovering the matching surplus based on the
observation of the matching patterns; this is the ``inverse problem'' to the
computation of the stable matching given the surplus. We show in Section~\ref%
{par:two-sided-identification} that these two problems are conjugate of each
other in the sense of convex duality. As a consequence, the matching surplus
is identified from the matching patterns given any distribution of errors
(Theorem~\ref{thm:Identification}).

\textbf{Fifth, we propose new computational methods for the equilibrium and
estimation problems}. The convexity of all of our objects allows for a
number of efficient computational methods to compute the stable matching
and/or recover the joint surplus. Section~\ref{sec:computation} shows how
this can be done by gradient descent, coordinate descent, and linear
programming. In particular, coordinate descent generates a very efficient
``IPFP'' algorithm for variants of the logit model;  we prove its
convergence in Theorem~\ref{thm:convergence}.

\medskip

Taken together, these results allow us to develop \textbf{a comprehensive
set of tools for the parametric estimation of the matching model.} We allow
for parameters both in the matching surplus and in the distribution of the
random utility. Section~\ref{sec:inference} first investigates the
properties of maximum likelihood estimation in that setting (Section~\ref%
{par:MLE}). We present an alternative method based on matching observed
moments of the distributions of matched pairs in Section~\ref%
{par:linearModel}. This is attractive as unlike maximum likelihood,  it
retains global convexity and has an intuitive interpretation. Finally,
\textbf{we test our approaches} in Section~\ref{sec:appli}, where we fit
several instances of separable models to the~\cite{choo-siow:06} dataset.

We have tried to keep the exposition intuitive in the body of the paper; all
proofs can be found in Appendix~\ref{Appx:proofs}. Appendix~\ref%
{App:Explicit} specializes our results to several common specifications of
unobserved heterogeneity. The paper is complemented by several online
appendices where we discuss the assumptions  that are relaxed or maintained
in the paper (Appendix~\ref{app:discussions-of-assumptions}); we provide
complementary results with equilibrium predictions (Appendix~\ref%
{app:additional-equilibrium});  we provide complementary estimation results
(Appendix~\ref{app:additional-estimation}); we  give pseudo-code for our
IPFP algorithm and give simulation results for this and  other algorithms
(Appendix~\ref{app:computation}); and we provide additional details on the
application of Section~\ref{sec:appli} (Appendix~\ref{Appx:CSdata}).
Finally, we provide Python and R code to estimate this class of
models at \url{https://bsalanie.github.io/.}

\section{Social surplus and identification in the one-sided case: discrete
choice models\label{sec:one-sided}}

As shown by \citet{csw:17}, separability reduces the two-sided matching
problem to a collection of one-sided discrete choice problems that are only
linked through a surplus-splitting formula. Men of a given group $x$ match
with women of different groups, since each man $i$ has idiosyncratic $%
\bm{\varepsilon
_{i\cdot}}$ shocks. \label{diffAmongSameTypes} But as a consequence of the
separability assumption, if a man of group $x$ matches with a woman of group
$y$, he would be equally well-off with any other woman of this group%
\footnote{%
Provided of course that she in turn ends up matched with a man of group $x$.}.

We now state this result more rigorously:

\begin{proposition}[Splitting the surplus]\label{prop:splitsurplus}
  Under Assumption~\ref{ass:separ}, there exist   $%
\bm{U}=(U_{xy})$ and $\bm{V}=(V_{xy})$ for $(x,y)\in A$, with $U_{x0}=V_{0y}=0$,  such that at any stable matching $%
(\mu_{xy})$,

(i) A man $i$ of group $x$ marries a woman of group $y^{\ast}\in \mathcal{Y}$
iff $y^\ast$  maximizes  $U_{xy}+\varepsilon_{iy}$ over $y\in \mathcal{Y}_{0}
$. If the maximum is achieved at $y=0$, this man remains  single. Man $i$'s
utility $\tilde{u}_i$ is the value of the maximum.

(ii) A woman $j$ of group $y$ marries a man of group $x^{\ast}\in \mathcal{X}
$  iff $x^\ast$  maximizes  $V_{xy}+\eta_{xj}$ over $x\in \mathcal{X}_{0}$.
If the maximum is achieved  at $x=0$, this woman remains  single. Woman $j$'s utility $\tilde{v}_j$ is the value of the maximum.

(iii) $U_{xy}+V_{xy} \geq \Phi_{xy}$ for all $(x,y)\in \mathcal{A}$, with
equality if $\mu_{xy}>0$.
\end{proposition}

Before we solve the two-sided matching problem, we need to derive results on
one-sided discrete choice problems. Since these results are of independent
interest, we step back from the matching problem and consider the classic
problem of an individual who chooses from a set of alternatives $y\in
\mathcal{Y}_0=\mathcal{Y}\cup \{0\}$ whose utilities are $U_y+\varepsilon_y$.
 We assume that the vector $\bm{\varepsilon}=(\varepsilon_y)_{y\in \mathcal{%
Y}_0}$ has a distribution $\mathbb{P}$; without loss of generality, we
impose $U_0=0$ and we denote $\bm{U}=(U_1,\ldots, U_{\abs{Y}})$.

\subsection{Social surplus in discrete choice models\label{par:ssdc}}

We first show that the ex-ante indirect surplus can be expressed as a sum of
two terms: the weighted sum of the mean utilities, and a \emph{generalized
entropy of choice} which stems from the unobservable heterogeneity. We
will provide two useful characterizations of the generalized entropy
function, one as the convex conjugate of the ex-ante indirect utility, and
the other one as the solution to an optimal transport problem \citep[see][for an introduction]{otme}. To the best of our knowledge, these
results are new.

The average utility of the agent is
\begin{equation}
G(\bm{U})=\mathbb{E}_{\bm{P}}\max_{y\in \mathcal{Y}_{0}}(U_{y}+%
\varepsilon_{y})  \label{eq:defG}
\end{equation}%
where the expectation is taken over the random vector $\bm{\varepsilon}%
=(\varepsilon _{0},\ldots ,\varepsilon _{\left\vert \mathcal{Y}\right\vert
})\sim \bm{P}$. The function $G$ is known as the \emph{Emax operator} in
the discrete choice literature\footnote{%
The Emax operator is available in closed-form in classical
instances like McFadden's generalized extreme value class \citep{McFadden:78}. In other cases, one needs
to use numerical integration; see~\citet{train} and references therein.
}.

\bigskip

Note that as the expectation of the maximum of linear functions of the $%
(U_{y})$, $G$ is a convex function of $\bm{U}$. Now consider a large
population of individuals who face the same mean utilities and draw
independent $\bm{\varepsilon}_i$ from $\mathbb{P}$.  Let $Y_{i}^\ast\in
\mathcal{Y}_{0}$ denote the optimal choice of  individual $i$; then
\begin{equation}
G(\bm{U})=\mathbb{E}_{\bm{P}} \left(
U_{Y_{i}^\ast}+\varepsilon_{i,Y_{i}^\ast} \right) =\sum_{y\in \mathcal{Y}}
\mu_y U_{y}+ \mathbb{E}_{\bm{P}}\left(\varepsilon_{i, Y_{i}^{\ast }}\right),
\label{probabilisticExpressionG}
\end{equation}
where $\mu_{y}$ is the proportion of individuals who choose alternative $y$.

\subsection{Generalized entropy of choice\label{par:gal-entropy-of-choice}}

Our analysis gives a prominent role to a classical concept in convex
analysis: the \emph{Legendre-Fenchel transform} of $G$. Let $\bm{\mu}%
=(\mu_1,\ldots,\mu_{\abs{\mathcal{Y}}})$; we define
\begin{equation}
G^{\ast}(\bm{\mu})=\sup_{\bm{\tilde{U}}=(\tilde{U}_{1},\ldots ,\tilde{U}_{%
\abs{\mathcal{Y}}})} \left(\sum_{y\in \mathcal{Y}}\mu_{y}\tilde{U}_{y}-G(%
\bm{\tilde{U}})\right)  \label{eq:constrG}
\end{equation}%
\label{pg:fenchel} whenever $\sum_{y\in \mathcal{Y}}\mu _{y}\leq 1$, and $%
G^{\ast }(\bm{\mu})=+\infty$ otherwise. Note that the domain of $G^{\ast}$
is the set of $\bm{\mu}$ that can be interpreted as vectors of choice
probabilities of alternatives in $\mathcal{Y}$. As the supremum of a set of
linear functions of $\bm{\mu}$, $G^{\ast}$ is a convex function.

We will see in Example~\ref{ex:nested-logit} that in the logit setting, $%
-G^{\ast }$ is the usual entropy function. This motivates the following
definition:

\begin{definition}\label{def:gcirc}
  We call the function $-G^{\ast }$ the \emph{generalized
entropy of choice}.
\end{definition}

\bigskip

The theory of convex duality implies that since $G$ is convex, it is
reciprocally the Legendre-Fenchel transform of $G^{\ast}$:
\begin{equation}
G(\bm{U})=\sup_{\bm{\tilde{\mu}}= ( \tilde{\mu}_{1},\ldots, \tilde{\mu}_{%
\abs{\mathcal{Y}}} ) } \left( \sum_{y\in \mathcal{Y}} \tilde{\mu}%
_{y}U_{y}-G^{\ast }(\bm{\tilde{\mu}}) \right).  \label{biconjugate}
\end{equation}

Assume that $\bm{\mu}$ attains the supremum in~\eqref{biconjugate}. Then
\begin{equation*}
G(\bm{U})+G^{\ast}(\bm{\mu})=\sum_{y\in \mathcal{Y}}\mu_{y}U_{y};
\end{equation*}%
and comparing with~\eqref{probabilisticExpressionG} shows that
\begin{equation}
G^{\ast }\left( \bm{\mu}\right) =-\mathbb{E}_{\bm{P}}\left( \varepsilon
_{iY_{i}^{\ast }}\right) .  \label{Fenchel}
\end{equation}

Therefore $-G^{\ast }\left(\bm{\mu}\right)$ is just the average
heterogeneity that is required to rationalize the conditional choice
probability vector $\bm{\mu}$. The following result goes beyond formula~%
\eqref{Fenchel} and allows us to provide a useful characterization of the
generalized entropy of choice. It shows that it can be computed by solving
an optimal transport problem.

\begin{theorem}[Characterization of the generalized entropy of choice]\label{thm:entropy-is-ot}
Let $\bm{\mu}=(\mu_1,\ldots,\mu_{\abs{\mathcal{Y}}%
})$ with $\sum_{y\in \mathcal{Y}}\mu_y \leq 1$, and denote $%
\mu_0=1-\sum_{y\in\mathcal{Y}}\mu_y$. Let $\mathcal{M}\left(\bm{\mu},\bm{P}%
\right)$ denote the set of probability distributions $\pi$ of the random
joint vector $\left( \bm{Y},\bm{\varepsilon}\right)$, where $\bm{Y}\sim
(\mu_0,\bm{\mu})$ is a random element of $\mathcal{Y}_{0}$, and $%
\bm{\varepsilon}\sim \bm{P}$ is a random vector of $\mathbb{R}^{%
\abs{\mathcal{Y}_{0}}}$.

Then $-G^{\ast }(\bm{\mu})$ is the value of the optimal transport problem
between the distribution $(\mu_0,\bm{\mu})$ of $\bm{Y}$ and the distribution
$\bm{P}$ of $\bm{\varepsilon}$, when the surplus is given by $\varepsilon _{%
\bm{Y}}$. That is,
\begin{equation}
-G^{\ast }(\bm{\mu})=\sup_{\pi \in \mathcal{M}\left(\bm{\mu },\bm{P}\right) }%
\mathbb{E}_{\pi}\left(\varepsilon _{\bm{Y}}\right) \text{.}  \label{assPb}
\end{equation}
\end{theorem}

Since a discretized version of problem~\eqref{assPb} can be solved by
efficient linear programming algorithms, it provides us with a practical
solution to the computation of generalized entropy for quite general
distributions of unobserved heterogeneity. Several applications of this result to useful classes of distributions are
given below in Section~\ref{par:example}.

\subsection{Identification of discrete choice models\label%
{par:ident-discrete-choice}}

The generalized entropy of choice function $-G^{\ast}$ is our gateway to
identifying the mean utilities. Let us first give the intuition of our
result. Assume that the distribution $\mathbb{P}$ is known and that it
generates functions $G$ and $G^{\ast}$ that are continuously differentiable
-- this is the case, in particular, when the distribution has a density with
full support. By the Daly-Zachary-Williams theorem\footnote{%
\citet{Williams:77} and \citet{DalyZachary:78}.}, we know that the
derivative of the average maximized utility of an agent with respect to $%
U_{y}$ is equal to the  probability that this agent chooses the
corresponding alternative $y$, that is%
\begin{equation}
\frac{\partial G}{\partial U_{y}}(\bm{U})=\mu_y.
\end{equation}%
This is simply an application of the envelope theorem to~\eqref{eq:defG}. We
can also use it on~\eqref{eq:constrG}; this gives
\begin{equation}
\frac{\partial G^{\ast}}{\partial \mu_{y}}(\bm{\mu} )=U_{y}  \label{eq:invG}
\end{equation}
where $U_y$ achieves the maximum in~\eqref{eq:constrG}. By the Fenchel
duality theorem\footnote{See e.g.\ \citet[p.\ 211]{hl:01}.}, these two sets
of conditions are equivalent. As a consequence, for any fixed distribution
of $\bm{\erps}$ (which determines the shape of $G$ and $G^\ast$), the mean
utilities $\bm{U}$ are identified from $\bm{\mu}$, the observed matching
patterns of the agents.

\begin{theorem}[Identifying the mean utilities]\label{thm:onesidedident}
    Let $\bm{\mu}=(\mu_1,\ldots,\mu_{\abs{\mathcal{Y}}%
})$ with $\sum_{y\in \mathcal{Y}}\mu_y \leq 1$; $U_0=0$; and $\bm{U}%
=(U_1,\ldots,U_{\abs{\mathcal{Y}}})$. If $\bm{P}$ has full support and is
absolutely continuous with respect to the Lebesgue measure, the following
statements are equivalent:

\begin{enumerate}
\item for every $y\in\mathcal{Y}$, $\mu _{y}={\displaystyle \frac{ \partial G%
}{\partial U_{y}}(\bm{U})}$

\item for every $y\in\mathcal{Y}$, $U_{y}={\displaystyle \frac{\partial
G^{\ast}}{\partial \mu _{y}}(\bm{\mu})}$

\item there exists a scalar function $u\left(\bm{\varepsilon}\right)$,
integrable with respect to $\bm{P}$, such that $\left(u,\bm{U}\right) $ are
the unique minimizers of the dual problem to~\eqref{assPb}, that is of:
\begin{eqnarray}
-G^{\ast }(\bm{\mu})=\min_{\bar{U},\bar{u}} &&\int \bar{u}\left(%
\bm{\varepsilon}\right) d\bm{P}\left( \bm{\varepsilon}\right) -\sum_{y\in
\mathcal{Y}}\mu _{y}\bar{U}_{y}  \label{mkPbForDC} \\
s.t.~ &&\bar{u}\left(\bm{\varepsilon}\right) -\bar{U}_{y}\geq \varepsilon
_{y}\;\;\;\; \forall y\in \mathcal{Y},\forall \bm{\varepsilon}\in \mathbb{R}%
^{\mathcal{Y}_{0}}  \notag \\
&&\bar{U}_{0}=0.  \notag
\end{eqnarray}
\end{enumerate}
\end{theorem}

Since the functions $G$ and $G^{\ast}$ are convex, they are differentiable
almost everywhere. Our assumption on $\bm{P}$ makes them continuously
differentiable. This is not essential to our approach\footnote{%
It holds in all popular specifications, including the multinomial logit
model of course.}, but it makes for simpler formul\ae\ and numerical
computations.

Part 1 of Theorem~\ref{thm:onesidedident} is well-known in the discrete
choice literature, and we only restate it for completeness. Parts~2 and~3 do
not seem to have appeared before our paper. The only related prior results
we could find are the inversion formul\ae\ of~\citet{HotzMiller:93} and~%
\citet{ArcidiaconoMiller:11} for dynamic discrete choice models; but their
scope is much more restricted since they only apply to multinomial logit and
to GEV models, respectively. In contrast, parts 2 and 3 provides a
constructive method to identify $U_{y}$ based on the conditional choice
probabilities $\bm{\mu}$, as the solution to a convex optimization problem
(part 2) which is in fact an optimal transport problem (part 3). The
intuition behind part~3 is simply that each observed choice probability $%
\mu_{y}$ must be matched to the values of idiosyncratic preference shocks $%
\bm{\varepsilon}_{i}\sim \bm{P}$ for which $y$ is the most preferred choice.
The $\bm{U}$ are the shadow prices that support this matching. %
\citet{ChiongGalichonShum} apply our approach to dynamic discrete-choice
models.

\subsection{Examples\label{par:example}}

\begin{example}[Logit and nested logit]\label{ex:nested-logit}
   The nested logit model is a well-known
generalization of the ubiquitous (multinomial) logit model. Consider a
two-layer nested logit model where alternative $0$ is alone in a nest and
each other nest $n\in \mathcal{N}$ contains alternatives $y\in \mathcal{Y}%
\left(n\right)$. The correlation of alternatives whithin nest $n$ is proxied
by $1-\lambda_{n}^2$ (with $\lambda _{0}=1$ for the nest made of alternative $0
$). Calculations detailed in Appendix~\ref{app:nestedLogit} show that
\begin{eqnarray}
G(\bm{U}) &=&\log \left[1+\sum_{n\in \mathcal{N}}\left( \sum_{y\in \mathcal{Y%
}\left(n\right)}\exp\left(\frac{U_{y}}{\lambda _{n}}\right)\right) ^{\lambda
_{n}}\right],  \label{MNL-G} \\
G^{\ast }(\bm{\mu }) &=&\mu_0\log\mu_0+\sum_{n\in \mathcal{N}}\left( \lambda
_{n}\sum_{y\in \mathcal{Y}\left( n\right) }\mu _{y}\log \mu _{y}+\left(
1-\lambda _{n}\right) \mu _{n}\log \mu _{n}\right)  \label{MNL-Gstar}
\end{eqnarray}%
where $\mu _{0}:=1-\sum_{y\in \abs{\mathcal{Y}}}\mu_y$ and $\mu
_{n}:=\sum_{y\in \mathcal{Y}\left(n\right) }\mu _{y}$.

Moreover, $U_{y}=\lambda _{n}\log \left( \mu _{y}/\mu _{0}\right) +\left(
1-\lambda _{n}\right) \log \left( \mu _{n}/\mu _{0}\right)$.

In particular, when $\lambda _{n}=1$ for every nest $n$, we recover the
multinomial logit model:
\begin{eqnarray}
G(\bm{U}) &=&\log \left(1+\sum_{y\in \mathcal{Y}}\exp (U_{y})\right)
\label{Glogit} \\
G^{\ast }(\bm{\mu}) &=&\mu _{0}\log \mu _{0}+\sum_{y\in \mathcal{Y}}\mu
_{y}\log \mu _{y}.  \label{Gstar-logit}
\end{eqnarray}%
along with $\mu _{y}=\exp \left( U_{y}\right) /\left( 1+\sum_{y^{\prime }\in
\mathcal{Y}}\exp (U_{y^{\prime }})\right) $ and $U_{y}=\log \left( \mu
_{y}/\mu _{0}\right) $.
\end{example}

\medskip

\begin{example}[Random coefficients multinomial logit and pure
characteristics model]\label{ex:mixed-logit}
 Now consider the random coefficient logit model which
underlies much of empirical industrial organization \citep{blp:1995}. In
this case, $\bm{\varepsilon} =\bm{Z}\bm{e} +T\bm{\eta}$, where $\bm{e}$ is a
random vector on $\mathbb{R}^{d}$ with distribution $\mathbf{P}_{e}$; $\bm{Z}
$ is a $\left\vert \mathcal{Y}_{0}\right\vert \times d$ matrix; $T>0$ is a
scalar parameter, and $\bm{\eta}$ is a vector of $\abs{\mathcal{Y}}$ extreme
value type-I (Gumbel) random variables that is independent from $e$.
Appendix~\ref{app:rcl} shows that $-G^{\ast }(\bm{\mu})$ is a solution to
the regularized optimal transport problem
\begin{equation}
-G^{\ast }(\bm{\mu})= \min_{U_0=0, \bm{U} \in \mathbb{R}^{\mathcal{Y}}} %
\left[ \int T \log\sum_{y\in \mathcal{Y}_0} \exp\left(\frac{U_{y}+\left(%
\bm{Z}\bm{e}\right)_{y}}{T}\right) d\bm{P}_{e}(\bm{e}) -\sum_{y\in \mathcal{Y%
}} \mu_{y}U_{y}\right]  \label{rcl-dual}
\end{equation}%
and the vector $\bm{U}$ that attains the minimum in~(\ref{rcl-dual}) is the
solution to the identification problem.

The case $T=0$ yields the pure characteristics model of~\citet{berrypakes:07}
described at greater length in Appendix~\ref{app:pure-char}. Then $%
\bm{\varepsilon}=\bm{Z}\bm{e}$; and
\begin{equation}
-G^{\ast }(\bm{\mu})=\min_{U_0=0, \bm{U}\in \mathbb{R}^{\mathcal{Y}}}\int
\max_{y\in \mathcal{Y}_{0}}\left\{\left(\bm{Z}\bm{e}\right)
_{y}+U_{y}\right\} d\bm{P}_{\epsilon }\left( \epsilon \right) -\sum_{y\in
\mathcal{Y}}\mu_y U_{y}  \label{pure-char-dual}
\end{equation}%
is the solution to the power diagram problem \citep[see][Chapter~5]{otme}.
\end{example}

%

%

\section{Social surplus and identification in the two-sided case: matching
models\label{sec:two-sided}}
  We now return to matching models.
Proposition~\ref{prop:splitsurplus} shows that a man $i$ of group $x$ can be
modeled as choosing a partner by maximizing $(U_{xy}+\erps_{iy})$ over $y\in
\mathcal{Y}_0$ (continuing with our convention that $U_{x0}=0$). Building on
our  results on one-sided discrete choice, we define $G_{x}$ to be the
corresponding Emax function:
\begin{equation*}
G_x(\bm{U}_{x\cdot})= E_{\bm{P}_x}\max_{y\in \mathcal{Y}_0}  \left(U_{xy}+%
\erps_{iy}\right)
\end{equation*}
and the Legendre-Fenchel transform
\begin{equation*}
G^\ast_x(\bm{\nu})= \max_{\bm{U}\in \mathrm{I\kern-.17em R}^{\mathcal{Y}}}
\left(  \sum_{y\in \mathcal{Y}} \nu_y U_y -G_x(\bm{U})\right)
\end{equation*}
for $\sum_{y\in \mathcal{Y}}\nu_y \leq 1$ (and $G^\ast_x(\bm{\nu})=+\infty$
otherwise). Given group numbers $\bm{n}=(n_x)$, the \emph{aggregate welfare
of men} is
\begin{equation}
G\left(\bm{U},\bm{n}\right) =\sum_{x\in \mathcal{X}}n_{x}G_{x}\left( \bm{U}
_{x\cdot }\right);  \label{def:GU}
\end{equation}%
for $\bm{\mu}={(\mu _{xy})}_{x\in \mathcal{X}, y\in \mathcal{Y}}$, we
denote its Legendre-Fenchel transform by
\begin{equation*}
G^{\ast }\left( \bm{\mu},\bm{n}\right) =\sup_{\bm{U}\in \mathbb{R}^{\mathcal{%
\ X}\times \mathcal{Y}}}\left( \sum_{x\in \mathcal{X}, y\in \mathcal{Y}}\mu
_{xy}U_{xy}-G\left( \bm{U},\bm{n}\right) \right)
\end{equation*}
which is (minus) the generalized entropy of choice of all men. Standard
calculations show that
\begin{equation*}
G^{\ast }\left( \bm{\mu},\bm{n}\right) =\sum_{x\in \mathcal{X}
}n_{x}G_{x}^{\ast }\left( \bm{\mu _{x \cdot} / n_x}\right).
\end{equation*}

We define $H_{y}(\bm{V}_{\cdot y})$ as the Emax function on women's side.
Given group numbers $\bm{m}=(m_y)$, the aggregate welfare of women is $H(%
\bm{V},\bm{m})$; the generalized entropy of choice of women of group $y$ and
of  all women are the respective Legendre-Fenchel transforms of $H_{y}$ and $%
H$.

\subsection{Social surplus, equilibrium and entropy of matching\label%
{par:ssid}}

It has been known since~\citet{SS:72} that under perfectly transferable
utility, the stable matching  maximizes the social surplus over the set of
feasible matchings. Theorem~\ref{MainThmSurplus} provides a simple
analytical expression for the value of the optimal social surplus. We start
with an intuitive derivation of this result.

The social surplus $\mathcal{W}$ is simply the sum of the aggregate welfare
of men and the aggregate welfare of women:
\begin{equation}
\mathcal{W}=G(\bm{U},\bm{n})+H(\bm{V},\bm{m})=\sum_{x\in \mathcal{X}%
}n_{x}G_{x}(\bm{U_{x \cdot}})+\sum_{y\in \mathcal{Y}}m_{y}H_{y}(\bm{V_{\cdot
y}}).  \label{exprW}
\end{equation}%
Let $\bm{\mu}=(\mu _{xy})_{x\in \mathcal{X}, y\in \mathcal{Y}}$ be the
stable matching that corresponds to $(\bm{U},\bm{V}=\bm{\Phi}-\bm{U})$.
Summing the expressions~(\ref{biconjugate}) over $x$ gives
\begin{equation*}
G\left(\bm{U},\bm{n}\right) =\sum_{x\in \mathcal{X}, y\in \mathcal{Y}}
\mu_{xy}U_{xy}-G^{\ast}\left(\bm{\mu},\bm{n}\right);
\end{equation*}
and similarly,
\begin{equation*}
H\left(\bm{V},\bm{m}\right) =\sum_{x\in \mathcal{X}, y\in \mathcal{Y}%
}\mu_{xy}V_{xy}-H^{\ast }\left(\bm{\mu},\bm{m}\right).
\end{equation*}

As a result, the value of the social welfare can be expressed as%
\begin{equation}
\mathcal{W}=\sum_{x\in \mathcal{X},y\in \mathcal{Y}}\mu _{xy}\Phi _{xy}+%
\mathcal{E}(\bm{\mu},\bm{n},\bm{m})  \label{exprWbis}
\end{equation}%
where we have defined the \emph{generalized entropy of matching }by%
\begin{equation}
\mathcal{E}(\bm{\mu},\bm{n},\bm{m}):=-G^{\ast }(\bm{\mu},\bm{n})-H^{\ast }(%
\bm{\mu},\bm{m}).  \label{defGalEntropy}
\end{equation}

\bigskip

To simplify the exposition, we will make sure that the $G,H, G^\ast$ and $%
H^\ast$ are continously differentiable everywhere.

\begin{assumption}[Full support]\label{ass:fullsupp}
  For all $x\in \mathcal{X}$ and $y\in \mathcal{Y}$, the
distributions $\bm{P}_x$ and $\bm{Q}_y$ have full support and are absolutely
continuous with respect to the Lebesgue measure.
\end{assumption}

Theorem~\ref{MainThmSurplus} shows that the values of the optimum social
welfare and the stable matching patterns emerge from the solution to simple
convex optimization problems:

\begin{theorem}[Social surplus at equilibrium]\label{MainThmSurplus}
   Under Assumptions~\ref{ass:separ} and~\ref%
{ass:fullsupp}, for any $\bm{\Phi}$ and $\bm{r}=(\bm{n},\bm{m})$ the stable
matching $\bm{\mu}$ maximizes the social gain over all feasible matchings $%
\bm{\mu}\in \mathcal{M}(\bm{r})$, that is
\begin{equation}
\mathcal{W}\left( \bm{\Phi},\bm{r}\right) =\max_{\bm{\mu}\in \mathbb{R}^{%
\mathcal{X}\times \mathcal{Y}}}\left( \sum_{x\in \mathcal{X}, y\in \mathcal{Y%
}}\mu _{xy}\Phi _{xy}+\mathcal{E}(\bm{\mu},\bm{r})\right).
\label{eq:socialWelfare}
\end{equation}%
Equivalently, $\mathcal{W}$ is given by its dual expression%
\begin{eqnarray}
\mathcal{W}\left( \bm{\Phi},\bm{r}\right) &=&\min_{\bm{U},\bm{V}\in \mathbb{R%
}^{\mathcal{X}\times \mathcal{Y}}}\left( G(\bm{U},\bm{n})+H(\bm{V},\bm{m}%
)\right)  \label{dualSocialWelfare} \\
\mbox{s.t.}~ &&U_{xy}+V_{xy}\geq \Phi _{xy}~\forall x\in \mathcal{X},y\in
\mathcal{Y}.  \notag
\end{eqnarray}%
The optimal solution $\bm{\mu}$ to~(\ref{eq:socialWelfare}) and the optimal
solution  $(\bm{U},\bm{V})$ to~(\ref{dualSocialWelfare}) are related by
\begin{equation}
\mu _{xy}=\frac{\partial G}{\partial U_{xy}}(\bm{U},\bm{n})=\frac{\partial H%
}{\partial V_{xy}}(\bm{V},\bm{m}).  \label{condmus}
\end{equation}
\end{theorem}

The proof of this result is given in Appendix~\ref{Appx:proofs}. It calls
for a few remarks.

\emph{Remark 1.} The right-hand side of equation~(\ref{eq:socialWelfare})
gives the value  of the social surplus when the matching patterns are $%
\bm{\mu}$. Its first  term $\sum_{xy}\mu _{xy}\Phi _{xy}$ reflects
\textquotedblleft systematic  preferences\textquotedblright\ on group
characteristics, while the second  term $\mathcal{E}(\bm{\mu},\bm{r})$
reflects the effect of idiosyncratic  preferences. The market equilibrium
trades off matching on group  characteristics and matching on unobserved
characteristics. If the first  term dominates, then one recovers the linear
programming problem of~\citet{SS:72}. If on the contrary, available data
were so poor that  unobserved heterogeneity dominates ($\bm{\Phi}\simeq 0$),
then the analyst  should observe something that looks like random  matching.
Information theory tells us that entropy is a natural measure of
statistical disorder; and as we will see in Example~\ref{Ex:ChooSiow}, in
the simple case analyzed by Choo and Siow the \textquotedblleft generalized
entropy of matching\textquotedblright\ $\mathcal{E}$ is just the usual
notion of entropy, which is why we chose this term.

\emph{Remark 2.} The dual problem~(\ref{dualSocialWelfare}) explains the
\emph{destination}  of the surplus shared at equilibrium between men and
women: $n_{x}G_{x}( \bm{U_{x \cdot}})$ is the total amount of utility going
to men of type $x$,  while $m_{y}H_{y}(\bm{V_{\cdot y}})$ is the total
amount of utility going to  women of type $y$. In contrast, the primal
problem (\ref{eq:socialWelfare})  accounts for the \emph{origin} of the
surplus: $\Phi _{xy}$ originates from  the part of the surplus that comes
from the interaction between observable  characteristics in pair $xy$, while
$\mathcal{E}(\bm{\mu},\bm{n},\bm{m})$  originates from unobservable
heterogeneities in tastes.

\emph{Remark 3.} Equations~\eqref{condmus} are the first-order conditions of~%
\eqref{dualSocialWelfare}. They can be rewritten as the equality  between
the demand of men of group $x$ for women of group $y$, and the  right-hand
side is the demand of women of group $y$ for men of group $x$. In
equilibrium these numbers must both equal the number of matches between
these two groups, $\mu_{xy}$.

\emph{Remark 4.} A wealth of comparative statics results and testable
predictions can be  deduced from Theorem~\ref{MainThmSurplus}; we explore
some of them in Appendices~ \ref{sub:testable_predictions} and~\ref%
{app:compStats}.

\bigskip

\textbf{Characterizing individual and systematic utilities}. We can now
offer a characterization of equilibrium utilities, both at the individual
level and aggregated over observable groups.

\begin{proposition}[Individual and group surpluses]\label{prop:IndivSurplus}
   Let $(\bm{U},\bm{V})$ solve~%
\eqref{dualSocialWelfare}, and $U_{x0}=V_{0y}=0$. Under Assumptions~\ref%
{ass:separ} and~\ref{ass:fullsupp},

(i) A man $i$ of group $x$ who marries a woman of group $y^{\ast}$ obtains
utility
\begin{equation*}
U_{xy^{\ast }}+\varepsilon _{iy^{\ast }}=\max_{y\in \mathcal{Y}_{0}}\left(
U_{xy}+\varepsilon _{iy}\right).
\end{equation*}

(ii) The average utility of men of group $x$ is
\begin{equation*}
u_{x}=G_{x}(\bm{U_{x\cdot}})=\frac{\partial \mathcal{W}}{\partial n_{x}}(%
\bm{\Phi},\bm{r}).
\end{equation*}

(iii) Parts (i) and (ii) transpose to the other side of the market with the
obvious changes; and $U_{xy}+V_{xy}=\Phi_{xy}$ for all $x, y$.
\end{proposition}

\subsection{Identification\label{par:two-sided-identification}}

Ideally, we would want to identify nonparametrically both the matrix $%
\bm{\Phi}$ and the distributions of the error terms. This is clearly out of
reach since we only observe the matching patterns $\bm{\mu}$. We focus in
this section on the case when the distributions of the  error terms are
(assumed to be) known. Section~\ref{sec:inference} will turn to parameric
inference.

Since Proposition~\ref{prop:IndivSurplus} allowed us to decompose the
matching problem into two collections of discrete choice problems, we can
now use Theorem~\ref{thm:onesidedident} in order to identify the matching
surplus matrix $\bm{\Phi}$ as s function of the corresponding stable
matching $\bm{\mu}$:

\begin{theorem}\label{thm:Identification}
  Under Assumptions~\ref{ass:separ} and~\ref{ass:fullsupp}:

\begin{enumerate}
\item $\bm{U}$ and $\bm{V}$ are identified from $\bm{\mu}$ by
\begin{equation}
\bm{U}=\frac{\partial G^\ast}{\partial \bm{\mu}}\left(\bm{\mu}\right)
\mbox{
  and } \bm{V}=\frac{\partial H^\ast}{\partial \bm{\mu}}\left(\bm{\mu}\right)
\label{UFromMu}
\end{equation}

\item The constraint in~(\ref{dualSocialWelfare}) is always saturated: $%
U_{xy}+V_{xy}=\Phi _{xy}$ for every $x\in \mathcal{X}$ and $y\in \mathcal{Y}$%
. As a result, the matching surplus $\bm{\Phi}$ is identified by
\begin{equation}
\Phi _{xy}=-\frac{\partial \mathcal{E}}{\partial \mu _{xy}}(\bm{\mu},\bm{r}),
\label{IdentPhi}
\end{equation}
which gives for any $x \in \mathcal{X}$ and $y\in \mathcal{Y}$:
\begin{equation}
\Phi _{xy}=\frac{\partial G_{x}^{\ast}}{\partial \bm{\mu}_{y\vert x}}\left( %
\bm{\mu _{\cdot \vert x}}\right) +\frac{\partial H_{y}^{\ast}}{\partial \mu
_{x\vert y}}\left(\bm{\mu_{\cdot \vert y}}\right),  \label{identEq}
\end{equation}
where $\mu_{xy}=\mu_{y\vert x} n_x=\mu_{x\vert y}m_y$.
\end{enumerate}
\end{theorem}

\medskip

Combining Theorem~\ref{thm:onesidedident} and~\ref{thm:Identification} shows
that all of the quantities in Theorem~\ref{MainThmSurplus} can be computed
by solving simple convex optimization problems.

\begin{example}[The Choo and Siow Specification]\label{Ex:ChooSiow}
  Assume that $\bm{P}_{x}$ and $\bm{Q}%
_{y}$ are the distributions of centered i.i.d.\ standard type~I extreme
value random variables. Then the generalized entropy is
\begin{equation}
\mathcal{E}=-\sum_{\substack{ x\in \mathcal{X}  \\ y\in \mathcal{Y}_{0}}}\mu
_{xy}\log \mu _{y|x}-\sum_{\substack{ y\in \mathcal{Y}  \\ x\in \mathcal{X}%
_{0}}}\mu _{xy}\log \mu _{x|y},  \label{eq:EChooSiow}
\end{equation}%
which is a standard entropy\footnote{%
The connection between the logit model and the classical entropy function is
well known; see e.g. \citet{ADT:88}.}.

Average utilities are linked to matching patterns by $u_{x}=-\log \mu _{0|x}$
and $v_{y}=-\log \mu _{0|y}$, and surpluses are related to matching patterns
by%
\begin{equation}
\Phi _{xy}=2\log \mu _{xy}-\log \mu _{x0}-\log \mu _{0y}.  \label{eq:homoCS}
\end{equation}%
This is \citet{choo-siow:06}'s identification result, which may be more
familiar as
\begin{equation}
\mu _{xy}=\sqrt{\mu _{x0}\mu _{0y}}\exp (\Phi _{xy}/2).  \label{eq:CSeq}
\end{equation}

Define
\begin{multline}
F(\bm{u},\bm{v};\bm{\Phi},\bm{r}):= \sum_{x\in \mathcal{X}}n_{x}\left(
u_{x}+e^{-u_{x}}-1\right) +\sum_{y\in \mathcal{Y}}m_{y}\left(
v_{y}+e^{-v_{y}}-1\right) \\
+2\sum_{\substack{ x\in \mathcal{X}  \\ y\in \mathcal{Y}}}\sqrt{n_{x}m_{y}}%
e^{\frac{\Phi _{xy}-u_{x}-v_{y}}{2}}  \label{ChooSiow:SocialSurplus}
\end{multline}%
As a sum of exponentials and of linear functions, it is a globally strictly
convex function of $(\bm{u},\bm{v})$. As proved in Appendix~\ref{Appx:proofs}%
, the social welfare $\mathcal{W}(\bm{\Phi};\bm{r})$ equals its minimum
value and at the minimum,
\begin{align*}
\mu _{x0}&=n_{x}\exp (-u_{x}) \\
\mu _{0y}&=m_y\exp (-v_{y}) \\
\mu_{xy}&=\sqrt{n_{x}m_{y}}\exp \left( (\Phi _{xy}-u_{x}-v_{y})/2\right).
\end{align*}
\end{example}

\section{Computation\label{sec:computation}}

We present two methods to compute the equilibrium: min-Emax (based on
gradient descent), and IPFP (based on coordinate descent). In Appendix~\ref%
{app:computation}, we benchmark them and present a third one: linear
programming based on simulated draws.

\subsection{Min-Emax method\label{sub:gradient_descent}}

Theorem~\ref{MainThmSurplus} gave two expressions for the social surplus.
Program~\eqref{eq:socialWelfare} solves for the equilibrium matching
patterns $\bm{\mu}$. Alternatively, program~\eqref{dualSocialWelfare} solves
for the $\bm{U}$ and $\bm{V}$ utility components. Since the generalized
entropy $\mathcal{E}$ is concave and the functions $G$ and $U$ are convex,
these two programs are globally convex, with linear inequality constraints.
Under Assumption~\ref{ass:fullsupp}, none of the constraints $\bm{\mu}\in
\mathcal{M}(\bm{n},\bm{m})$ in the first program bind at the optimum since
all $\mu _{x0}$ and $\mu _{0y}$ are positive; and by part (ii) of
Proposition~\ref{thm:Identification}, the constraints $\bm{U}+\bm{V}\geq %
\bm{\Phi}$ in the second program are all saturated at the optimum. Therefore
by Theorem~\ref{MainThmSurplus}, we can obtain the equilibrium matching
patterns by solving the globally concave unconstrained maximization problem~(%
\ref{eq:socialWelfare}), and we can obtain the $\bm{U}$ and $\bm{V}$
matrices by solving its dual, the globally convex unconstrained minimization
problem
\begin{equation}
\min_{\bm{U}\in \mathrm{I\kern-.17emR}^{\mathcal{X} \mathcal{Y}}}\left(G(%
\bm{U},\bm{n})+H(\bm{\Phi}-\bm{U},\bm{m})\right) .  \label{compGD2}
\end{equation}%
Since $G=\sum n_{x}G_{x}$, where $G_{x}$ is the average value of the maximum
utility of men of group $x$, we call the method based on~\eqref{compGD2} the
\emph{min-Emax} method. Problem~\eqref{compGD2} has dimension $%
\abs{\mathcal{X}}\times \abs{\mathcal{Y}},$ is unconstrained, and has a very
sparse structure: it is easy to see that the Hessian of the objective
function contains a large number of zeroes. It only requires evaluating the $%
G_{x}$ and $H_{y}$, which is often available in closed-form; when not, we
will show later (in Appendix~\ref{app:linprogr}) how to use simulation and
linear programming to approximate the problem. As~\eqref{compGD2} is
globally convex, a descent algorithm converges nicely under weak conditions%
\footnote{%
As would other algorithms---see~\cite{boydvdb:04}.}; each of its iterations
consists of updating $\bm{U}$ so as to reduce the excess demand of $x$ for $y
$ for instance by decreasing $U_{xy}$, or equivalently increasing the price $%
V_{xy}=\Phi _{xy}-U_{xy}$ of women of group $y$ for men of group $x$.
Solving \eqref{compGD2} therefore replicates a Walrasian t\^{a}tonnement
process; we need not be concerned about its convergence since global
convexity guarantees it\footnote{%
Anorher way to see it is that the demand for partners satisfies the global
substitutes property.}.

In some cases, such as the Choo and Siow specification, the sparse structure
of the problem can be exploited very easily to reduce the dimensionality
further. The function $F$ of \eqref{ChooSiow:SocialSurplus} only has $\abs{X}%
+\abs{Y}$ arguments, rather than the $\abs{X}\times\abs{Y}$ of $G$ and $H$.
This speeds up the search for a minimum considerably---see Appendix~\ref%
{app:computation}.

\subsection{IPFP\label{sub:ipfp}}

In some applications, the number of groups $\abs{\mathcal{X}}$ and $%
\abs{\mathcal{Y}}$ is large and solving for equilibrium by minimizing (\ref%
{compGD2}) may not be a practical option. We develop here an algorithm that
extends the Iterative Projection Fitting Procedure (IPFP); it can provide a
very efficient solution if the generalized entropy $\mathcal{E}$ is easy to
evaluate.

The idea that underlies the algorithm is that the average utilities $(u_x)$
and $(v_y)$ of the groups of men and women play the role of prices that
equate demand and suppply. Accordingly, we adjust the prices alternatively
on each side of the market. First we fix the prices $(v_{y})$ and we find
the prices $(u_{x})$ such that the demands of women for partners clear the
markets for men of each group, in the sense that $\sum_{y\in \mathcal{Y}%
}\mu_{xy}+\mu_{x0}=n_x$ for each $x\in \mathcal{X}$. Then we fix these new
prices $(u_{x})$ and we find the prices $(v_{y})$ such that the demands of
men for partners clear the markets for women of each group $y\in \mathcal{Y}$%
; and we iterate. This is a \emph{coordinate descent} procedure. As its name
indicates, the Iterative Projection Fitting Procedure was designed to find
projections on intersecting sets of constraints, by projecting iteratively
on each constraint\footnote{%
It is used for instance to impute missing values in data (and known for this
purpose as the RAS method.)}. We describe the algorithm in full detail in
Appendix~\ref{app:computation}, and we prove its convergence there.

\begin{theorem}
\label{thm:convergence} Under Assumptions~\ref{ass:separ} and \ref%
{ass:fullsupp}, the IPFP algorithm converges to the solution $\bm{\mu}$ of~%
\eqref{eq:socialWelfare} and to the corresponding average utilities $\bm{u}$
and $\bm{v}$.
\end{theorem}

\bigskip

In the case of the multinomial logit Choo-Siow model of Example~\ref%
{Ex:ChooSiow} for instance, we show in Appendix~\ref{appx:compIPFP} that the
algorithm boils down to%
\begin{equation}
\left\{
\begin{array}{l}
\mu _{x0}^{\left( 2k+1\right) }=\left( \sqrt{n_{x}+\frac{A_{x}^{2}}{4}}-%
\frac{A_{x}}{2}\right) ^{2}\text{ with }A_{x}=\sum_{y\in \mathcal{Y}}\exp
\left( \frac{\Phi _{xy}}{2}\right) \sqrt{\mu _{0y}^{(2k)}} \\
\mu _{0y}^{\left( 2k+2\right) }=\left( \sqrt{m_{y}+\frac{B_{y}^{2}}{4}}-%
\frac{B_{y}}{2}\right) ^{2}\text{ with }B_{y}=\sum_{x\in \mathcal{X}}\exp
\left( \frac{\Phi _{xy}}{2}\right) \sqrt{\mu _{x0}^{(2k+1)}}%
\end{array}%
\right.  \label{CS-ipfp-update}
\end{equation}

We tested the performance of our proposed algorithms on an instance of the
Choo and Siow model; we report the results in Appendix~\ref{app:computation}%
. The IPFP algorithm is extremely fast compared to standard optimization or
equation-solving methods. The min-Emax method of~\eqref{compGD2} is slower
but it still works very well for medium-size problems, and it is applicable
to all separable models.

\section{Parametric Inference\label{sec:inference}}

We assume in this section that all observations concern a single matching
market; we briefly discuss approaches that use several markets in Appendix~%
\ref{app:tradeoff}. While the formula in Theorem~\ref{MainThmSurplus} (i)
gives a straightforward estimator of the systematic surplus function $%
\bm{\Phi}$, with multiple payoff-relevant observed characteristics $x$ and $y
$ it is likely to result in large standard errors when matching patterns are
estimated from data on a finite number of matches. In addition, we do not
know the distributions $\bm{P}_{x}$ and $\bm{Q}_{y}$. Both of these remarks
point to the need for a parametric model in most applications. Such a model
would be described by a family of joint surplus functions $\Phi _{xy}^{%
\bm{\lambda}}$ and distributions $\bm{P}_{x}^{\bm{\lambda}}$ and $\bm{Q}%
_{y}^{\bm{\lambda}}$ for $\bm{\lambda}$ in some finite-dimensional parameter
space $\Lambda $.

In matching markets, the sample may be drawn from the population at the
individual level or at the household level. In the former case, each man or
woman in the population is a sampling unit; in the latter, all individuals
in a household are sampled. Household-based sampling is the norm in
population surveys and we will assume it here: our sample consists of a
predetermined number $H$ of households, some of which consist of a single
man or woman and some of which consist of a married couple. Such a sample
will have $\hat{S}=\sum_{x}\hat{N}_{x}+\sum_{y}\hat{M}_{y}$ individuals,
where $\hat{N}_{x}$ (resp.\ $\hat{M}_{y}$) denotes the number of men of
group $x$ (resp.\ women of group $y$) in the sample. Since sampling is at the
household level, for any given value of $H$ the numbers $\hat{\bm{N}}$ and $%
\hat{\bm{M}}$ of men and women of each group the sample are random: if for
instance we happen to draw many households with single men, then the number
of men in the sample will be large.

We will denote $\hat{n}_{x}=\widehat{N}_{x}/\hat{S}$ and $\hat{m}_{y}=%
\widehat{M}_{y}/\hat{S}$ the respective empirical frequencies of types of
men and women. We group them in $\bm{\hat{r}}=(\bm{\hat{n}},\bm{\hat{m}})$;
and we let $\hat{\mu}_{xy}$ denote the observed number of matches between
men of group $x$ and women of group $y$, which satisfy the usual margin
equations
\begin{equation}
\left\{
\begin{array}{l}
\sum_{y\in \mathcal{Y}}\mu _{xy}^{\bm{\lambda}}+\mu _{x0}^{\bm{\lambda}}=%
\hat{n}_{x} \\
\sum_{x\in \mathcal{X}}\mu _{xy}^{\bm{\lambda}}+\mu _{0y}^{\bm{\lambda}}=%
\hat{m}_{y}%
\end{array}%
\right.  \label{eq:MarginConstraints}
\end{equation}
We assume that this dataset is drawn from a population where matching was
generated by the parametric model above, with true parameter vector $%
\bm{\lambda}_{0}$. Recall the expression of the social surplus:%
\begin{equation*}
\mathcal{W}(\bm{\Phi} ^{\bm{\lambda} },\hat{\bm{r}})=\max_{\bm{\mu} \in
\mathcal{M}\left(\hat{\bm{r}}\right) }\left(\sum_{x,y}\mu _{xy}\Phi _{xy}^{%
\bm{\lambda} }+\mathcal{E}^{\bm{\lambda} }\left(\bm{\mu},\hat{\bm{r}}%
\right)\right).
\end{equation*}

Let $\bm{\mu}^{\bm{\lambda}}(\hat{\bm{r}})$ be the stable matching for
parameters $\bm{\lambda}$ and margins $\hat{\bm{r}}$. We have shown in
Section~\ref{sec:computation} how it can be computed efficiently. We now
focus on statistical inference on $\bm{\lambda}$. We propose three methods:
maximum likelihood, a moment matching method, and a minimum distance
estimator.

\subsection{Maximum Likelihood estimation\label{par:MLE}}

Estimation requires that we first compute the optimal matching with
parameters $\bm{\lambda}$ for given populations of men and women. To do
this, we take the numbers $\hat{n}_x$ and $\hat{m}_y$ as fixed; that is, we
impose the constraints~\eqref{eq:MarginConstraints}. The simulated number of
households
\begin{equation*}
H^{\bm{\lambda}} \equiv \sum_{(x,y)\in \mathcal{X}\times\mathcal{Y}} \mu^{%
\bm{\lambda}}_{xy}+\sum_{x\in \mathcal{X}}\mu^{\bm{\lambda}}_{x0}+
\sum_{y\in \mathcal{Y}}\mu^{\bm{\lambda}}_{0y}= \sum_{x\in \mathcal{X}}\hat{n%
}_x+ \sum_{y\in \mathcal{Y}} \hat{m}_y - \sum_{(x,y)\in \mathcal{X}\times%
\mathcal{Y}} \mu^{\bm{\lambda}}_{xy}
\end{equation*}
depends on the values of the parameters. Let $\hat{\mu}_{x0}$ (resp.\ $\hat{%
\mu}_{0y}$) be the number of single men (resp.\ women) of observed
characteristics $x$ (resp.\ $y$) in the sample; and $\hat{\mu}_{xy}$ the
number of $(x,y)$ couples\footnote{%
By construction, $\sum_{(x,y)\in \mathcal{X}\times\mathcal{Y}} \hat{\mu}%
_{xy}+\sum_{x\in \mathcal{X}}\hat{\mu}_{x0}+ \sum_{y\in \mathcal{Y}}\hat{\mu}%
_{0y}=H$.}. It is easy to see that the log-likelihood of this sample can be
written as
\begin{equation*}
\log L\left(\bm{\lambda}\right) =\sum_{x\in \mathcal{X}}\sum_{y\in \mathcal{Y%
}}\hat{\mu}_{xy}\log \frac{\mu _{xy}^{\bm{\lambda}}}{H^{\bm{\lambda}}}
+\sum_{x\in \mathcal{X}}\hat{\mu}_{x0}\log\frac{\mu_{x0}^{\bm{\lambda}}}{H^{%
\bm{\lambda}}} +\sum_{y\in \mathcal{Y}}\hat{\mu}_{0y}\log\frac{\mu_{0y}^{%
\bm{\lambda}}}{H^{\bm{\lambda}}}.  \label{logLikelihood}
\end{equation*}

\label{page:ExMLE}The maximum likelihood estimator $\hat{\bm{\lambda}}^{MLE}$
given by the maximization of $\log L$ is consistent, asymptotically normal,
and asymptotically efficient under the usual set of assumptions.

\subsection{Moment-based estimation in semilinear models\label%
{par:linearModel}}

Maximum likelihood estimation allows for joint parametric estimation of the
surplus function and of the unobserved heterogeneity. However, the
log-likelihood may have several local extrema and it may be hard to
maximize. We now introduce an alternative method, which is computationally
very efficient but can only be used under two additional conditions. First,
the distribution of the unobservable heterogeneity must be
parameter-free---as it is in \citet{choo-siow:06} for instance; or at least
we conduct the analysis for fixed values of its parameters. Second, the
parametrization of the $\bm{\Phi}$ matrix must be linear in the parameter
vector:
\begin{equation}
\Phi _{xy}^{\bm{\lambda}}=\sum_{k=1}^{K}\lambda _{k}\phi _{xy}^{k}
\label{ParamModel}
\end{equation}%
where the parameter $\bm{\lambda}\in \mathbb{R}^{K}$, and $\bm{\tilde{\phi}}%
:=(\bm{\phi^{1}},\ldots ,\bm{\phi^{K}})$ are $K$ known linearly independent
\emph{basis surplus vectors}. If the number of basis surplus vectors is rich
enough, this can approximate any surplus function. The \emph{moment-matching
estimator} of $\bm{\lambda}$ we propose in this section simply matches the
moments predicted by the model with the empirical moments; that is, it
solves the system
\begin{equation}
\sum_{\substack{ x\in \mathcal{X}  \\ y\in \mathcal{Y}}}\hat{\mu}_{xy}\phi
_{xy}^{k}=\sum_{\substack{ x\in \mathcal{X}  \\ y\in \mathcal{Y}}}\mu
_{xy}^{\lambda }\phi _{xy}^{k}\;\mbox{for all}\;k.
\label{eq:equatingMoments}
\end{equation}%
Then the moment-matching estimator is
\begin{equation}
\hat{\bm{\lambda}}^{MM}:=\arg \max_{\bm{\lambda}\in \mathbb{R}^{K}}\left(
\sum_{\substack{ x\in \mathcal{X}  \\ y\in \mathcal{Y}}}\hat{\mu}_{xy}\Phi
_{xy}^{\bm{\lambda}}-\mathcal{W}\left( \bm{\Phi}^{\bm{\lambda}},\hat{\bm{r}}%
\right) \right).  \label{maxProgr}
\end{equation}%
Since $\mathcal{W}$ is convex in $\bm{\Phi}$ and $\bm{\Phi^\lambda}$ is
linear in $\bm{\lambda}$, the objective function in this program is globally
concave. Moreover, equation~\eqref{eq:socialWelfare} shows that the
derivative of $\mathcal{W}$ with respect to $\Phi_{xy}$ is the corresponding
$\mu_{xy}$. It follows that the first-order conditions associated with~%
\eqref{maxProgr} are~\eqref{eq:equatingMoments}. Appendix~\ref%
{app:additional-estimation} shows how to derive a specification test from
this program.

We show in \cite{gs-glm} that in the case of the~\cite{choo-siow:06} model,
the moment matching estimator can be reformulated as a generalized linear
model and estimated by a Poisson regression with two-sided fixed effects.

\subsection{Minimum distance estimation}

Finally, one can use~\eqref{IdentPhi} as the basis for a minimum distance
estimator. That is, we write a mixed hypothesis as
\begin{equation*}
\exists \bm{\lambda}, \; \; \bm{D}^{\bm{\lambda}} \equiv \bm{\Phi}^{%
\bm{\lambda}}+\frac{\partial \mathcal{E}^{\bm{\lambda}}}{\partial \bm{\mu}};
\end{equation*}
and we choose $\hat{\bm{\lambda}}$ to minimize $\norm{\bm{D}^{\bm{\lambda}}}%
^2_{\bm{\Omega}}$ for some positive definite matrix $\bm{\Omega}$. If we
make the efficient choice $\bm{\Omega} = \left(V\bm{D}^{\bm{\lambda}%
}\right)^{-1}$, the minimized value of the squared norm follows a $\chi^2(p)$
if the model is well-specified, where $p=\abs{X}\times \abs{Y}-\mbox{dim}(%
\bm{\lambda})$.

This is a particularly appealing strategy if the distributions $\mathbb{P}_x$
and $\mathbb{Q}_y$ are parameter-free and the surplus matrix $\bm{\Phi}^{%
\bm{\lambda}}$ is linear in the parameters, as the minimum distance
estimator can then be implemented by linear least-squares.

\section{Empirical Application\label{sec:appli}}

We tested our methods on Choo and Siow's original dataset, which they used
to evaluate the impact of the \textit{Roe vs Wade\/} 1973 Supreme Court
abortion ruling on marriage patterns and on both genders' marriage market
surpluses. A detailed description of the data can be found in Appendix~\ref%
{Appx:CSdata}. \cite{choo-siow:06} exploited two waves of surveys: one from
the years 1970 to 1972, and one for 1980 to 1982. They distinguished those
states in which abortion was already liberalized (the ``reform states'')
from those where the Supreme Court ruling implied major legal changes. Our
focus here is not on reexamining the effect of the ruling. We aim to test
their chosen specification (a fully flexible surplus $\bm{\Phi}$ and iid
type I EV errors) against some of the many other specifications that our
analysis allows for. To do this, we select one of their subsamples. We chose
to work with the 1970s wave, when couples married younger. This allows us to
focus on the age range 16 to 40 with little loss\footnote{\cite{choo-siow:06}
allowed for marriage from ages 16 to 75. Our sample is 12\% smaller.}. We
use the ``non-reform states'' subsample, which has 224,068 observations
representing 13.3m individuals.

Our Proposition~\ref{thm:Identification} implies that if we let the surplus $%
\bm{\Phi}$ be non-parametric as in \cite{choo-siow:06}, all separable models
achieve an exact fit to the data. In that sense, there is no way to choose
between say a nested logit model and a Random Scalar Coefficients model. To
circumvent this issue, we proceed in two steps. First, we keep Choo and
Siow's choice of error distribution but we fit several hundred parametric
models of surplus to the data, using the semilinear model described in~\ref%
{par:linearModel}. We use the Bayesian Information Criterion (BIC) to select
a set of basis functions $(\phi^k_{xy})$, as described in Appendix~\ref%
{Appx:estims}. We then fit alternative specifications to the data, using
this set of basis functions and different distributions for the error terms.

\subsection{Heteroskedastic Logit Models\label{par:heteroskedastic-appli}}

We focused on specifications that allow for parameterized distributions of
the error terms $\bm{\varepsilon}$ and $\bm{\eta}$. These parameters cannot
be estimated by moment matching, which can only be used to estimate the
coefficients of the basis functions for given values of the distributional
parameters. One could maximize the resulting profile log-likelihood.
Alternatively, the moment-matching equalities can be imposed as constraints
in an MPEC approach. We have found that in practice, maximizing the
log-likelihood over all parameters (distributional and coefficients of basis
functions) worked well. This is the approach we use in the rest of this
section\footnote{%
The one difficulty we faced is in inverting the information matrix to
compute the standard errors: the matrix has one or two very small
eigenvalues that corresponds to two coefficients of the interactions of $y$
and $y^{2}$ with $D=\mathrm{1\kern-.40em1}(x\geq y)$. We held them fixed
when computing the standard errors.}.

We explored several ways of adding heteroskedasticity to our benchmark
model, while maintaining the scale normalization that is required in this
two-sided discrete choice problem\footnote{%
We normalize the standard error of $\varepsilon$ to be 1 for a man of age
28---the midpoint in our sample.}. As reported in Appendix~\ref{Appx:estims}%
, adding heteroskedasticity across genders barely improves the fit, and
deteriorates the BIC value. On the other hand, we found that introducing
heteroskedasticity on both gender and age does improve the value of the BIC.
Our preferred model in this class replaces the term $\varepsilon_{iy}
+\eta_{xj}$ with $\sigma_x \varepsilon_{iy} +\tau_y\eta_{xj}$, with $%
\sigma_x = \exp(\sigma_1 x)$, and $\tau_y = \exp(\tau_0)$. This still quite
parsimonious model yields a noticeable improvement in the fit: $+37.5$
points of loglikelihood, and $+25.2$ points on BIC. The two distributional
parameters are precisely estimated.

Our estimates give $\tau _{y}=0.47$ and a $\sigma _{x}$ that increases from $%
0.19$ at age 16 to $5.29$ at age 40; or, to focus on more likely ages at
marriage for men in the early 1970s\footnote{%
Recall that \textquotedblleft age\textquotedblright\ is as recorded in 1970,
while marriage occurs in 1971 or 1972.}, from $0.28$ at age 18 to $0.72$ at
age 25. This large relative variation directly impacts the shares of surplus
that each partner can expect to get in a match. Simple calculations show
that in this heteroskedastic version of the \cite{choo-siow:06} model, the
average share of the man in an $(x,y)$ match is
\begin{equation*}
\frac{u_{x}}{u_{x}+v_{y}}=\frac{\sigma _{x}\log \mu _{0|x}}{{\sigma _{x}\log
\mu _{0|x}}+\tau _{y}\log \mu _{0|y}}.
\end{equation*}

\begin{figure}[tbp]
\caption{Men's Share of the Marriage Surplus in the Logit Models}
\label{fig:MenShares}
\begin{center}
\includegraphics[width=15cm]{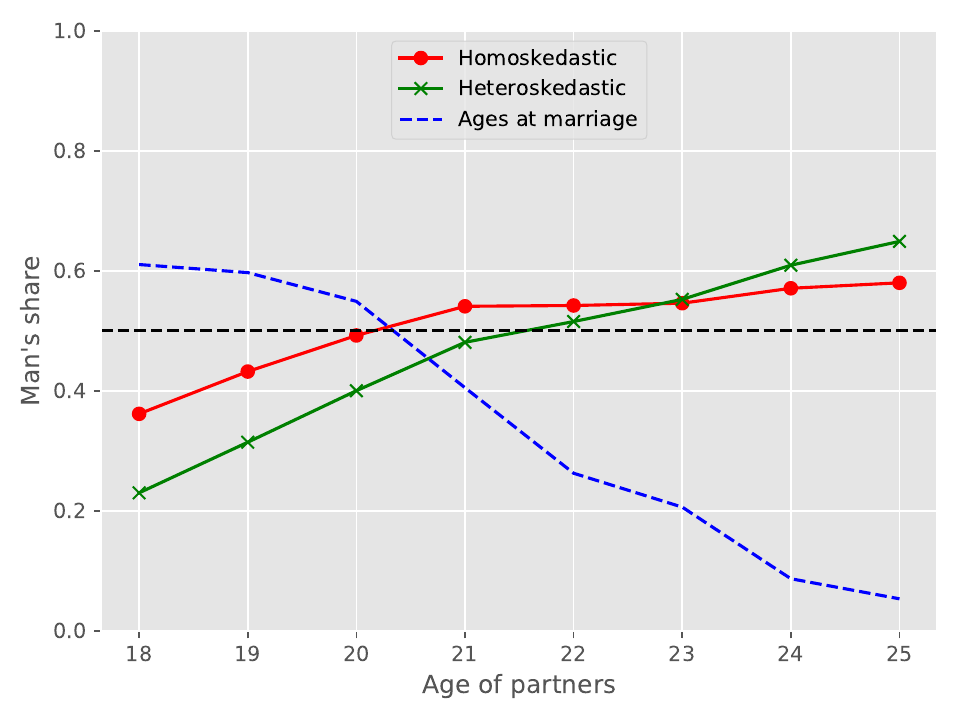} \subcaption*{The dashed blue line
indicate the number of same-age marriages. The dashed black line corresponds
to equal sharing of the surplus.}
\end{center}
\end{figure}

Figure~\ref{fig:MenShares} plots this ratio in the homoskedastic and in the
heteroskedastic models for same-age couples ($x=y$). The surplus share of
men clearly increases much more with age at marriage in the heteroskedastic
version. Since the heteroskedastic model fits the data better, this suggests
caution in interpreting the results of \cite{choo-siow:06} on the effect of
Roe vs Wade on the average utilities of men and women in marriage.

\subsection{Flexible Multinomial Logit Models\label{subsec:mnlestims}}

Nested logit models assign equal correlation between all the alternatives in
a given nest. This is not well-suited to the kind of correlations we would
like to capture\footnote{%
We did estimate a simple two-level nested logit, and we found that the
likelihood barely improves---see Appendix~\ref{Appx:estims}.}. What we need
is a specification in which the preference shock for a partner of say age 22
is more positively correlated with the preference shock for a partner of age
23 than it is with the preference shock for a partner of age 29. In order to
capture ``age-local'' correlations, we turned to the Flexible Coefficient
Multinomial Logit (FC-MNL) model of \cite{DavisSchiraldiRand2014}\footnote{%
We thank Gautam Gowrisankaran for suggesting that we use this model.}. This
specification belongs to the class of Generalized Extreme Values models that
we discussed in Appendix~\ref{Appx:GEV}. It allows for much more general
substitution patterns between the different choices of partners, and in
particular for ``age-local'' substitution patterns that we expect to find on
the marriage market.

We estimated a few models of this family, along the lines suggested by \cite%
{DavisSchiraldiRand2014}. All specifications we tried gave similar results;
we present here the results we obtained where the matrix $\bm{b}$ that
drives substitution patterns is given by
\begin{equation*}
b^x_{y,y^\prime} =
\begin{cases}
\frac{b_m(x)}{\abs{y-y^\prime}} & \mbox{ if } y\neq y^\prime \\
1 & \mbox{ if } y=y^\prime;%
\end{cases}%
\end{equation*}
where $b_m(x)$ is an affine function of the man's age. We used a similar
specification on women's side, with an affine function $b_w(y)$ divided by $%
\abs{x-x^\prime}$.

\begin{figure}[tbp]
\begin{center}
\subcaptionbox{Choo-Siow} {%
\includegraphics[width=12cm]{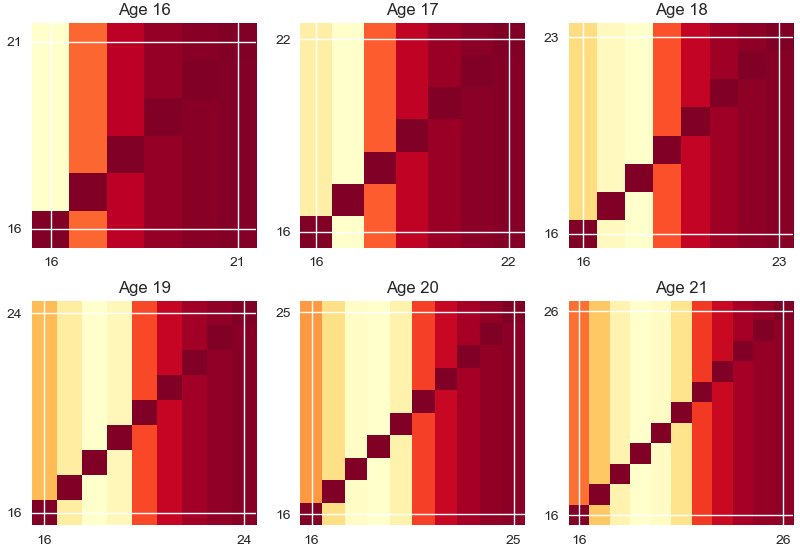}} %
\subcaptionbox{FC-MNL} {%
\includegraphics[width=12cm]{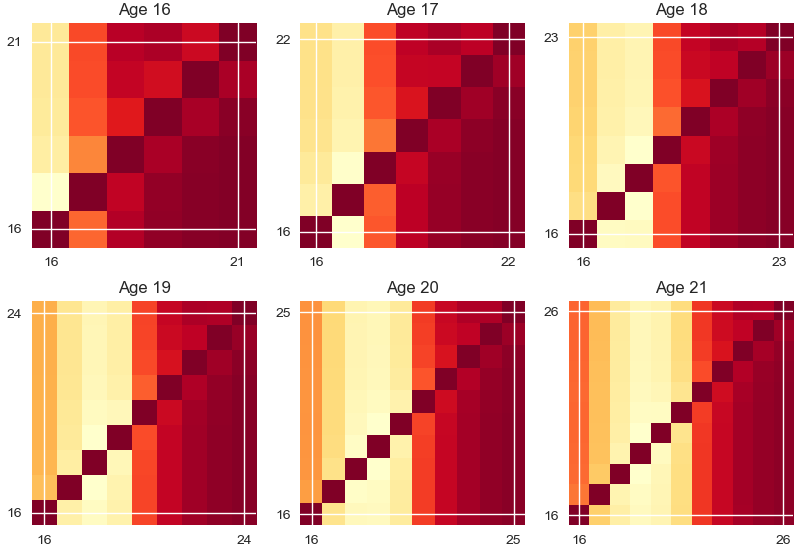}}
\end{center}
\caption{Semi-elasticities of substitution across partners: men}
\label{fig:Fcmnl_men}
\end{figure}

The maximum likelihood estimator of this model achieves a meager gain of $%
0.5 $ point of the total loglikelihood over the basic Choo and Siow model.
The affine functions are zero for the older men and women. Their estimated
values for young men and women are positive but small\footnote{%
See Appendix~\ref{Appx:estims}.}. Still, they do suggest more subtle
patterns of substitution between partners than the Choo and Siow model
allows for. We illustrate this on Figures~\ref{fig:Fcmnl_men} and~\ref%
{fig:Fcmnl_women}. Figure~\ref{fig:Fcmnl_men} for instance plots the
``demand semi-elasticities'': $\partial \log \mu_{t\vert x} / \partial V_y$
for men whose age $x$ goes from 16 (in 1970) to 21. The horizontal and
vertical axes represents partner's ages $y$ and $t$ (five on each side of $x
$, with the obvious truncation.)

\begin{figure}[tbp]
\begin{center}
\subcaptionbox{Choo-Siow} {%
\includegraphics[width=12cm]{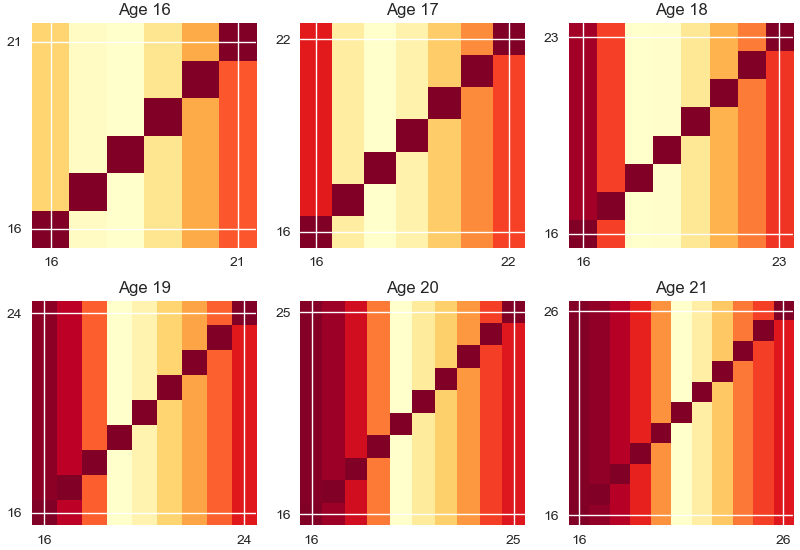}} %
\subcaptionbox{FC-MNL} {%
\includegraphics[width=12cm]{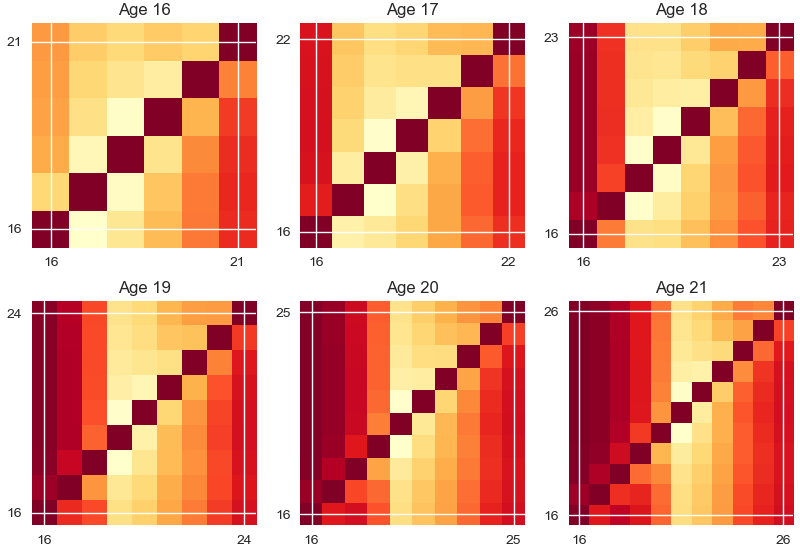}}
\end{center}
\caption{Semi-elasticities of substitution across partners: women}
\label{fig:Fcmnl_women}
\end{figure}

In the Choo and Siow model, the semi-elasticities are given by the usual
formula:
\begin{equation*}
\frac{\partial \log \mu_{t\vert x}}{\partial V_y} = \mathrm{1\kern-.40em 1}%
(y=t)-\mu_{y\vert x}.
\end{equation*}
Aside from the diagonal $y=t$, the semi-elasticities do not depend on $t$.
This appears as the vertical bands in the upper panel of Figure~\ref%
{fig:Fcmnl_men}. The lower panel shows the same semi-elasticities for the
FC-MNL model. Even with the small values of the $b$ coefficients we
estimate, richer substitution patterns appear. Figure~\ref{fig:Fcmnl_women}
tells a similar story for women.

\section*{Concluding Remarks}

Several assumptions made in our paper, in particular the separability
assumption and the large market assumption are tested on simulations by \cite%
{cns:19}. We find these simulation results reassuring about the assumptions
we have maintained in the present paper. Other assumptions we made in the
present paper can also be dispensed with. In particular, one challenge is to
extend our analysis to the case where the observable characteristics of the
partners may be continuous. This issue is addressed by \citet{dg:14} for the
Choo and Siow model, using the theory of extreme value processes; they also
propose a test of the number of relevant dimensions for the matching
problem. Our results also open the way to applications beyond the bipartite,
one-to-one matching framework of this paper. \citet{cgs:roommates} for
instance describe a formal analogy between the \textquotedblleft
roommate\textquotedblright\ (non-bipartite) problem and the bipartite
one-to-one model. We expect that this framework should also prove useful in
the study of trading on networks, when transfers are allowed (thus providing
an empirical counterpart to \citet{hatfieldkom:12} and \citet{hatfieldkom5}).
Finally, our assumption that utility is fully transferable without frictions
can be relaxed. \citet{GKW} study models with imperfectly transferable
utility and separable logit heterogeneity, while \citet{galichon-hsieh} look
at models with nontransferable utility and a similar form of heterogeneity.

\bibliography{MatchingCupid}

\newpage

\section*{Appendix\protect\appendix}

\section{Proofs\label{Appx:proofs}}

\subsection{Proof of Proposition~\protect\ref{prop:splitsurplus}}

Denote by $\left(\tilde{u}_{i}\right)$ and $\left(\tilde{v}_{j}\right)$ the
equilibrium utilities of men and women. Stability requires that for all $%
(i,j)$,

\begin{itemize}
\item $\tilde{u}_i\geq \tilde{\Phi}_{i0}$, with equality if $i$ is single

\item $\tilde{v}_j \geq \tilde{\Phi}_{0j}$, with equality if $j$ is single

\item $\tilde{u}_i+\tilde{v}_j \geq \tilde{\Phi}_{ij}$, with equality if $i$
and $j$ are matched.
\end{itemize}

Let us focus on man $i$ in group $x$. This man must be single or matched. If
he is matched, then $\tilde{u}_{i}=\max_{j}\left(\tilde{\Phi}_{ij}-\tilde{v}%
_{j}\right)$; and by Assumption~\ref{ass:separ}, we have $\tilde{\Phi}%
_{ij}=\Phi _{xy_{j}}+\varepsilon _{iy_{j}}+\eta _{xj}$ so that
\begin{equation*}
\tilde{u}_{i} =\max_{y}  \left(  \Phi _{xy}+\varepsilon  _{iy}+\max_{j: y_j
= y}\left(\eta _{xj}-\tilde{v}_{j}\right)  \right).
\end{equation*}
If he is single, then $\tilde{u}_i=\tilde{\Phi}_{i0}=\varepsilon_{i0}$.

Let $V_{xy}=\inf_{j:y_{j}=y}\left( \tilde{v}_{j}-\eta _{xj}\right)$ and $%
V_{x0}=0$. Then
\begin{equation*}
\tilde{u}_{i}=\max \left( \max_{y\in \mathcal{Y}}\left(\Phi_{xy}-V_{xy}+%
\varepsilon _{iy}\right) ,\varepsilon _{i0}\right)=\max_{y\in \mathcal{Y}_0}
\left(\Phi_{xy}-V_{xy}+\varepsilon _{iy}\right).
\end{equation*}
Considering women would lead us to define $U_{xy}=\inf_{i: x_i=x} (\tilde{u}%
_i-\varepsilon_{iy})$ and $U_{0y}=0$. Since $\tilde{\Phi}_{ij}
=\Phi_{x_iy_j}+\varepsilon_{i y_j}+\eta_{x_i j}$ cannot be larger than $%
\tilde{u}_i+\tilde{v_j}$, we obtain
\begin{equation}  \label{eq:appxphibound}
\Phi_{x_iy_j} \leq (\tilde{u}_i-\varepsilon_{iy_j})+(\tilde{v}_{j}-\eta
_{xj});
\end{equation}
taking lower bounds gives $\Phi_{xy}\leq U_{xy}+V_{xy}$. Finally, if $%
\mu_{xy}>0$ then there is a couple $(i,j)$ with $x_i=x, y_j=y$ for which~%
\eqref{eq:appxphibound} is an equality, so that $\Phi_{xy}=U_{xy}+V_{xy}$.
\qed

\subsection{Proof of Theorem~\protect\ref{thm:entropy-is-ot}}

Replacing the expression of $G$ given by \eqref{eq:defG} in formula~%
\eqref{eq:constrG} for $G^{\ast}$ gives
\begin{equation*}
-G^{\ast}(\bm{\mu})=\inf_{\bm{\tilde{U}}}\left( -\sum_{y\in \mathcal{Y}%
_{0}}\mu _{y}\tilde{U}_{y}+\mathbb{E}_{\bm{P}}\max_{y\in \mathcal{Y}%
_{0}}\left( \varepsilon _{y}+\tilde{U}_{y}\right) \right)
\end{equation*}%
where the minimization is over $\bm{\tilde{U}}$ such that $\tilde{U}_{0}=0$.
The first term in the minimand can be seen as the expectation of the random
variable $-\tilde{U}_{Y}$ under the distribution $Y\sim \mu _{Y}$. The term $%
\max_{y\in \mathcal{Y}_{0}}\left( \varepsilon _{y}+\tilde{U}_{y}\right)$ is
the maximized utility of a man with mean utilities $\bm{\tilde{U}}$ and
random taste shocks $\bm{\varepsilon}$. Alternatively, it is the value of
the problem
\begin{equation*}
\min \tilde{u} \; \mbox{ s.t.} \; \; \tilde{u}\geq \tilde{U}_y+\varepsilon_y %
\mbox{ for all } y\in\mathcal{Y}_0, \mbox{ with one equality}.
\end{equation*}
Therefore
\begin{eqnarray*}
-G^{\ast}(\bm{\mu})=\inf_{\bm{\tilde{U}, \tilde{u}} }\left( -\sum_{y\in
\mathcal{Y}_{0}}\mu _{y}\tilde{U}_{y}+\mathbb{E}_{\bm{P}}\tilde{u}(%
\bm{\varepsilon})\right) \\
\; \mbox{ s.t.} \; \; \tilde{u}(\bm{\varepsilon})\geq \tilde{U}%
_y+\varepsilon_y \mbox{ for all } y\in\mathcal{Y}_0, \bm{\varepsilon}.
\end{eqnarray*}%
Setting $V_{y}=-U_{y}$, we finally have
\begin{eqnarray*}
-G^{\ast}(\bm{\mu}) &=&\inf_{\bm{V},\tilde{u}}\left( \mathbb{E}_{\mu
_{Y}}V_{Y}+\mathbb{E}_{\bm{P}}\tilde{u}(\bm{\varepsilon})\right) \\
\mbox{s.t.}~V_{0} &=&0\;\mbox{ and }\; V_{y}+\tilde{u}(\bm{\varepsilon})\geq
\varepsilon _{y}\;\; \; \forall y\in \mathcal{Y}_{0}, %
\bm{\varepsilon} \in \mbox{supp}(\bm{P}).
\end{eqnarray*}%
This is exactly the value of the dual of an optimal transport problem in
which the margins are $\mu_{Y}$ and $\bm{P}$ and the surplus $\varepsilon_{y}
$ is split into $V_y$ and $\tilde{u}(\bm{\varepsilon})$. By the equivalence
of the primal and the dual, this yields expression~(\ref{assPb}). \qed

\subsection{Proof of Theorem~\protect\ref{thm:onesidedident}}

Since $\bm{P}$ has full support and is absolutely continuous, each $y$
achieves the maximum with positive probability; the function $G$ is strictly
convex and by the envelope theorem, it is continuous differentiable and $%
\frac{\partial G}{\partial U_y}(\bm{U})$ is the probability that $y$
achieves the maximum. This is just the classical Daly-Zachary-Williams
theorem. By the same token, $G^{\ast}$ is also strictly convex and
continuously differentiable. The general theory of convex duality---or a
straightforward application of the envelope theorem---tells us that $%
\mu_{y}=\left(\partial G/\partial \mu _{y}\right) (\bm{U})$ if and only if $%
U_{y}=\left( \partial G^{\ast }/\partial \mu _{y}\right)(\bm{\mu
  })$, which proves Part~2.

Now consider the strictly convex function $\bm{\tilde{U}}\longmapsto  G\left(%
\bm{\tilde{U}}\right) -\sum_{y\in \mathcal{Y}}\mu  _{y}\tilde{U}_{y}$.
Part~3 follows from the fact that by the envelope theorem, $\bm{U}$
minimizes the value of this function if and only if $U_{y}=\left(\partial
G^{\ast }/\partial \mu_{y}\right)(\bm{\mu})$. Since $G(\bm{U})= \mathbb{E}_{%
\bm{P}}\max_{y\in \mathcal{Y}_0} (U_y+\varepsilon_y)$, defining $\tilde{u}(%
\bm{\varepsilon})$ as in our proof of Theorem~\ref{thm:entropy-is-ot}
yields~(\ref{mkPbForDC}). \qed

\subsection{Proof of Theorem~\protect\ref{MainThmSurplus}}

In this proof we denote $\tilde{n}$ the distribution of $\left(x,%
\bm{\varepsilon}\right)$ when the distribution of $x$ is $\bm{n}$ and the
distribution of $\bm{\varepsilon}$ conditional on $x$ is $\bm{P}_{x}$.
Formally, for $S\subseteq \mathcal{X}\times \mathbb{R}^{\mathcal{Y}_{0}}$,
we get
\begin{equation*}
\tilde{n}\left(S\right) =\sum_{x}n_{x}\int_{\mathbb{R}^{\mathcal{Y}_{0}}}%
\mathrm{1\kern-.40em 1}\left( x,\bm{\varepsilon}\in S\right) d\bm{P}%
_{x}\left( \bm{\varepsilon} \right).
\end{equation*}
We define $\tilde{m}$ in the same way.

By the dual formulation of the matching problem (see \citet{GretskyOZ:1992}%
), the value of total welfare in equilibrium is obtained by solving
\begin{eqnarray}
\mathcal{W}=\inf_{\tilde{u},\tilde{v}} &&\left( \int \tilde{u}\left( x,%
\bm{\varepsilon} \right) d\tilde{n}\left( x,\bm{\varepsilon}\right) +\int
\tilde{v}\left( y,\bm{\eta} \right) d\tilde{m}\left( y,\bm{\eta} \right)
\right)  \label{dualWelfare} \\
\mbox{s.t.}~ &&\tilde{u}\left( x,\bm{\varepsilon} \right) +\tilde{v}\left( y,%
\bm{\eta} \right) \geq \Phi _{xy}+\varepsilon _{y}+\eta _{x}\;\;\forall (x,y,%
\bm{\varepsilon}, \bm{\eta})  \notag \\
&&\tilde{u}\left( x,\bm{\varepsilon} \right) \geq \varepsilon_{0}\;\;\forall
(x,\bm{\varepsilon})  \notag \\
&&\tilde{v}\left( y,\bm{\eta} \right) \geq \eta _{0}\;\;\forall (y,\bm{\eta}%
).  \notag
\end{eqnarray}%
Fix any $\tilde{u},\tilde{v}$ that satisfies all constraints in this
program. As in the proof of Proposition~\ref{prop:splitsurplus}, for $x\in
\mathcal{X}$ and $y\in \mathcal{Y}$ we define
\begin{equation*}
U_{xy}=\inf_{\bm{\varepsilon}}\left\{ \tilde{u}\left( x,\bm{\varepsilon}
\right) -\varepsilon _{y}\right\} \mbox{ and }V_{xy}=\inf_{\bm{\eta}}\left\{
\tilde{v}\left(y,\bm{\eta} \right) -\eta _{x}\right\};
\end{equation*}%
and we let $U_{x0}=V_{0y}=0$. Then $\tilde{u}\left( x,\bm{\varepsilon}%
\right) \geq \max_{y\in \mathcal{Y}_{0}}\left\{ U_{xy}+\varepsilon
_{y}\right\} $ and $\tilde{v}\left( y,\bm{\eta} \right) \geq \max_{x\in
\mathcal{X}_{0}}\left\{ V_{xy}+\eta _{x}\right\} $; and the first constraint
in~\eqref{dualWelfare} is simply $U_{xy}+V_{xy}\geq \Phi _{xy}$.
Reciprocally, assume that $U_{x0}=V_{0y}=0$ and $U_{xy}+V_{xy}\geq \Phi _{xy}
$ for all $x\in \mathcal{X} $ and $y\in \mathcal{Y}$, and define
\begin{equation*}
\tilde{u}\left( x,\bm{\varepsilon}\right) =\max_{y\in \mathcal{Y}%
_{0}}\left\{ U_{xy}+\varepsilon _{y}\right\} \;\mbox{ and }\;\tilde{v}\left(
y,\bm{\eta} \right) \geq \max_{x\in \mathcal{X}_{0}}\left\{ V_{xy}+\eta
_{x}\right\} ;
\end{equation*}%
Then $(\tilde{u},\tilde{v})$ satisfies all constraints. Therefore we can
rewrite the whole program as:
\begin{eqnarray*}
\mathcal{W} &=&\min_{U,V}\left( \int \max_{y\in \mathcal{Y}_{0}}\left\{
U_{xy}+\varepsilon _{y}\right\} d\tilde{n}\left( x,\bm{\varepsilon} \right)
+\int \max_{x\in \mathcal{X}_{0}}\left\{ V_{xy}+\eta _{x}\right\} d\tilde{m}%
\left( y,\bm{\eta} \right) \right) \\
\mbox{s.t.}~ &&U_{xy}+V_{xy}\geq \Phi _{xy}~\forall x\in \mathcal{X},y\in
\mathcal{Y} \\
\mbox{and}~ &&U_{x0}=V_{0y}=0~\forall x\in \mathcal{X},y\in \mathcal{Y}.
\end{eqnarray*}%
Now remember that we defined $G_{x}(\bm{U}_{x\cdot })=\int \max_{y\in
\mathcal{Y}_{0}}(U_{xy}+\varepsilon _{y})dP_{x}(\bm{\varepsilon})$ and $G(%
\bm{U},\bm{n})=\sum_{x}n_{x}G_{x}(\bm{U}_{x\cdot }).$ Under Assumption~\ref%
{ass:separ},
\begin{equation*}
\left\vert \max_{y\in \mathcal{Y}_{0}}\left( U_{xy}+\varepsilon _{y}\right)
\right\vert \leq \max_{y\in \mathcal{Y}_{0}}\abs{U_{xy}}+\max_{y\in \mathcal{%
Y}_{0}}\abs{\varepsilon_y}
\end{equation*}%
is integrable, so that $G_{x}$ is well-defined. It follows that
\begin{eqnarray*}
\mathcal{W} &=&\min_{U,V}\left( G\left( \bm{U},\bm{n}\right) +H\left( \bm{V},%
\bm{m}\right) \right) \\
\mbox{s.t.}~ &&U_{xy}+V_{xy}\geq \Phi _{xy}~\forall x\in \mathcal{X},y\in
\mathcal{Y}
\end{eqnarray*}%
which is expression~(\ref{dualSocialWelfare}). Introducing multipliers $(\mu
_{xy})$, this convex minimization problem can be written in a minimax form as%
\begin{eqnarray*}
\mathcal{W} &=&\min_{\bm{U},\bm{V}}\max_{\mu \geq 0}\left( G\left( \bm{U},%
\bm{n}\right) +H\left( \bm{V},\bm{m}\right) +\sum_{xy}\mu _{xy}\Phi
_{xy}-\sum_{xy}\mu _{xy}U_{xy}-\sum_{xy}\mu _{xy}V_{xy}\right) \\
&=&\max_{\bm{\mu}\geq 0}\left( \sum_{xy}\mu _{xy}\Phi _{xy}-\max_{\bm{U},%
\bm{V}}\left( \sum_{xy}\mu _{xy}U_{xy}+\sum_{xy}\mu _{xy}V_{xy}-G\left( %
\bm{U},\bm{n}\right) -H\left( \bm{V},\bm{m}\right) \right) \right)
\end{eqnarray*}%
which is (\ref{eq:socialWelfare}); and~(\ref{condmus}) are its first-order
conditions. \qed

\subsection{Proof of Proposition~\protect\ref{prop:IndivSurplus}}

Part (i) and $U_{xy}+V_{xy}=\Phi _{xy}$ restate Proposition~\ref%
{prop:splitsurplus} (since Assumption~\ref{ass:fullsupp} guarantees that $%
\mu_{xy}>0$ for all $(x,y)\in \mathcal{A}$). For part (ii), note that
applying the envelope theorem twice,
\begin{equation*}
\frac{\partial \mathcal{W}}{\partial n_{x}}=-\frac{\partial G^{\ast }}{%
\partial n_{x}}=\frac{\partial G}{\partial n_{x}}
\end{equation*}%
which equals $G_{x}$ by the definition~\eqref{def:GU}. Part (iii) is
similar. \qed

\subsection{Proof of Theorem~\protect\ref{thm:Identification}}

Part (i) follows from Theorem~\ref{thm:onesidedident} (ii). Moreover, the $%
\bm{\mu}$'s are the multipliers in (\ref{dualSocialWelfare}); since they are
all positive, the constraints must be saturated, proving (ii). \qed

\subsection{Extending the Entropy}

\label{appx:extendingentropy} Lemma~\ref{lemma:dualuv} below is instrumental
in the derivation of an efficient algorithm in Section~\ref{sub:ipfp}.

While the generalized entropy $\mathcal{E}$ defined in~\eqref{defGalEntropy}
is concave in the matching patterns $\bm{\mu}$, it is only strictly concave
when $\bm{\mu}$ has the margins $\bm{r}$ (otherwise $\mathcal{E}$ is
infinite). We will need to extend it to a function that is strictly concave
everywhere.

\begin{definition}[Extended Entropy]
Let $\mathcal{\ E}(\bm{\mu};\bm{r})$ be the generalized entropy of matching.
We say that a function $E$ \emph{extends\/} $\mathcal{E}$ if it is a
strictly concave function of $\bm{\mu}$ that coincides with over the set of
feasible matchings $\bm{\mu}\in \mathcal{M}(\bm{r})$.
\end{definition}

There are many ways of extending a given generalized entropy function $%
\mathcal{E}$. Any choice of
\begin{equation*}
E\left( \bm{\mu};\bm{r}\right) =\mathcal{E}\left( \bm{\mu},\sum_{y}\mu
_{xy}+\mu _{x0},\sum_{x}\mu _{xy}+\mu _{0y}\right) +K\left( \bm{\mu};\bm{r}%
\right)
\end{equation*}%
will work, where%
\begin{equation}
K\left( \bm{\mu};\bm{r}\right) =\sum_{x}\left\{ A_{x}\left( \sum_{y}\mu
_{xy}+\mu _{x0}\right) -A_{x}\left( n_{x}\right) \right\} +\sum_{y}\left\{
B_{y}\left( \sum_{x}\mu _{xy}+\mu _{0y}\right) -B_{y}\left( m_{y}\right)
\right\} ,  \label{defK}
\end{equation}%
and $A_{x}$ and $B_{y}$ are concave functions from $\mathbb{R}$ to $\mathbb{R%
}$. Defining $E$ in this way ensures that it coincides with $\mathcal{E}(%
\bm{\mu},\bm{r})$ for any feasible matching; and adding the term $K$ makes $%
E $ strictly concave in $\bm{\mu}$.

\begin{lemma}\label{lemma:dualuv}
   Let $E$ extend $\mathcal{E}$. For $\bm{u}\in \mathrm{I%
\kern-.17emR}^{\mathcal{X}}$ and $\bm{v}\in \mathrm{I\kern-.17emR}^{\mathcal{%
Y}}$, define $S(\bm{u},\bm{v})$ as the value of
\begin{equation}
\max_{\bm{\mu}}\left( E(\bm{\mu};\bm{r})+\sum_{x,y\in \mathcal{X}\times
\mathcal{Y}}\mu _{xy}(\Phi _{xy}-u_{x}-v_{y})+\sum_{x\in \mathcal{X}%
}(n_{x}-\mu _{x0})u_{x}+\sum_{y\in \mathcal{Y}}(m_{y}-\mu _{0y})v_{y}\right)
.  \label{eq:defS}
\end{equation}%
Then $S$ is a convex function of $(\bm{u},\bm{v})$. The social welfare $%
\mathcal{W}$ is its minimum value; the minimizers $\bm{u}$ and $\bm{v}$ are
the average utilities of the different types of men and women in
equilibrium; and the solutions $\bm{\mu}$ to~\eqref{eq:defS} at $(\bm{u},%
\bm{v})$ are the equilibrium matching patterns.
\end{lemma}

\begin{varproof}
\label{sub:proof_of_lemma_ref_dualuv} Recall from equation (\ref%
{eq:socialWelfare}) that the equilibrium matching $\bm{\mu}$ maximizes $%
\sum_{x,y}\mu _{xy}\Phi _{xy}+\mathcal{E}(\bm{\mu},\bm{r})$ over $\mu $ in $%
\mathbb{R}^{\mathcal{X}\times \mathcal{Y}}$. This can be rewritten as%
\begin{eqnarray}
\max_{\bm{\mu}} &&\sum_{x,y\in \mathcal{X}\times \mathcal{Y}}\mu _{xy}\Phi
_{xy}+E(\bm{\mu};\bm{r})  \label{maxWelfare2-primal} \\
s.t.~ &&\mu _{x0}+\sum_{y\in \mathcal{Y}}\mu _{xy}=n_{x}  \notag \\
&&\mu _{0y}+\sum_{x\in \mathcal{X}}\mu _{xy}=m_{y}.  \notag
\end{eqnarray}

Denote $a_{x}$ and $b_{y}$ the multipliers of the constraints. The
Lagrangian of~\eqref{maxWelfare2-primal} can be written as
\begin{align*}
\mathcal{L}& =\max_{\bm{\mu}}\min_{\bm{a},\bm{b}}\left(
\begin{array}{c}
\sum_{x,y\in \mathcal{X}\times \mathcal{Y}}\mu _{xy}\Phi _{xy}+E(\bm{\mu};%
\bm{r}) \\
-\sum_{x\in \mathcal{X}}a_{x}\left( \mu _{x0}+\sum_{y\in \mathcal{Y}}\mu
_{xy}-n_{x}\right) -\sum_{y\in \mathcal{Y}}b_{y}\left( \mu _{0y}+\sum_{x\in
\mathcal{X}}\mu _{xy}-m_{y}\right)%
\end{array}%
\right) \\
& =\max_{\bm{\mu}}\min_{\bm{a},\bm{b}}\left(
\begin{array}{c}
\sum_{x,y\in \mathcal{X}\times \mathcal{Y}}\mu _{xy}\left( \Phi
_{xy}-a_{x}-b_{y}\right) +E(\bm{\mu};\bm{r}) \\
+\sum_{x\in \mathcal{X}}a_{x}\left( n_{x}-\mu _{x0}\right) +\sum_{y\in
\mathcal{Y}}b_{y}\left( m_{y}-\mu _{0y}\right)%
\end{array}%
\right) .
\end{align*}%
Interchanging $\min $ and $\max $ gives $\mathcal{L}=\min_{\bm{a},\bm{b}}S(%
\bm{a},\bm{b};\bm{\Phi},\bm{r})$, where $S$ is defined in the corollary. It
is a maximum of linear functions of $\bm{a},\bm{b}$ and therefore convex.
Since the constraints are binding at the optimum, $\mathcal{W}=\mathcal{L}$.
Moreover, by the envelope theorem $\frac{\partial \mathcal{W}}{\partial n_{x}%
}=\frac{\partial S}{\partial n_{x}}=a_{x}$. By Proposition~\ref%
{prop:IndivSurplus}, this gives $a_{x}=u_{x}$; and the $\mu $'s are the
corresponding matching patterns.
\end{varproof}

\subsection{Proof of Theorem \protect\ref{thm:convergence}}\label{proofIPFP}
 We start by extending the generalized entropy $\mathcal{E}$
to a strictly concave function $E$ as explained in~\ref%
{sub:proof_of_lemma_ref_dualuv}. For notational simplicity, we now drop the
arguments $\bm{r}$ and $\bm{\Phi}$. Proposition~\ref{lemma:dualuv} shows
that the value of the matching problem is $\min_{\bm{a},\bm{b}}S(\bm{a},%
\bm{b})$. We solve for the minimum iteratively by coordinate descent. At
step $2k$, we first fix $\bm{b}=\bm{b}^{(2k)}$ and we solve the convex
minimization problem over $\bm{a}$ only:
\begin{equation*}
 \bm{a}^{(2k+1)}\equiv \arg \min_{\bm{a}}S(\bm{a},\bm{b}^{(2k)}).
\end{equation*}%
Then we keep $\bm{a}=\bm{a}^{(2k+1)}$ fixed at this new value and we solve
the minimization problem over $\bm{b}$:
\begin{equation*}
\bm{b}^{(2k+2)}\equiv \arg \min_{\bm{b}}S(\bm{a}^{(2k+1)},\bm{b}).
\end{equation*}%
We stop the iterations when $\bm{b}^{(2k+2)}$ and $\bm{b}^{(2k)}$ are close
enough. We take $\bm{u}^{(2k+1)}$ and $\bm{v}^{(2k+2)}$ to be the average
utilities, and the associated $\bm{\mu}$ to be the equilibrium matching
patterns.

\bigskip

Let us now prove that the algorithm converges to the global minimum $(\bm{u},%
\bm{v})$ of $S$. We rely on results in \citet{bauschkeborwein97}, which
builds on \citet{Csiszar}. The map $\bm{\mu}\rightarrow -E(\bm{\mu})$ is
smooth and strictly convex; hence it is a ``Legendre function'' in their
terminology. Introduce the associated ``Bregman divergence'' $D$ as
\begin{equation*}
D\left( \bm{\mu},\bm{\bar{\nu}}\right) =E\left( \bm{\bar{\nu}}\right)
-E\left( \bm{\mu}\right) +\left\langle \nabla E\left( \bm{\bar{\nu}}\right) ,%
\bm{\mu}-\bm{\bar{\nu}}\right\rangle ,
\end{equation*}%
where $\nabla $ denotes the gradient wrt $\bm{\bar{\nu}}$; and define the
linear subspaces $\mathcal{L}\left( \bm{n}\right) $ and $\mathcal{L}\left( %
\bm{m}\right) $ by
\begin{equation*}
\mathcal{L}\left( \bm{n}\right) =\{\bm{\mu}\geq 0:\forall x\in \mathcal{X}%
,~\sum_{y\in \mathcal{Y}_{0}}\mu _{xy}=n_{x}\}\text{ and }\mathcal{L}\left( %
\bm{m}\right) =\{\bm{\mu}\geq 0:\forall y\in \mathcal{Y},~\sum_{x\in
\mathcal{X}_{0}}\mu _{xy}=m_{y}\}
\end{equation*}%
so that $\mathcal{M}(\bm{r})=\mathcal{L}\left( \bm{n}\right) \cap \mathcal{L}%
\left( \bm{m}\right) $. It is easy to see that $\bm{\mu}^{(k)}$ results from
iterative projections with respect to $D$ on the linear subspaces $\mathcal{L%
}(\bm{n})$ and $\mathcal{L}(\bm{m})$:
\begin{equation}
\bm{\mu}^{(2k+1)}=\arg \min_{\bm{\mu}\in \mathcal{L}\left( \bm{n}\right)
}D\left( \bm{\mu},\bm{\mu}^{(2k)}\right) \text{ and }\bm{\mu}^{(2k+2)}=\arg
\min_{\bm{\mu}\in \mathcal{L}\left( \bm{m}\right) }D\left( \bm{\mu},\bm{\mu}%
^{(2k+1)}\right) .  \label{oddIt}
\end{equation}

By Theorem 8.4 of Bauschke and Borwein, the iterated projection algorithm
converges to the projection $\bm{\mu} $ of $\bm{\mu} ^{(0)}$ on $\mathcal{M}(%
\bm{r})$, which is also the maximizer $\bm{\mu} $ of (\ref{eq:socialWelfare}%
).

\medskip

As mentioned earlier, there are many possible ways of extending $\mathcal{E}$
to $E$, depending on the choice of the functions $A_{x}$ and $B_{y}$ in~(\ref%
{defK}). In practice, good judgement should be exercised, as the choice of
an extension $E$ that makes it easy to solve the systems in~\ref{oddIt} is
crucial for the performance of the algorithm.

\bigskip

\section{Examples of random utility models\label{App:Explicit}}

\subsection{The Generalized Extreme Value Framework\label{Appx:GEV}}

Consider a function $g:\mathbb{R}^{\mathcal{Y}_{0}}\rightarrow \mathbb{R}$
that (i) is positive homogeneous of degree one; (ii) goes to $+\infty $
whenever any of its arguments goes to $+\infty $; (iii) has partial
derivatives (outside of $\bm{0}$) at any order $k$ of sign $\left( -1\right)
^{k}$; (iv) is such that the function defined by $F\left( w_{0},\ldots,w_{%
\abs{\mathcal{Y}}}\right) =\exp \left( -g\left( e^{-w_{0}},\ldots,e^{-w_{%
\abs{\mathcal{Y}}}}\right) \right)$ is a multivariate cumulative
distribution function associated to some distribution, which we denote $%
\bm{P}$. Then introducing utility shocks $\bm{\varepsilon}\sim \bm{P}$, we
have by a theorem of \citet{McFadden:78}:%
\begin{equation}
G(\bm{w})=\mathbb{E}_{\bm{P}}\left[ \max_{y\in \mathcal{Y}_{0}}\left\{
w_{y}+\varepsilon _{y}\right\} \right] =\log g\left( e^{\bm{w}}\right)
+\gamma  \label{GEV}
\end{equation}%
where $\gamma $ is the Euler constant $\gamma \simeq 0.577$.

For any vector $\bm{p}\in \mathrm{I\kern-.17em R}^{\mathcal{Y}}$ such that $%
\sum_{y\in \mathcal{Y}}p_{y}=1$, we denote $\bm{\bar{p}}=(p_1,\ldots,p_{%
\abs{\mathcal{Y}}})$. Then
\begin{equation*}
G^{\ast }\left(\bm{\bar{p}} \right) =\log g\left(e^{\bm{w}\left(\bm{p}%
\right) }\right) +\gamma -\sum_{y\in \mathcal{Y}_{0}}p_{y}w_{y}\left(\bm{p}%
\right),
\end{equation*}

where the vector $\bm{w}\left(p\right)$ solves the system of equations
\begin{equation}
p_{y}=\frac{\partial \log g}{\partial w_{y}} \left(e^{\bm{w}}\right) \;%
\mbox{ for  all }\;y\in \mathcal{Y}_{0}.  \label{eq:GEVp}
\end{equation}

Now take a vector   $\bm{\mu}=(\mu_{y})_{y \in \mathcal{Y}}$ such that $%
\sum_{y\in \mathcal{Y}}\mu_{y}\leq 1$. The generalized entropy of choice
arising from this heterogeneity is
\begin{equation}
G^{\ast}(\bm{\mu})=\log g\left(e^{\bm{w}\left(\bm{\mu}\right)}\right)
-\sum_{y\in \mathcal{Y}_{0}}\mu _{y}w_{y}\left( \bm{\mu}\right)+\gamma.
\label{eq:GEVGcirc}
\end{equation}%
Applying the envelope theorem, the derivative of this expression with
respect to $\mu_{y}$ is $-w_{y}\left(\bm{\mu}\right)$. Therefore the $\bm{U}$
vector is identified by
\begin{equation}
U_{y}=w_{y}\left(\bm{\mu}\right).  \label{U:GEV}
\end{equation}

\subsection{The nested logit model\label{app:nestedLogit}}
We consider the two-layer nested logit model of Example~\ref{ex:nested-logit}: alternative $0$ is alone in a nest and each other nest $n\in \mathcal{N}$
contains alternatives $y\in \mathcal{Y}\left( n\right)$. The correlation of
alternatives whithin nest $n$ is proxied by $(1-\lambda _{n}^2)$.

\subsubsection{The entropy of choice of the one-sided nested logit model}
 It is
well-known that\footnote{%
We omit the Euler constant $\gamma$ from now on, as it plays no role in any
of our calculations.}
\begin{equation*}
G(\bm{U})=\log \left(1+ \sum_{n\in \mathcal{N}}\exp(I_n(\bm{U}))\right)
\end{equation*}%
where $I_n(\bm{U}) \equiv \lambda_n \log \left(\sum_{y\in \mathcal{Y}%
\left(n\right)} \exp(U_{y}/\lambda _{n})\right)$ is the \emph{inclusive
value\/} of nest $n$. For $y\in \mathcal{Y}_{n}$, this gives
\begin{equation*}
\mu_{y}=\frac{\partial G}{\partial U_{y}}(\bm{U})=\mu_{n} \times \frac{%
\exp(U_{y}/\lambda_n)}{\exp\left(I_n(\bm{U})/\lambda_n\right)},
\end{equation*}
where
\begin{equation*}
\mu _{n}:=\sum_{y\in \mathcal{Y}\left( n\right) }\mu _{y}  = \frac{
\exp\left(I_n(\bm{U})\right)}{1+\sum_{m\in \mathcal{N}} \exp\left(I_m(\bm{U}%
)\right)}.
\end{equation*}
As a result, $\log \mu _{n} = I_n(\bm{U}) - G(\bm{U})$ and $\log \mu _{y} =
\log \mu _{n} + (U_{y}-I_n(\bm{U}))/\lambda _{n}$. Moreover,
\begin{equation*}
\mu_{0} = 1-\sum_{n\in \mathcal{N}} \mu_{n}= \exp(-G(\bm{U})),
\end{equation*}
so that we can solve for
\begin{align}
G(\bm{U}) &= -\log \mu_{0}  \notag \\
I_n(\bm{U}) &= \log (\mu_{n}/\mu_{0})  \notag \\
U_{y}&=\lambda _{n}\log \frac{\mu _{y}}{\mu _{0}}+\left( 1-\lambda
_{n}\right) \log \frac{\mu _{n}}{\mu _{0}}.  \label{ident-nl}
\end{align}
Since $G^\ast(\bm{\mu})=\sum_{y\neq 0} \mu_y U_y - G(\bm{U})$ at the
optimum, this gives
\begin{align*}
G^\ast(\bm{\mu}) &=\sum_{n\in \mathcal{N}}\lambda_n \sum_{y\in \mathcal{Y}%
(n)}\mu_y \log \mu_y-\left(\sum_{y\neq 0} \mu_y\right)\log \mu_0 \\
&+ \sum_{n\in \mathcal{N}}(1-\lambda_n) \left(\sum_{y\in \mathcal{Y}%
(n)}\mu_y\right) \log \mu_n+\log \mu_0;
\end{align*}
using $\sum_{y\neq 0} \mu_y=1-\mu_0$ and $\sum_{y\in \mathcal{Y}%
(n)}\mu_y=\mu_n$, we get the generalized entropy of choice
\begin{equation*}
G^\ast(\bm{\mu}) =\sum_{n\in \mathcal{N}}  \left(  \lambda_n \sum_{y\in
\mathcal{Y}(n)}\mu_y \log \mu_y  +(1-\lambda_n) \mu_n \log \mu_n
\right)+\mu_0\log\mu_0.
\end{equation*}

\subsubsection{The two-sided nested logit model}

Now suppose that the above (indexed by $x$ as $\lambda_n^x, \mathcal{N}^x,
\mathcal{Y}^x(n)$) describes the structure of errors for men of group $x$,
and that women of group $y$ have a similar error structure with parameters $%
\nu_n^y, \mathcal{N}^y, \mathcal{X}^y(n)$. We denote $n(y;x)$ the nest of
partner group $y$ for men of group $x$, and $n(x;y)$ the nest of partner
group $x$ for women of group $y$. Then the matrix $\bm{U}$ is identified as
\begin{equation*}
U_{xy}=\lambda^x _{n(y;x)}\log \frac{\mu _{xy}}{\mu _{x0}}+\left(
1-\lambda^x _{n(y;x)}\right) \log \frac{\mu _{x,n(y;x)}}{\mu _{x0}}.
\end{equation*}
Along with the corresponding formula for $\bm{V}$, this identifies the joint
surplus as
\begin{align*}
\Phi_{xy} &= (\lambda^x _{n(y;x)}+\nu^y _{n(x;y)})\log\mu_{xy}
-\log\mu_{x0}-\log\mu_{0y}  \\
&+\left(1-\lambda^x  _{n(y;x)}\right) \log \mu
_{x,n(y;x)} + \left(1-\nu^y _{n(x;y)}\right) \log\mu _{n(x;y),y}
\end{align*}
for any given values of the parameters of the nested logit errors.

\subsection{The random coefficients logit model\label{app:rcl}}
Recall that Example~\ref{ex:mixed-logit} had $\bm{\varepsilon} =\bm{Z}\bm{e}
+T\bm{\eta}$, where $\bm{e}$ is a random vector on $\mathbb{R}^{d}$ with
distribution $\mathbf{P}_{\epsilon }$; $\bm{Z}$ is a $\left\vert \mathcal{Y}%
_{0}\right\vert \times d $ matrix; $T>0$; and $\bm{\eta}$ is an extreme
value type-I (Gumbel) random variable i.i.d. on $\mathbb{R}^{\mathcal{Y}_{0}}
$ and independent from $\bm{e}$.

By the law of iterated expectations, making use of the independence of $%
\bm{e}$ and $\bm{\eta}$, we get
\begin{eqnarray}
G\left( \bm{U}\right)  &=&\mathbb{E}\left[ \mathbb{E}\left[ \max_{y\in
\mathcal{Y}_{0}}\left\{ U_{y}+\left( \bm{Z}\bm{e}\right) _{y}+T\eta
_{y}\right\} |\bm{e}\right] \right]  \\
&=&\int G_{0}\left( U+\bm{Z}\bm{e}\right) f\left( e\right) de \label{eq:blp-as-regularized-ot}
\end{eqnarray}%
where
\begin{equation*}
G_{0}\left( \bm{U}\right) =T\log \sum_{y\in \mathcal{Y}_{0}}\exp \left(
\frac{U_{y}}{T}\right)
\end{equation*}%
is the Emax operator associated with the plain multinomial logit model. It is easy to compute its convex conjugate: $G_{0}^{\ast }\left( \bm{\pi}\right) =
  T\sum_{y}\pi _{y}\log \pi _{y}$
  if  $\sum_{y}\pi _{y}=1$, and $+\infty $ otherwise.

  We will use two well-known properties of convex conjugates  \citep[see e.g.][part E]{hl:01}:
\begin{itemize}
  \item the convex conjugate of a translated function $\bm{x}\to g_t(\bm{x})\equiv g(\bm{x}+\bm{t})$ is $g_t^\ast(\bm{y})=g^\ast(\bm{y})+\bm{y}\cdot \bm{t}$
  \item the convex
  conjugate of a sum of convex functions is the infimum-convolution of their convex conjugates:
  \[
    (f_1+f_2)^\ast(\bm{y})=\inf_{\bm{y}_1+\bm{y}_2=\bm{y}}
    \left(f_1^\ast(\bm{y}_1)+f_2^\ast(\bm{y}_2)\right).
    \]
\end{itemize}
Together, they imply that%
\begin{equation}
G^{\ast}\left(\bm{\mu}\right) =\inf_{\bm{\pi}(\cdot)\geq 0}
\left\{
\int G_{0}^{\ast}\left(\bm{\pi}(\bm{e})\right)
d\bm{P}_{\bm{e}}(\bm{e})
-\sum_{y}\int \left(Ze\right)_{y}
\pi _{y}\left(\bm{e}\right) d\bm{P}_{\bm{e}}(\bm{e}): \; \int_{\bm{e}}\pi _{y}\left(\bm{e}\right)
d\bm{P}_{\bm{e}}(\bm{e})=\mu _{y} \; \forall y
\right\}.\label{Gstar-blp}
\end{equation}%
It  follows that
\begin{eqnarray*}
-G^{\ast }\left( \bm{\mu}\right)  &=&\max_{\bm{\pi}(\cdot)\geq 0}\left\{
\sum_{y}\int \left(Ze\right) _{y}\pi _{y}\left(\bm{e}\right)
-T\sum_{y}
\pi _{y}\left(\bm{e}\right) \log \pi _{y}\left(
\bm{e}\right) \right\}  d\bm{P}_{\bm{e}}(\bm{e}) \\
s.t. &&\int\pi _{y}\left(\bm{e}\right) d\bm{P}_{\bm{e}}(\bm{e})=\mu _{y}  \; \forall y\\
&&\sum_{y}\pi_{y}\left(\bm{e}\right) =1 \; \forall \bm{e}.
\end{eqnarray*}

This  is an optimal transport problem with entropic regularization, \citep[see][Chapter 7]{otme}. In the absence of the second term in the
objective function,  it would be an optimal transport problem between the
discrete random variable $Y\sim \bm{\mu}$ and the continuous random vector $%
\bm{e}\sim \bm{P}_{\bm{e}}$, with transport surplus $\left( y,\bm{e}\right)
\rightarrow -\left( \bm{Z}\bm{e}\right) _{y}$. The second term is an
entropic regularization.

\subsection{The pure characteristics model\label{app:pure-char}}
The second part of Example~\ref{ex:mixed-logit} is obtained by setting $T=0$
in~(\ref{eq:blp-as-regularized-ot}). The regularization term in~(\ref{Gstar-blp}) disappears, and
\begin{equation}
G^{\ast}(\bm{\mu})=\max_{\bm{\pi} \in \mathcal{M}}\sum_{y\in \mathcal{Y}%
_{0}}\mu_y \int_{\bm{e} \in \mathbb{R}^{d}}-\left( Z\bm{e} \right) d\bm{P}_{%
\bm{e} }\left( \bm{e} \right)  \label{pure-char-as-unregularized-ot}
\end{equation}
which is a standard optimal transport problem (this time without the entropic regularization) between a discrete
random variable on $\mathbb{R}^{d}$ $\tilde{z}$ such that $ \tilde{z}_i=Z_{\tilde{y}i}$ where $\tilde{y} \sim \mu$, and the
continuous random variable $\bm{e} \sim \bm{P}$, where the transport surplus
is now the scalar product $\left( z ,\bm{e} \right) \rightarrow z
^{\top }\epsilon $. This is exactly the power diagram situation described in
Chapter 5 of~\citet{otme}.

\subsection{The FC-MNL Model\label{Appx:FC-MNL}}

\citet{DavisSchiraldiRand2014} introduced a flexible GEV specification which
they called the Flexible Coefficients-Multinomial Choice Model.

\begin{example}[FC-MNL]
\label{ex:FC-MNL} The function $g$ that appears in~(\ref{GEV}) takes the
following form:
\begin{equation*}
g(\bm{t})=\sum_{(y,y^{\prime })\in \mathcal{Y}_{0}^{2}}b_{y,y^{\prime
}}\left( \frac{t_{y}^{1/\sigma }+t_{y^{\prime }}^{1/\sigma }}{2}\right)
^{\tau \sigma }
\end{equation*}%
where $(b_{y,y^{\prime }})$ is a non-negative symmetric matrix, and the
parameters satisfy the inequalities $0<\sigma <1$, $\tau >1$, $\tau \sigma
\leq 1$. We can set $b_{yy}=1$ for every $y$. Note that we recover the
standard multinomial logit model when $\bm{b}$ is the identity matrix.
\end{example}

We followed \citet{DavisSchiraldiRand2014} in making $g$ a $\tau $%
-homogeneous function, rather than 1-homogeneous. This is a harmless
normalization. It gives
\begin{equation*}
G(\bm{U})=\frac{1}{\tau }\left( \log \sum_{(y,y^{\prime })\in \mathcal{Y}%
_{0}^{2}}b_{y,y^{\prime }}\left( \frac{\exp (U_{y}/\sigma )+\exp
(U_{y^{\prime }}/\sigma )}{2}\right) ^{\tau \sigma }\right) +\gamma .
\end{equation*}%
While this may look forbidding, it is easy to evaluate and it yields simple
demands:
\begin{equation*}
\mu _{y}=\frac{1}{g}\exp (U_{y}/\sigma )\sum_{y^{\prime }\in \mathcal{Y}%
_{0}}b_{y,y^{\prime }}\left( \frac{\exp (U_{y}/\sigma )+\exp (U_{y^{\prime
}}/\sigma )}{2}\right) ^{\tau \sigma -1}.
\end{equation*}%
It is apparent from the formul\ae\ that the \textquotedblleft cross-price
elasticities\textquotedblright\ (the dependence of $\bm{\mu}$ on $\bm{U}$
are largely driven by the matrix $\bm{b}$.) In fact %
\citet{DavisSchiraldiRand2014} show that for any fixed $\sigma $ and $\tau $%
, $\bm{b}$ can be chosen to replicate any given set of own- and cross-price
elasticities.

\pagebreak

\part*{Suggested online appendices}

\pagebreak

\section{More on the assumptions [online]\label%
{app:discussions-of-assumptions}}

In this online appendix, we discuss the separability assumption (which we
maintain throughout), and the type I extreme value assumption of~\cite%
{choo-siow:06}  (which we relax).

\subsection{The separability assumption\label{Appx:separability}}

Assumption~\ref{ass:separ} imposes that the matching surplus $\tilde{\Phi}$
be separable in the sense that
\begin{equation*}
\tilde{\Phi}_{ij}=\Phi _{xy}+\varepsilon _{iy}+\eta _{xj}.
\end{equation*}
It is easy to see that Assumption~\ref{ass:separ} is equivalent to the
follwing:

\begin{assumption}[Separability restated]
If two men $i$ and $i^{\prime }$ belong to the same group $x$, and their
respective partners $j$ and $j^{\prime }$ belong to the same group $y$, then
the total surplus generated by these two matches is unchanged if partners
are shuffled:
\begin{equation*}
\tilde{\Phi}_{ij}+\tilde{\Phi}_{i^{\prime }j^{\prime }}=\tilde{\Phi}%
_{ij^{\prime }}+\tilde{\Phi}_{i^{\prime }j}.
\end{equation*}
\end{assumption}

It should be clear from this equivalent definition that we need not adopt
Choo and Siow's original interpretation, in which $\bm{\varepsilon}$ was a
vector of preference shocks of the husband and $\bm{\eta}$ was a vector of
preference shocks of the wife. More precisely, they assumed that the utility
of a man $i$ of group $x$ who marries a woman $j$ of group $y$ was given by
\begin{equation}
\alpha _{xy}+\tau +\varepsilon _{iy},  \label{utMM}
\end{equation}%
where $\alpha _{xy}$ was the ``systematic'' part of the surplus; $\tau$
represented the utility transfer (possibly negative) that the husband gets
from his partner in equilibrium; and $\varepsilon _{iy}$ was a standard
type~I extreme value random term\footnote{%
For a single, $\alpha_{x0}=\tau=0$.}. The utility of this man's wife would
be written as
\begin{equation}
\gamma _{xy}-\tau +\eta _{xj}.  \label{utMW}
\end{equation}%
This formulation clearly implies separability, but it is much stronger than
we need. To take an extreme example, assume that men are indifferent over
partners and are only interested in the transfer they receive; while women
also care about some attractiveness characteristic of men, in a way that may
depend on the woman's group. In a marriage between man $i$ of group $x$ and
woman $j$ of group $y$, if the wife transfers $\tau$ to the husband his net
utility would be $\tau$, and hers would be $(\varepsilon _{iy}-\tau $).
Since the joint surplus is $\varepsilon_{iy}$, it clearly satisfies
Assumption~\ref{ass:separ}. All of our results would apply in this case.
Since there is a continuum of women in each group $y$, but only one man $i$,
he must capture all joint surplus if he marries a woman of group $y$: his
net utility must be $\varepsilon_{iy}$, and hers zero. In other words, this
man will receive a transfer $\tau _{i}=\max_{y\in \mathcal{Y}}\varepsilon
_{iy}$, which depends on his unobservable characteristic. In contrast,
in \citeauthor{choo-siow:06}'s preferred interpretation equilibrium
transfers only depend on characteristics that are observed by the analyst.
Once again, this is a matter of modelling choice and not a logical necessity
since the $\bm{\varepsilon}$ and $\bm{\eta}$ terms are observed by all agents.

\subsection{The logit assumption\label{Appx:IIA}}

\label{page:discussionCS} A second major assumption in the Choo and Siow
model states that the distribution of the unobserved heterogeneity terms $%
\varepsilon_{iy}$ and $\eta_{xj}$ are distributed as type I extreme value
iid random vectors. This brings in familiar but restrictive features of the
logit model, and in particular, the Independence of Irrelevant Alternatives
(IIA) property.

The literature on single-agent discrete choice models has long stressed the
links between the type I-EV specification and IIA. In his famous discussion
of~\citet{Luce:59}, \citet{Debreu:60} showed that given IIA, introducing
irrelevant attributes would change choice probabilities. Matching markets
are two-sided by their very nature, and defining IIA is less straightforward
than in single-agent models---we propose two definitions and draw out their
implications in~\cite{galsalmatchingiia:19}. Still, it is not hard to
construct illustrations similar to~\citeauthor{Debreu:60}'s example within
the~\citeauthor{choo-siow:06} model.

Let $x$ and $y$ consist of education, with two levels $C$ (college) and $N$
(no college). Now suppose that the analyst distinguishes two types of
college graduates: those whose Commencement fell on an even-numbered day $%
C_e $ and those for whom it was on an odd-numbered day $C_o$. Assume that
this difference in fact is payoff-irrelevant: the joint surplus of any match
does not depend on whether the college graduates in it (if any) had
Commencement on an even day. We show in \citet{galsalmatchingiia:19} that
adding the Commencement distinction to the model changes equilibrium
marriage patterns: it reduces the number of singles, and it increases the
number of matches between college graduates while reducing the number of
matches between non-graduates. These are clearly unappealing properties:
since the Commencement date is irrelevant to all market participants, a more
reasonable model would imply none of these changes.

The \citeauthor{choo-siow:06} model has other stark comparative statics
predictions. Since $u_{x}=-\log (\mu _{x0}/n_{x})$ in this framework,
average utilities are in a one-to-one relationship with the probabilities of
singlehood. Property~\eqref{eq:dudn} becomes a statement on
semi-elasticities of these probabilities. Moreover, the equilibrium equation~%
\eqref{eq:CSeq} implies that for any 4-tuple of characteristics $%
(x,y,x^{\prime },y^{\prime }),$
\begin{equation*}
\frac{\mu _{y|x}\mu _{y^{\prime }|x^{\prime }}}{\mu _{y|x^{\prime }}\mu
_{y^{\prime }|x}}=\exp ((\Phi _{xy}+\Phi _{x^{\prime }y^{\prime }}-\Phi
_{x^{\prime }y}-\Phi _{xy^{\prime }})/2).
\end{equation*}%
Therefore the log-odds ratio $(\mu _{y|x}\mu _{y^{\prime }|x^{\prime
}})/(\mu _{y|x^{\prime }}\mu _{y^{\prime }|x})$ should only depend on the
joint surplus matrix $\bm{\Phi}$, and not on the availability of different
types $\bm{n},\bm{m}$. It is easy to see that none of the other
specifications we study in this section has this invariance property. It is
in principle testable, given data for several markets which can be assumed
to have the same surplus function. This property was first pointed out by %
\citet{Graham:13}, who also describes other predictions of the Choo and Siow
framework\footnote{\citet{MourifieSiow} and \citet{Mourifieet19} extend this
and other results of \citet{Graham:13} to models with peer effects.}.

%
%
%

\section{Some properties of the stable matching [online]\label%
{app:additional-equilibrium}}

We now state additional results which took too much space to fit into the
main text.

\subsection{Symmetry}

\label{sub:testable_predictions} Recall from Proposition~\ref%
{prop:IndivSurplus} that the partial derivative of the social surplus $%
\mathcal{W}(\bm{\Phi},\bm{r})$ with respect to $n_{x}$ is $u_{x}$. It
follows immediately that
\begin{equation}
\frac{\partial u_{x}}{\partial n_{x^{\prime }}}=\frac{\partial u_{x^{\prime
}}}{\partial n_{x}}.  \label{eq:dudn}
\end{equation}%
Hence the \textquotedblleft unexpected symmetry\textquotedblright\ result
proven by \citet{Deckermccannetal:12} for \citeauthor{choo-siow:06} model is
a direct consequence of the symmetry of the Hessian of $\mathcal{W}$; and it
holds for \emph{all\/} separable models.

Our second corollary states some properties of the objective function $%
\mathcal{W}$, as a direct implication of Theorem~\ref{MainThmSurplus}.

\begin{corollary}
\label{cor:compstats}

The function $\mathcal{W}\left(\bm{\Phi},\bm{n},\bm{m}\right)$ is convex in $%
\bm{\Phi}$. It is homogeneous of degree 1 and concave in $\bm{r}=(\bm{n},%
\bm{m})$.
\end{corollary}

\begin{varproof}
The convexity of $\mathcal{W}$ w.r.t. $\Phi $ follows immediately from %
\eqref{eq:socialWelfare}; the concavity of $\mathcal{W}$ w.r.t. $(\bm{r})$
similarly follows from \eqref{dualSocialWelfare}. Since $G(\bm{U},\bm{n})$
is 1-homogeneous in $\bm{n}$ and $H(\bm{V},\bm{m})$ is 1-homogeneous in $%
\bm{m}$, the dual program shows that $\mathcal{W}$ is 1-homogeneous in $%
\bm{r}=(\bm{n},\bm{m})$.
\end{varproof}

Corollary~\ref{cor:compstats} entails further consequences. Since the
function $\mathcal{W}(\bm{\Phi},\bm{r})$ is concave in $\bm{r}$, the matrix $%
\partial ^{2}\mathcal{W}/\partial \bm{r}\partial \bm{r}^{\prime }$ must be
semidefinite negative. This implies the symmetry result above, and much
more---including sign constraints on the minors\footnote{%
The most obvious one implies that the expected utility of a type must
decrease with the mass of its members:
\begin{equation*}
\frac{\partial u_{x}}{\partial n_{x}}=\frac{\partial ^{2}\mathcal{W}}{%
\partial n_{x}^{2}}\leq 0.
\end{equation*}%
}. Similarly, since $\mathcal{W}$ is convex in $\bm{\Phi}$ the matrix of
general term $\partial ^{2}\mathcal{W}/\partial \Phi _{xy}\partial \Phi
_{zt} $ must be semi-definite positive, which implies certain symmetry and
determinant sign constraints. \citet{GalSalpp17} studies the comparative
statics of separable models in more detail.

Finally, the homogeneity of $\mathcal{W}$ in $\bm{r}$ implies that all
utilities (e.g. $U_{xy}$ and $v_{t}$) and all conditional matching
probabilities $\mu _{y|x}$ must be homogeneous of degree 0 in $\bm{r}$. In
that sense, all separable models exhibit constant returns to scale. This
property distinguishes separable models from those in~\cite{dagsvik:00} or~%
\cite{Menzel:15}. It can be viewed either as a feature or as a bug. %
\citet{MourifieSiow} and \citet{Mourifieet19} argue for a class of
\textquotedblleft Cobb-Douglas marriage matching
functions\textquotedblright\ that extends the multinomial logit
specification of \citet{choo-siow:06} beyond separable models and allows for
scale and peer effects.

\subsection{Other comparative statics results\label{app:compStats}}

Theorem~\ref{MainThmSurplus} can be used to show that other comparative
statics results of~\citet{Deckermccannetal:12} extend beyond the logit model
to our generalized framework, beyond those stated in Subsection~\ref%
{sub:testable_predictions}. Many of these results are collected in~%
\citet{GalSalpp17}, but we recall some here for completeness. From the
results of Section~\ref{par:ssid}, recall that $\mathcal{W}\left(\bm{\Phi} ,%
\bm{r}\right) $ is given by the dual expressions%
\begin{eqnarray}
\mathcal{W}\left(\bm{\Phi} ,\bm{r}\right) &=& \max_{\mu \in \mathcal{M}\left(%
\bm{r}\right)} \left( \sum_{xy}\mu _{xy}\Phi _{xy}+\mathcal{E}\left(\bm{\mu},%
\bm{r}\right)\right) \text{, and}  \label{envPrimal} \\
\mathcal{W}\left(\bm{\Phi} ,\bm{r}\right) &=&\min_{U_{xy}+V_{xy}=\Phi _{xy}}
\left(\sum n_{x}G_{x}\left(U_{xy}\right) +\sum
m_{y}H_{y}\left(V_{xy}\right)\right);  \label{envDual}
\end{eqnarray}
and that
\begin{equation*}
\frac{\partial \mathcal{W}}{\partial \Phi _{xy}}=\mu _{xy},\text{ }\frac{%
\partial \mathcal{W}}{\partial n_{x}}=G_{x}\left(U_{xy}\right) =u_{x},\text{
}\text{and }\frac{\partial \mathcal{W}}{\partial m_{y}}=H_{y}\left(
V_{xy}\right) =v_{y}.
\end{equation*}

By the same logic as the one that obtained~\eqref{eq:dudn}, the
cross-derivative of $\mathcal{W}$ with respect to $n_{x^{\prime }}$ and $%
\Phi _{xy}$ yields
\begin{equation}
\frac{\partial \mu _{xy}}{\partial n_{x^{\prime }}}=\frac{\partial ^{2}%
\mathcal{W}}{\partial n_{x^{\prime }}\partial \Phi _{xy}}=\frac{\partial
u_{x^{\prime }}}{\partial \Phi _{xy}}  \label{crossdiff2}
\end{equation}%
which is proven (again in the case of the multinomial logit Choo and Siow
model) in \citet[section~3]{Deckermccannetal:12}. The effect of an increase
in the matching surplus between groups $x$ and $y$ on the surplus of
individual of group $x^{\prime }$ equals the effect of the mass of
individuals of group $x^{\prime }$ on the mass of matches between groups $x $
and $y$. Let us provide an interpretation for this result. Assume that
groups $x$ and $y$ are men and women with a PhD, and that $x^{\prime }$ are
men with a college degree. Suppose that $\partial \mu _{xy}/\partial
n_{x^{\prime }}<0$, so that an increase in the mass of men with a college
degree causes the mass of matches between men and women with a PhD to
decrease. This suggests that men with a college degree or with a PhD are
substitutes for women with a PhD. Hence, if there is an increase in the
matching surplus between men and women with a PhD, men with a college degree
will become less of a substitute for men with a PhD. Therefore their share
of surplus will decrease, and $\partial u_{x^{\prime }}/\partial \Phi
_{xy}<0 $.

Finally, differentiating $\mathcal{W}$ twice with respect to $\Phi _{xy}$
and $\Phi _{x^{\prime }y^{\prime }}$ yields%
\begin{equation}
\frac{\partial \mu _{xy}}{\partial \Phi _{x^{\prime }y^{\prime }}}=\frac{%
\partial ^{2}\mathcal{W}}{\partial \Phi _{xy}\partial \Phi _{x^{\prime
}y^{\prime }}}=\frac{\partial \mu _{x^{\prime }y^{\prime }}}{\partial \Phi
_{xy}}.  \label{crossdiff3}
\end{equation}%
The interpretation is the following: if increasing the matching surplus
between groups $x$ and $y$ has a positive effect on marriages between groups
$x^{\prime }$ and $y^{\prime }$, then increasing the matching surplus
between groups $x^{\prime }$ and $y^{\prime }$ has a positive (and equal) effect on
marriages between groups $x$ and $y$.
 Again, all  comparative statics results derived in this section
hold in \emph{any} model  that satisfies our assumptions.

\section{Additional results on estimation [online]\label%
{app:additional-estimation}}

\subsection{Moment matching\label{app:moment-matching}}


Assume that the specification of the joint surplus $\bm{\Phi}^{\bm{\lambda}}$
is linear in $\bm{\lambda}$ and that the distributions of the unobserved
heterogeneity terms $\bm{P}_{x}$ and $\bm{Q}_{y}$ are known. Let $%
(\phi^k_{xy}) $ be the basis functions, and define the \emph{comoments} $%
C^k(\mu)=\sum_{xy} \mu_{xy} \phi^k_{xy}$ for any matching $\mu$. This
appendix proves the following result:

\begin{theorem}[Comoments and a specification test]
\label{thm:comoments} Denote $\hat{\bm{\lambda}}^{MM}$ the moment-matching
estimator defined by~\eqref{maxProgr}.

\begin{enumerate}
\item It makes predicted comoments equal to observed comoments: $C^{k}(\hat{\bm{\mu}})=C^{k}(\bm{\mu}^{\bm{\lambda}})$ for all $k$ when $\bm{\lambda} =\hat{
\bm{\lambda}}^{MM}$.

\item It is also the vector of Lagrange multipliers of the comoment equality
constraints in the program
\begin{equation}
\mathcal{E}_{\max}\left(\hat{\bm{\mu}},\hat{\bm{r}}\right) =\max_{\bm{\mu}
\in \mathcal{M}(\hat{\bm{r}})}\left(\mathcal{E} \left(\bm{\mu},\hat{\bm{r}}%
\right) :C^{k}(\bm{\mu} )=C^{k}(\hat{ \bm{\mu}}) \; \forall k\right).
\label{maxEntropy}
\end{equation}

\item The value of $\mathcal{E}_{\max}\left(\hat{\bm{\mu}},\hat{\bm{r}}%
\right)$ is $\mathcal{E}\left(\bm{\mu}^{\hat{\bm{\lambda}}^{MM}},\hat{\bm{r}}
\right)$. Moreover, $\mathcal{E}\left( \hat{\bm{\mu}},\hat{\bm{r}}\right)
\leq \mathcal{E}_{\max }\left( \hat{\bm{\mu}},\hat{\bm{r}}\right) $, with
equality if and only if there is a value $\bm{\lambda}$ of the parameter
such that $\bm{\Phi}^{\bm{\lambda}}=\bm{\Phi}$.
\end{enumerate}
\end{theorem}

\begin{varproof}
We denote $\hat{\bm{\lambda}}:=\hat{\bm{\lambda}}^{MM}$ to simplify the
notation.

\begin{enumerate}
\item By definition, $\sum_{x,y}\hat{\mu}_{xy}\phi _{xy}^{k}=\left( \partial
\mathcal{W }/\partial \lambda _{k}\right) (\bm{\Phi}^{\hat{\bm{\lambda}}},%
\bm{\hat{r}})$. Applying the envelope theorem to~\eqref{eq:socialWelfare}
shows that
\begin{equation*}
\frac{\partial \mathcal{W}}{\partial \lambda _{k}} (\bm{\Phi}^{\hat{%
\bm{\lambda}}}, \bm{\hat{r}})= \sum_{x,y}\mu_{xy}^{\hat{\bm{\lambda}}}
\phi_{xy}^{k}.
\end{equation*}
Therefore $\sum_{xy} \mu_{xy}^{\hat{\bm{\lambda}}}  \phi_{xy}^{k}=\sum_{xy}
\hat{\mu}_{xy}  \phi_{xy}^{k}$.

\item Given~\eqref{eq:socialWelfare}, the program~(\ref{maxProgr}) can be
rewritten as%
\begin{equation*}
\max_{\bm{\lambda}\in \mathbb{R}^{K}}\min_{\bm{\mu}\in \mathcal{M}(%
\bm{\hat{r}})}\left( \sum_{k}\lambda _{k}\sum_{x,y}\left( \hat{\mu}_{xy}-\mu
_{xy}\right) \phi _{xy}^{k}-\mathcal{E}\left( \bm{\mu},\bm{\hat{r}}\right)
\right).
\end{equation*}%
Since the objective function is convex in $\bm{\mu}$ and linear in $%
\bm{\lambda}$, we can exchange the $\max $ and the $\min $. Consider a value
of $\bm{\mu}$ such that $\sum_{x,y}\left( \hat{\mu}_{xy}-\mu _{xy}\right)
\phi _{xy}^{k}\neq 0$ for some $k$; then minimizing over $\bm{\lambda}$
gives $-\infty $. Therefore these $K$ equalities must hold at the optimum,
and $\bm{\mu}$ minimizes $\mathcal{E}$ over the set of $\bm{\mu}\in \mathcal{%
M}(\bm{\hat{r}})$ such that $\sum_{x,y}\left( \hat{\mu}_{xy}-\mu
_{xy}\right) \phi _{xy}^{k}=0$ for all $k$.

\item Since $\bm{\mu}^{\hat{\bm{\lambda}}}$ maximizes $\sum_{x,y}\mu
_{xy}\Phi _{xy}^{\hat{\bm{\lambda}}}+\mathcal{E}(\bm{\mu};\bm{\hat{r}})$
over $\bm{\mu}$,
\begin{equation*}
\mathcal{E}\left(\bm{\mu}^{\hat{\bm{\lambda}}},\bm{\hat{r}}\right) -\mathcal{%
E}\left(\hat{\bm{\mu}},\bm{\hat{r}}\right) \geq \sum_{x,y}\left(\hat{\mu}%
_{xy}-\mu _{xy}^{\hat{\bm{\lambda}}}\right) \Phi _{xy}^{\hat{\bm{\lambda}}}
\end{equation*}%
and the inequality is strict unless $\bm{\mu}^{\hat{\bm{\lambda}}}=\hat{%
\bm{\mu}}$, since $\mathcal{E}$ is strictly concave in $\bm{\mu}$. By
part~1, the RHS is zero. Therefore $\mathcal{E}_{\mbox{max}}\left( \hat{%
\bm{\mu}},\bm{\hat{r}}\right) =\mathcal{E}(\bm{\mu}^{\hat{\bm{\lambda}}},%
\bm{\hat{r}})\geq \mathcal{E}\left( \hat{\bm{\mu}},\bm{\hat{r}}\right) $,
with equality if and only if $\bm{\mu}^{\hat{\bm{\lambda}}}=\hat{\bm{\mu}}$.

If $\bm{\Phi}=\bm{\Phi}^{\bm{\lambda}}$, then $\bm{\hat{\mu}}$ maximizes $%
\sum_{x,y}\mu _{xy}\Phi_{xy}^{\bm{\lambda}}+\mathcal{E}(\bm{\mu},\bm{\hat{r}}%
)$, and $\bm{\mu}^{\bm{\lambda}}=\bm{\hat{\mu}}$. Therefore $\mathcal{E}(%
\bm{\hat{\mu}},\bm{\hat{r}})=\mathcal{E}_{\mbox{max}}(\hat{\bm{\mu}},%
\bm{\hat{r}})$.
\end{enumerate}
\end{varproof}

Comparing the values of $\mathcal{E}(\bm{\hat{\mu}},\bm{\hat{r}})$ and $%
\mathcal{E}\left(\bm{\mu}^{\hat{\bm{\lambda}}^{MM}},\hat{\bm{r}} \right)$
gives a simple specification test. Its critical values can be obtained by
bootstrapping for instance. One could also run the test for different
specifications of the distributions of heterogeneities and invert it to
obtain confidence intervals for the parameters of $\bm{P}_x$ and $\bm{Q}_y$.

\subsection{Geometric interpretation of the estimation procedure\label%
{app:Geom}}

The approach to inference we describe in Section~\ref{par:linearModel} has a
simple geometric interpretation. In this appendix, we fix the distributions $%
\bm{P}_x$ and a specification $(\phi_{xy}^k)_{k=1,\ldots,K}$ of the linear
model of surplus $\bm{Q}_y$; and we vary the parameter vector $\bm{\lambda}$%
. Now consider the set of moments associated to all feasible matchings:
\begin{equation*}
\mathcal{F}=\left\{ \left(C^{1},...,C^{K}\right) :C^{k}=\sum_{xy}\mu
_{xy}\phi _{xy}^{k},~\bm{\mu} \in \mathcal{M}\left(\hat{\bm{r}}%
\right)\right\}
\end{equation*}

\begin{figure}[tbp]
\begin{center}
\includegraphics[width=17cm]{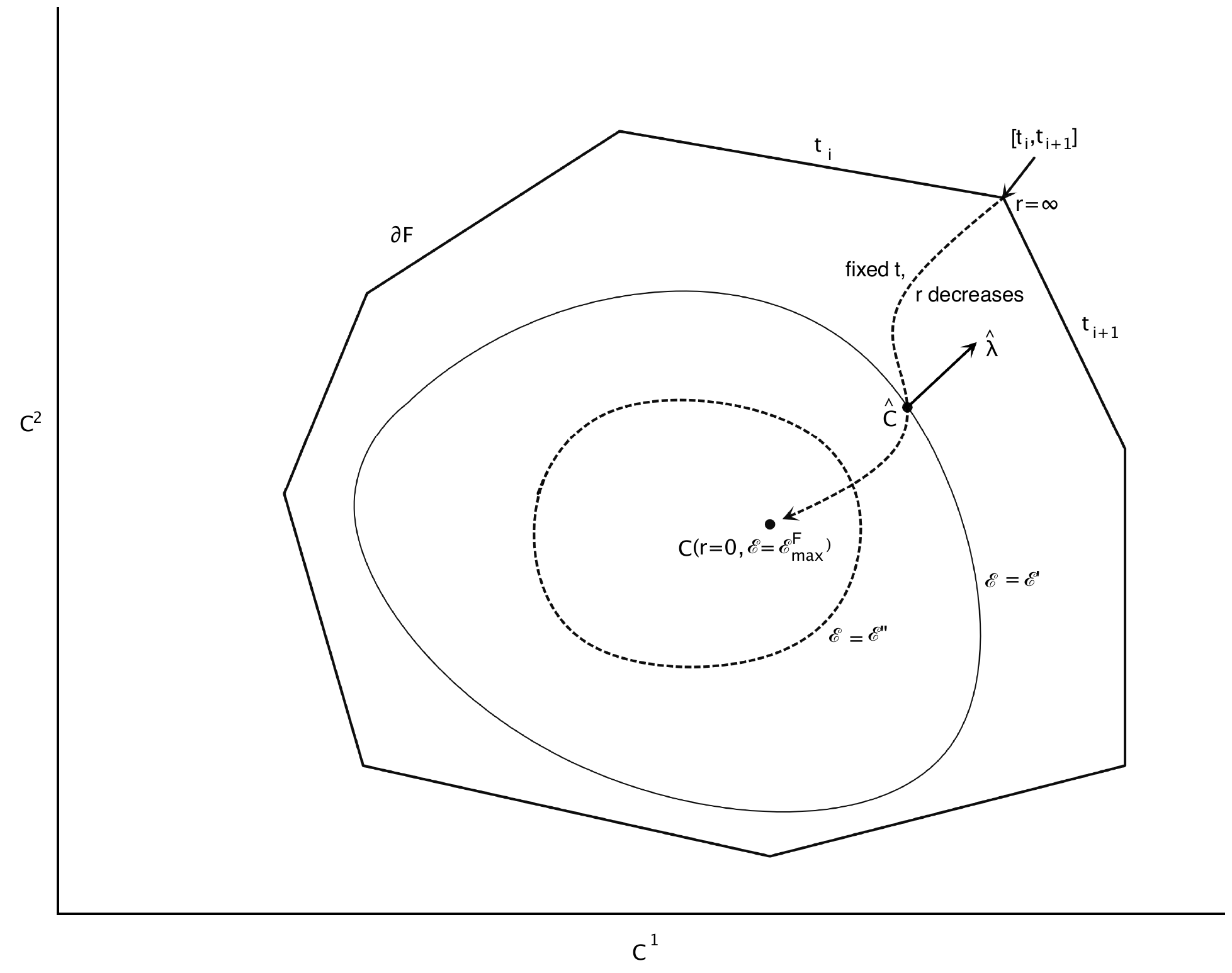}
\end{center}
\caption{The covariogram and related objects}
\label{fig:covario}
\end{figure}

This is a convex polyhedron, which we call the \emph{covariogram}. It
includes the observed commoments $\hat{\bm{C}}$, as well as the vector of
moments $C^{\bm{\lambda}}$ generated by the optimal matching $\bm{\mu}^{%
\bm{\lambda}}$ for any value of the parameter vector $\bm{\lambda}$. Each
feasible matching $\bm{\mu}$ also has a generalized entropy $\mathcal{E}(%
\bm{\mu}, \bm{\hat{r}})$; we denote $\mathcal{E}^{\bm{\lambda} }\equiv
\mathcal{E}(\bm{\mu} ^{\bm{\lambda}},\bm{\hat{r}})$ the generalized entropy
associated with parameter vector $\bm{\lambda}$. Since the vectors $\phi$
are linearly independent, the mapping $\bm{\lambda} \longrightarrow C^{%
\bm{\lambda} }$ is invertible on the covariogram. Denote $\bm{\lambda}(C)$
its inverse. The corresponding optimal matching has generalized entropy $%
\mathcal{E}_{r}\left(C\right) =\mathcal{E}^{\bm{\lambda} \left(C\right)}$.
The level sets of the function $\mathcal{E}_{r}$ are \emph{isoentropy
surfaces} in the covariogram.

Figure~\ref{fig:covario} illustrates these concepts. It assumes $K=2$ basis
functions, so that the covariogram is a convex polyhedron in $(C^1,C^2)$
plane. Since $\bm{\lambda}$ also is two-dimensional, it can be represented
in polar coordinates. Let the data be generated by $\bm{\lambda} =r\exp (it)$%
. For $r=0$, the model is uninformative: matching is random and generalized
entropy takes its maximum possible value $\mathcal{E}^F_{\mbox{max}}$ among
all possible matchings. We denote $C_0$ the corresponding moments. At the
other extreme, the boundary $\partial F$ of the covariogram corresponds to $%
r=\infty$. There is no unobserved heterogeneity; generically over $t$, the
moments generated by $\bm{\lambda}$ must belong to a finite set of vertices,
so that $\bm{\lambda} $ is only set-identified.

As $r$ decreases for a given $t$, the corresponding moments follow a
trajectory indicated by the dashed line on Figure~\ref{fig:covario}, from
the boundary $\partial F$ to the point $C_{0}$. The entropy $\mathcal{E}^{%
\bm{\lambda}}$ increases as this trajectory crosses contours of higher
entropy ($\mathcal{E}^{\prime}$ then $\mathcal{E}^{\prime\prime}$ on the
figure.)

We know from Theorem~\ref{thm:comoments}.2 that the moment-matching
estimator $\hat{\bm{\lambda}}^{MM}$ is the vector of multipliers of the
program that maximizes entropy over the matchings that generate the observed
values of the moments. Therefore $\partial \mathcal{E}_{r}(\hat{C})/\partial
C^{k}=\hat{\lambda}_{k}^{MM}$; and the moment-matching estimator lies on the
normal to the isoentropy contour that goes through the observed moments $%
\bm{\hat{C}}$. This is shown as $\hat{\lambda}$ on Figure~\ref{fig:covario}.

\subsection{Parameterization, testing, and multimarket data\label%
{app:tradeoff}}

Proposition \ref{thm:Identification} shows that, given a specification of
the distribution of the unobserved heterogeneities $\bm{P}_{x}$ and $\bm{Q}%
_{y}$, there is a one-to-one correspondence between $\bm{\mu}$ and $\bm{\Phi}
$. To put it differently: any matching on a single market can be
rationalized by exactly one model that satisfies Assumptions~\ref{ass:separ}
and~\ref{ass:fullsupp}, for any such vector of distributions. This has
several consequences for analysts using data on a single market. Without
further restrictions, it is impossible to test separability, even assuming
perfect knowledge of the distributions of unobserved heterogeneity. It is
also impossible to discriminate between separable models based on different
distributions. \label{page:IdentDiscussion} One way out of this conundrum is
to incorporate credible restrictions (inspired by theoretical restrictions,
or by institutional features of the market) into both the surplus matrix $%
\bm{\Phi}$ and the distributions of unobservable heterogeneity $\bm{P}_{x}$
and $\bm{Q}_{y}$. To take a simple example, suppose that we know that there
is no interaction between partner characteristics $x^{k}$ and $y^{l}$ in the
production of joint surplus: for fixed values of the other characteristics, $%
\Phi _{xy}$ is additive in $x^{k}$ and $y^{l}$. Given our identification
formula~\eqref{IdentPhi} and observed matching patterns, this translates
into a set of constraints on the derivatives of the generalized entropy, and
therefore on the distributions $\bm{P}_{x}$ and $\bm{Q}_{y}$. Adding
constraints on the distributions would make the model testable\footnote{%
As a trivial illustration, finding that $\log \hat{\mu}_{xy}$ is not
additive in $x^{k}$ and $y^{l}$ would reject the Choo and Siow model in this
example.}. As another example, consider a semiparametric specification in
the spirit of \citet{EHN:04}: $\Phi _{xy}=\bm{b}_{y}^{\prime}\bm{\phi} _{x}$, with
known $d$-dimensional vectors $\bm{\phi}_{x}$ and unknown vectors $\bm{b}_{y}$. If $d<%
\abs{Y}$, this would restrict the number of degrees of freedom in $\bm{\Phi}$%
, freeing parameters to specify the distributions of heterogeneity and/or to
test the model. An alternative empirical strategy is to use multiple markets
with restricted parametric variation in the joint surplus $\bm{\Phi}$ and
the distributions of unobserved heterogeneity $\bm{P}_{x}$ and $\bm{Q}_{y}$.
The variations in the group sizes $\bm{n}$ and $\bm{m}$ across markets then
generate variation in optimal matchings that can be used to overidentify the
model and generate testable restrictions. \citet{csw:17} relied on a variant
of this approach.

\section{Computational Methods and Benchmarks [online]}

\label{app:computation}

Section~\ref{sec:computation} described two classes of methods to compute
the equilibrium matching patterns: min-Emax, and IPFP. Min-Emax is more
generally applicable than IPFP; on the other hand, IPFP is much faster. To
document these claims, we present in this appendix a small simulation of the
Choo and Siow model that explores the computational performance of four
different methods: a general-purpose equation solver, the min-Emax method,
the minimization of the function $F$ expressed in~\eqref%
{ChooSiow:SocialSurplus}, and IPFP.

In the second part of this appendix, we show how linear programming
techniques can be used to solve and estimate a model with discretized error
distributions.

\subsection{Benchmarks}

We simulated ten cases, with a number of categories $|\mathcal{X}|=|\mathcal{%
Y}|$ that goes from 100 to 5,000. For each of these ten cases, we draw the $%
n_{x}$ and $m_{y}$ uniformly in $\{1,\ldots ,100\}$; and for each $(x,y)$
match we draw $\Phi _{xy}/2$ from $\mathcal{N}(0,1).$

\subsubsection{Minpack}

We applied the Levenberg-Marquardt solver Minpack to the system of $(%
\mathcal{X}|+|\mathcal{Y})$ equations that characterizes the optimal
matching (see Section~\ref{sec:computation}). Minpack is probably the
most-used solver in scientific applications; it underlies many statistical
and numerical packages.

\subsubsection{Min-Emax}

The min-Emax method we described in Section~\ref{sec:computation} minimizes $%
\left( G(\bm{U},\bm{n})+H(\bm{\Phi}-\bm{U},\bm{m})\right) $ over the $|%
\mathcal{X}|\times |\mathcal{Y}|$ object $\bm{U}=(U_{xy}).$ In the
particular case of the Choo and Siow model, the function $G$ is given by
\begin{equation*}
G(\bm{U},\bm{n})=\sum_{x\in \mathcal{X}}n_{x}\log \left( 1+\sum_{y\in
\mathcal{Y}}\exp (U_{xy})\right)
\end{equation*}%
and $H$ has the same form.

We used the Knitro optimizer\footnote{%
See \citet{knitropaper}.} to obtain the solution.

\subsubsection{Minimizing $F$}

Formula \eqref{ChooSiow:SocialSurplus} provides us with an alternative
method that works on the smaller object $(u_{x},v_{y})$ of group average
utilities. Here again we used the Knitro optimizer.

\subsubsection{IPFP}

\label{appx:compIPFP}

Consider the logit model of Choo and Siow.

Fix a value of $\bm{\lambda}$ and drop it from the notation: let the joint
surplus function be $\bm{\Phi}$, with optimal matching $\bm{\mu}.$ Formula~(%
\ref{eq:homoCS}) can be rewritten as
\begin{equation}
\mu _{xy}=\exp \left( \frac{\Phi _{xy}}{2}\right) \sqrt{\mu _{x0}\mu _{0y}}.
\label{eq:homoipfp}
\end{equation}%
As noted by \citet{Deckermccannetal:12} we could just plug this into the
feasibility constraints $\sum_{y}\mu _{xy}+\mu _{x0}=n_{x}$ and $\sum_{x}\mu
_{xy}+\mu _{0y}=m_{y}$ and solve for the masses of singles $\mu _{x0}$ and $%
\mu _{0y}.$ This results in a system of $\abs{\mathcal{X}}+\abs{\mathcal{Y}}$
equations:
\begin{align}
\mu _{x0}+\left( \sum_{y\in \mathcal{Y}}\exp \left( \frac{\Phi _{xy}}{2}%
\right) \sqrt{\mu _{0y}}\right) \sqrt{\mu _{x0}}& =n_{x}  \label{eq:CSquadra}
\\
\mu _{0y}+\left( \sum_{x\in \mathcal{X}}\exp \left( \frac{\Phi _{xy}}{2}%
\right) \sqrt{\mu _{x0}}\right) \sqrt{\mu _{0y}}& =m_{y}.
\end{align}%
Taking the unknowns to be $\sqrt{\mu _{x0}}$ and $\sqrt{\mu _{0y}}$, each of
these equations is quadratic in the unknowns. IPFP simply consists of
solving the system~\eqref{eq:CSquadra} iteratively. Starting from an
arbitrary guess $\mu _{0y}^{(0)}$, at step $(2k+1)$ we find the following
updating equation~(\ref{CS-ipfp-update}). These equations are easily solved
in closed form. The pseudo-code in Algorithm~\ref{algo:IPFPpseudo} gives a
detailed implementation. \ Note that since in the Choo and Siow model the
shadow prices $u_{x}$ and $v_{y}$ are simply minus the logarithms of the
corresponding $\mu _{x0}$ and $\mu _{0y}$, this algorithm in fact operates
on $u_{x}$ and $v_{y}$.

\begin{myalgorithm}
\textbf{Solving for the optimal matching by IPFP}
\begin{algorithmic}
    \Require two non-negative vectors $\bm{n}$ and $\bm{m}$ (sizes $M$ and $N$);
    a matrix $\bm{\Phi}$ of size $(M,N)$
    \Require a tolerance $\tau$ and a maximum number of iterations $I$
\Ensure the matrix $\bm{\mu}$ of size $(M,N)$ holds the marriage patterns at the optimal matching
\State $X \gets \mbox{size}(\bm{n})$
\State $Y \gets \mbox{size}(\bm{m})$
\State $\bm{K} \gets \exp(\bm{\Phi}/2)$
\State $\delta \gets \infty, i \gets 0$
\State $\bm{T} \gets 0_Y$
\While{$\delta > \tau$ and $i < I$}
    \State $\bm{S} \gets \bm{K} \bm{T}$
    \Comment{Project on $\bm{n}$ margins}
    \State $\bm{t} \gets \left(\sqrt{\bm{S}^2+4\bm{n}}-\bm{S}\right)/2$
    \State $S \gets \bm{K}^{\prime}  \bm{t}$
    \Comment{Project on $\bm{m}$ margins}
    \State $\bm{T} \gets \left(\sqrt{\bm{S}^2+4\bm{m}}-\bm{S}\right)/2$
\State $\delta_1 \gets \max \; \abs{\bm{t}^2 +\bm{t} \odot \bm{K}\bm{T}-\bm{n}}$
\Comment{Error on $\bm{n}$ margins}
\State $\delta_2 \gets \max \; \abs{\bm{T}^2 +\bm{T} \odot \bm{K}^\prime\bm{t}-\bm{m}}$
\Comment{Error on $\bm{m}$ margins}
\State $\delta \gets \max (\delta_1,\delta_2)$
\State $i \Leftarrow i+1$
\EndWhile
    \If {$i\geq I$}
        \State  Failed to achieve requested precision
    \Else
        \State $\bm{\mu}\gets \bm{K} \odot
        (\bm{t} \otimes \bm{T})$
        \Comment{$\otimes$ denotes outer product and $\odot$ element-wise product}
    \EndIf
    \end{algorithmic}
\label{algo:IPFPpseudo}
\end{myalgorithm}

\subsubsection{Results}

\begin{figure}[tbp]
\begin{center}
\includegraphics[width=16cm,height=18cm]{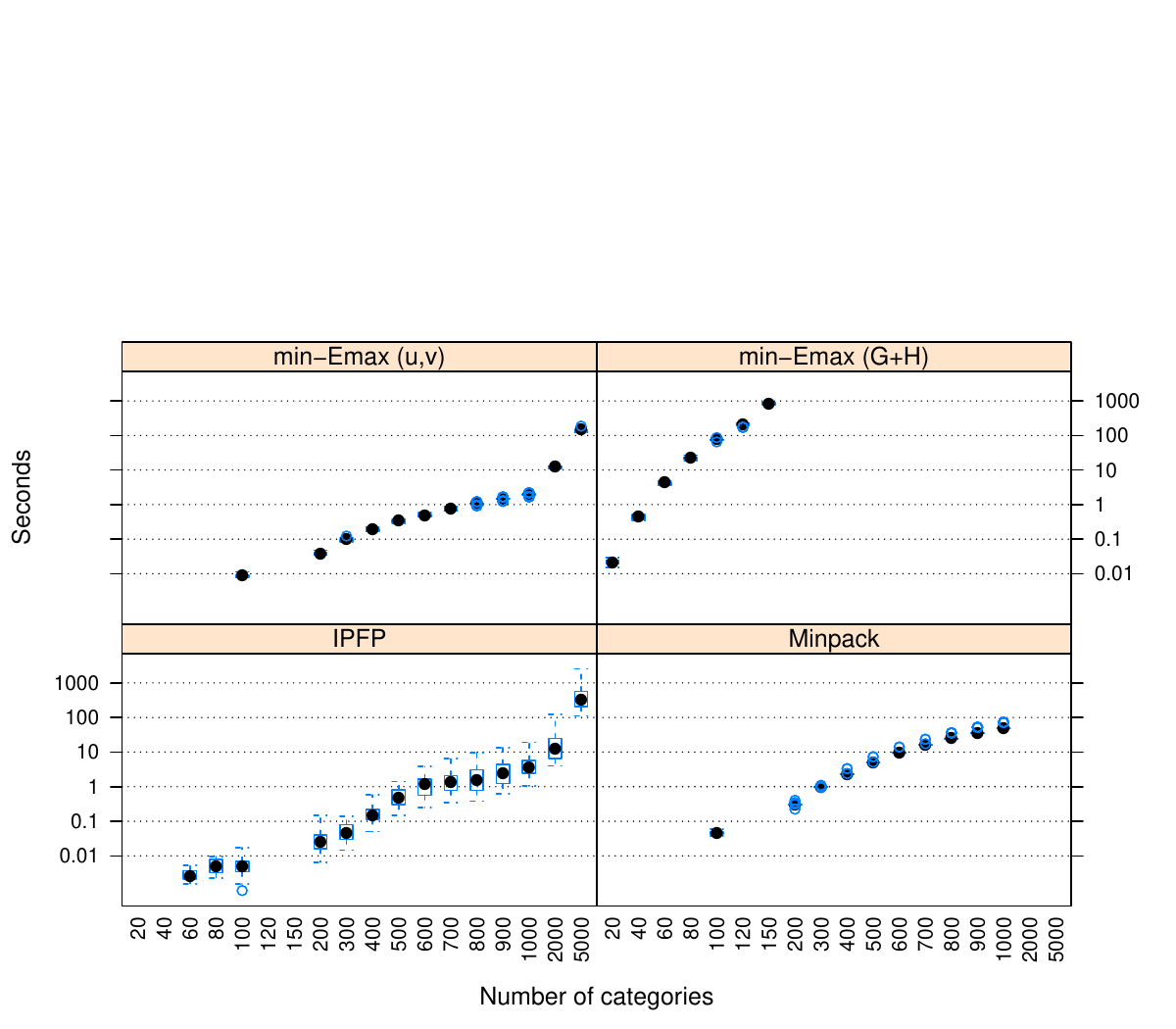}
\end{center}
\caption{Solving for the optimal matching}
\label{fig:compareTimes}
\end{figure}

For all four methods, we used \texttt{C/C++} programs run on a single
processor of a Mac desktop. We set the convergence criterion for all methods
as a relative estimated error of $10^{-6}.$ This is not as straightforward
as one would like: both Knitro and Minpack rescale the problem before
solving it, while we did not attempt to do it for IPFP. Still, varying the
tolerance within reasonable bounds hardly changes the results, which we
present in Figure~\ref{fig:compareTimes}. Each panel gives the distribution
of CPU times for one of the four methods, in the form of a Tukey
box-and-whiskers graph\footnote{%
The box goes from the first to the third quartile; the horizontal bar is at
the median; the lower (resp.\ upper) whisker is at the first (resp.\ third)
quartile minus (resp.\ plus) $1.5$ times the interquartile range, and the
circles plot all points beyond that.}.

There are three things to note about these graphs. First, distances on the $%
x $-axis are not drawn to scale, except for the smaller number of
categories; second the $y$-axis is logarithmic; third, for some methods we
only report results on the lower range of categories. The reasons are
obvious from the graphs. Minpack solving not scale up well. The min-Emax
method that minimizes $(G(\bm{U})+H(\bm{\Phi}-\bm{U}))$ is even worse: in
this ``logit'' case, it is not competitive beyond 100 categories as it
minimizes in a high-dimensional space. On the other hand, the min-Emax
method that optimizes over $\bm{u}$ and $\bm{v}$ and the IPFP algorithm both
perform remarkably well, even with several thousands of categories.

Choo and Siow only used 60 categories in their application (ages from 16 to
75). For such numbers, all four methods work well, but IPFP and min-Emax on $%
(u,v)$ again clearly dominate. We should emphasize that only the special
structure of the Choo and Siow model allowed us to reduce the dimensionality
by minimizing over $\bm{u}$ and $\bm{v}$. IPFP, on the other hand, can be
used in a broader class of models. While IPFP has more variability than the
other methods (perhaps because we did not rescale the problem beforehand),
even the slowest convergence times for each problem size are at least three
times smaller than those of Minpack. This is all the more remarkable that
IPFP does not require any calculation of derivatives; by comparison, we fed
the code for the Jacobian of the system of equations into Minpack. IPFP also
compares very well with the min-Emax method on $(u,v)$, even though we fed
the Jacobian and the Hessian into Knitro.

Finally, while we have run these experiments on a single processor, it is
clear that IPFP is much more amenable to parallel implementation than the
optimization methods, since each iteration solves $|\mathcal{X}|$ or $|%
\mathcal{Y}|$ equations that are independent of each other.

\subsection{An additional method: linear programming}

\label{app:linprogr}

Min-Emax and IPFP both exploit the structure of the separable matching
problem. A more ``brute-force'' alternative is to simply solve the
underlying linear programming problem. This requires discretizing the
distribution of the error terms. We now explain how it can be done, and we
extend it to obtain the moment-based estimator in a semilinear model.

\subsubsection{Equilibrium via linear programming}

\label{ssub:simulations_linear_programming_and_discrete_support} Now suppose
that the vectors $\varepsilon$ and $\eta$, instead of having full support,
only take a finite number of values: these are analogous to the unobserved
``types'' of many structural econometric models. We define $%
(\varepsilon_y^{xk})_{y\in\mathcal{Y}_0, k=1,\ldots,K_x}$ to be the points
of support of $\mathbf{P}_x$, and $(r_x^k)$ their probabilities; and we
define $(\eta_x^{yl})$ and $(s^l_y)$ similarly. In this case, $G(\bm{U},%
\bm{n})$ is $\sum_x n_x E_{\mathbf{P}_x}\max_y(U_{xy}+\varepsilon_y)$, that
is
\begin{equation*}
G(\bm{U},\bm{n})=\sum_{x\in\mathcal{X}} n_x \sum_{k=1}^{K_x} r^k_x
\max\left(\varepsilon_0^{xk}, \max_{y\in\mathcal{Y}}(U_{xy}+%
\varepsilon_y^{xk})\right).
\end{equation*}
Define $u_x^k=\max\left(\varepsilon_0^{xk}, \max_{y\in\mathcal{Y}%
}(U_{xy}+\varepsilon_y^{xk})\right)$, and $v_y^l=\max\left(\eta_0^{yl},
\max_{x\in\mathcal{X}}(V_{xy}+\eta_x^{yl})\right)$. By construction,
\begin{eqnarray}
u_{x}^{k} &\geq &U_{xy}+\varepsilon _{y}^{xk}\;\;\forall y\;\;\mbox{ and }%
\;u_{x}^{k}\geq \varepsilon _{0}^{xk}  \label{eq:constronu} \\
v_{y}^{l} &\geq &V_{xy}+\eta _{x}^{yl}\;\;\forall x\;\;\mbox{ and }%
\;v_{y}^{l}\geq \eta _{0}^{yl}.  \label{eq:constronv}
\end{eqnarray}%
It follows from \eqref{dualSocialWelfare} that we minimize the objective
function and given the constraint $U_{xy}+V_{xy}\geq \Phi_{xy}$, it is easy
to see that the optimal matching solves
\begin{equation*}
\mathcal{W}(\bm{\Phi},\bm{n},\bm{m})=\min_{\bm{u},\bm{v},\bm{U}}
\left(\sum_{x\in\mathcal{X}}n_x \sum_{k=1}^{K_x} r^k_x u^k_x+\sum_{y\in%
\mathcal{Y}}m_y \sum_{l=1}^{L_y} s^l_y v_y^l\right)
\end{equation*}
subject to the constraints \eqref{eq:constronu} and \eqref{eq:constronv}
with $V_{xy}=\Phi_{xy}-U_{xy}.$ Note that the objective function and the
constraints are linear in the variables. Therefore solving for equilibrium
with finite types boils down to a linear programming problem, for which very
fast algorithms are available (even with many variables). The multipliers of
the constraints at the optimum give the matching patterns for each type in
each group, and can be averaged over types to yield the $\mu_{xy}$. This
idea can be taken further: any distributions $\mathbf{P}_x$ and $\mathbf{Q}%
_y $ can be discretized. Solving the program above for a given
finite-support approximation of the distributions gives an approximation
that can be shown to converge to the optimum for the limit of the discrete
distributions, by adapting an argument of \citet[Theorem~3.1]{ChernoGHH}.
Hence the approach described in this subsection is applicable to any
separable model.


\subsubsection{Computing the moment-matching estimator}\label{ssub:estimating_the_linear_model}
The linear programming approach of Subsection~\ref{ssub:simulations_linear_programming_and_discrete_support} can be
extended in order to compute the moment-matching estimator in the semilinear
models of Section~\ref{par:linearModel}. Equation~(\ref{maxProgr}) shows
that the moment-matching estimator minimizes $\min_{\bm{\lambda}}\left(
\mathcal{W}(\bm{\lambda}^{\prime }\bm{\tilde{\phi}},\hat{\bm{r}})-%
\bm{\lambda}^{\prime }\hat{C}\right) $ over $\lambda $. This suggests a
general approach to the estimation of separable models with known
distributions of heterogeneity. First, specify a linear surplus function and
distributions of unobservable heterogeneity $\mathcal{P}_{x}$ and $\mathcal{Q%
}_{y}$. Second, discretize the latter distributions. Third, solve the
following linear program:
\begin{equation*}
\min_{\bm{u},\bm{v},\bm{U},\bm{\lambda}}\left( \sum_{x\in \mathcal{X}}\hat{n}%
_{x}\sum_{k=1}^{K_{x}}r_{x}^{k}u_{x}^{k}+\sum_{y\in \mathcal{Y}}\hat{m}%
_{y}\sum_{l=1}^{L_{y}}s_{y}^{l}v_{y}^{l}-\bm{\lambda}^{\prime }\hat{C}\right)
\end{equation*}%
under the constraints~\eqref{eq:constronu} and~\eqref{eq:constronv},
replacing $V_{xy}$ with $\bm{\lambda}^{\prime }\bm{\tilde{\phi}_{xy}}-U_{xy}$%
. The objective and the constraints are still linear with respect to all
variables, which now also include $\bm{\lambda}$. Once again, this program
can be solved efficiently by linear programming algorithms, yielding both
the moment-matching estimator and the corresponding average utilities and
matching patterns.

\subsection*{A summary}

Each computational method has pros and cons. The min-Emax method can be
applied quite generally. It requires many evaluations of $G$ and $H$
however, which may be costly for large $|\mathcal{X}|,|\mathcal{Y}|$. Linear
programming is attractive in semilinear models, at the price of
discretization. IPFP requires no discretization, provides easy estimation of
linear model, and is highly scalable. It does require evaluating the
extended entropy $E$ of Section~\ref{appx:extendingentropy}, which is
straightforward in logit-type models.

\section{Application to Choo and Siow's data [online]}\label{Appx:CSdata}

Our empirical application uses the data \citet{choo-siow:06} put together,
with some minor changes. We also put more emphasis on the treatment of those
$(x,y)$ cells that have zero observations.

\subsection{The data}\label{sub:the_data}

Choo and Siow used data from the Census to evaluate the numbers $\bm{n}$ and
$\bm{m}$ of men and women of every age in every state; and they relied on
National Center for Health Statistics data to estimate the number of
marriages by state and by age cell. They were kind enough to share with us
their samples and programs; the description that follows is very similar to
that in their paper, and in fact quotes freely from it.

\subsubsection{The populations}

\label{ssub:the_populations} Data on the populations of men and women of
every age and state were extracted from the Integrated Public-Use Microdata
Sample files of the U.S. Census \citep[see][]{ipums:ref}. Choo and Siow used
data from the 1970 and 1980 U.S. Census to construct population vectors:

\begin{quote}
The samples used were the 5 percent state samples for 1980 and the 1 percent
Form 1 and Form 2 samples for 1970. The 1970 data sets were appropriately
scaled to be comparable with the 1980 files\footnote{%
State of residence in the 1970 census files can be identified only in the
state samples (Form 1 and Form 2 samples, both of which are 1 percent
samples). This is the reason that the other samples were not used for 1970
calculations. Further, the age of marriage variable is available only in
Form 1 samples in 1970, which meant that only one sample, the Form 1 state
sample, was used for calculations involving married couples in the 1970
Census.}.

[\ldots]

We use the \texttt{marst} variable in the census to identify a person as
either never married, currently married, or previously married (divorced or
widowed). To calculate the number of available individuals, we simply add
the never marrieds and previously married.
\end{quote}

Choo and Siow kept all individuals aged 16 to 75. Since the number of first
marriages in which either partner is older than 40 is rather small in the
70s and 80s, we decided to focus on the populations aged 16 to 40 instead.
The ``state'' of an individal is defined as his/her place of residence.


\subsubsection{The marriages}

\label{ssub:the_marriages} Choo and Siow obtained data on marriages from the
Vital Statistics reports that many states send to the National Center for
Health Statistics (NCHS):

\begin{quote}
Marriage records from the 1971/72 and 1981/82 Vital Statistics were used to
construct the bivariate distributions of marriages. A state has to report
the number of marriages to the National Center for Health Statistics to be
in the sample.
\end{quote}

We deviated from their paper in two respects.

\begin{itemize}
\item To be consistent with our age window for populations in the basis year
we only keep marriages in which either partner is at most 41 (in the Census
year+1) or 42 (in the Census year+2). We corrected a small mistake in the
construction of the  data---\citet{choo-siow:06} did not update the ages of the subjects
between Census year+1 and Census year+2. This does not affect their main
conclusions.

\item The list of states we include in our application is slightly
different. They excluded Iowa, Minnesota, and South Carolina which we do use
since they reporteed to the NCHS in both waves. Colorado only reported to
the NCHS after 1980. Choo and Siow excluded it from their study; we keep it
in the 1980s wave. Choo and Siow also excluded New York City from New York
State. We eventually decided to exclude both.
\end{itemize}

A ``reform'' state is one in which the Roe v.\ Wade Supreme Court decision
affected the legal status of abortion. Our list of reform states comprises
Alaska, California, Delaware, Florida, Georgia, Hawaii, Kansas, Maryland,
and (in the 1980s only) Colorado. Our non-reform states are Alabama,
Connecticut, Idaho, Illinois, Indiana, Iowa, Kentucky, Louisiana, Maine,
Massachusetts, Michigan, Mississippi, Missouri, Montana, Nebraska, New
Hampshire, New Jersey, Ohio, Pennsylvania, Rhode Island, South Dakota,
Tennessee, Utah, Vermont, West Virginia, Wyoming, and the District of
Columbia. We exclude from our study Arizona, Arkansas, Nevada, New Mexico,
New York, North Dakota, Oklahoma, Texas, Washington, and (in the 1970s)
Colorado.


\subsubsection{Merging availables and marriages}

\label{ssub:merging_availables_and_marriages}

Table~\ref{tab:numGenders} describes our data on the populations of men and
women. The numbers between parenthesis refer to the population, the other
numbers to the sample. With a total of $2.19$m observations representing $%
58.67$m individuals, our universe of men and women is about 40\% smaller
than Choo and Siow's. This is a direct consequence of our focus on younger
ages. The reform states have $34.6$\% of the population in 1970 and $37.9$\%
in 1980. The sample is much larger in 1980, as the ACS dataset we use had a
better sampling rate then.

\begin{table}[ht]
\centering
\begin{tabular}{|r|r|r|r|r||}
\hline
\multicolumn{2}{|c|}{Census} & \multicolumn{1}{|c|}{1970} &
\multicolumn{1}{|c|}{1980} & Population increase \\ \hline
\multirow{2}{40mm}{Reform states} & Men & $81,260$ ($4.32$m) & $351,231$ ($%
7.20$m) & $66.7$\% \\ \cline{2-5}
& Women & $66,920$ ($3.63$m) & $308,808$ ($6.37$m) & $76.2$\% \\ \hline
\multirow{2}{40mm}{Non-reform states} & Men & $150,887$ ($7.82$m) & $566,460$
($11.51$m) & $47.2$\% \\ \cline{2-5}
& Women & $137,839$ ($7.16$m) & $524,741$ ($10.68$m) & $49.2$\% \\
\hline\hline
\multirow{2}{40mm}{Total} & Men & $232,147$ ($12.14$m) & $917,691$ ($18.71$m)
& $54.2$\% \\ \cline{2-5}
& Women & $204,759$ ($10.77$m) & $833,549$ ($17.05$m) & $58.3$\% \\ \hline
\end{tabular}%
\caption{\textbf{Numbers of men and women}}
\label{tab:numGenders}
\end{table}

Table~\ref{tab:numMarrs} describes our subsample from the NCHS dataset. In
this table $(rt,N)$ for instance refers to marriages in which the husband
lists a reform state as his residence, and the wife lists a non-reform
state. In more than 95\% of marriages, husband and wife list a state with
the same ``reform status''. This is not surprising since a large majority of
marriages in fact unite partners from the same state. As in Choo and Siow,
the number of marriages increased much more in reform states than in
non-reform states; but also less than the general population.

\begin{table}[ht]
\centering
\begin{tabular}{|c|c|c|r||}
\hline
Wave & 1971--72 & 1981--82 & Population increase \\ \hline
(r,R) & $138,483$ ($838,140$) & $424,416$ ($1.00$m) & $19.4$\% \\
\hline\hline
(r,N) & $5,866$ ($38,518$) & $10,383$ ($32,952$) & $-14.5$\% \\ \hline
(n,R) & $6,108$ ($33,440$) & $10,182$ ($24,530$) & $-26.6$\% \\ \hline
(n,N) & $216,428$ ($1.70$m) & $506,953$ ($1.79$m) & $4.9$\% \\ \hline\hline
Total & $366,885$ ($2.61$m) & $951,934$ ($2.84$m) & $8.9$\% \\ \hline
\end{tabular}%
\caption{\textbf{Numbers of marriages}}
\label{tab:numMarrs}
\end{table}

Finally, Table~\ref{tab:ageMarrs} shows that the average age at marriage
increased by two years, quite uniformly across reform status and genders. As
a consequence, the age difference also did not change, with husbands two
years older than their wives.

\begin{table}[ht]
\centering
\begin{tabular}{|c|c|c|r|r|}
\hline
\multicolumn{2}{|c|}{Wave} & 1971--72 & 1981--82 & Increase \\ \hline
\multirow{2}{40mm}{Reform states} & Men & $23.0$ & $25.1$ & $2.1$ \\
\cline{2-5}
& Women & $20.9$ & $23.0$ & $2.1$ \\ \hline
\multirow{2}{40mm}{Non-reform states} & Men & $22.7$ & $24.7$ & $2.0$ \\
\cline{2-5}
& Women & $20.6$ & $22.6$ & $2.0$ \\ \hline
\end{tabular}%
\caption{\textbf{Ages at marriage}}
\label{tab:ageMarrs}
\end{table}



\subsection{Zero cells}

\label{sub:zero_cells} Like much discrete-valued economic data, the Choo and
Siow data contains a small but non-negligible percentage of $(x,y)$ cells
with no observed match---up to 3\%, depending on the subsample\footnote{%
Trade is another area where matching methods have become popular in recent
years \citep[see][]{Costinot-Vogel:15}; and trade data also has typically
many zero cells.}. The CS specification by construction rules out zero
probability cells, and \citet[footnote 15, p.\ 186]{choo-siow:06} used
kernel smoothers to impute positive probabilities in these ``zero cells''.
More generally, no separable model with full support can simulate zero cells
(see our discussion of Assumption~\ref{ass:fullsupp}).

This is not an issue with unrestricted estimation, since we only need to
assign a value of $-\infty $ to the corresponding $\Phi _{xy}$. With
parametric inference, maintaining Assumption~\ref{ass:fullsupp} implies that
the model is misspecified. This is a minor consideration in practice, as the
estimated probabilities of these cells turn out to be very small. A cleaner
alternative is to specify error distributions $\bm{P}_{x}$ and $\bm{Q}_{y}$
that do not have full support, either because their supports have lower
dimension and/or because their supports are bounded.


\subsection{Detailed Estimation Results}

\label{Appx:estims}

\subsubsection{Selecting Basis Functions}

We used our moment matching method to estimate 625 semilinear versions of
the original \cite{choo-siow:06} specification, which we will call ``the
homoskedastic logit model''. They all include the two basis functions $%
\phi^1_{xy}\equiv 1$ and $\phi^2_{xy}=D_{xy}\equiv \mathrm{1\kern-.40em 1}%
(x\geq y)$, where $x$ is the age of the husband and $y$ that of the
wife---both linearly transformed to be in [-1,1]. The $D$ term accounts for
possible jumps or kinks in surplus when the wife is older than the husband ($%
D=0$). In addition to these two basis functions, we include a varying set of
functions of the form $x^i y^j$ and $x^i y^j D$. Our richest candidate
specification has 98 basis functions; note that the nonparametric model has
625 (as many as marriage cells.)

\begin{figure}[tbp]
\begin{center}
\includegraphics[width=15cm]{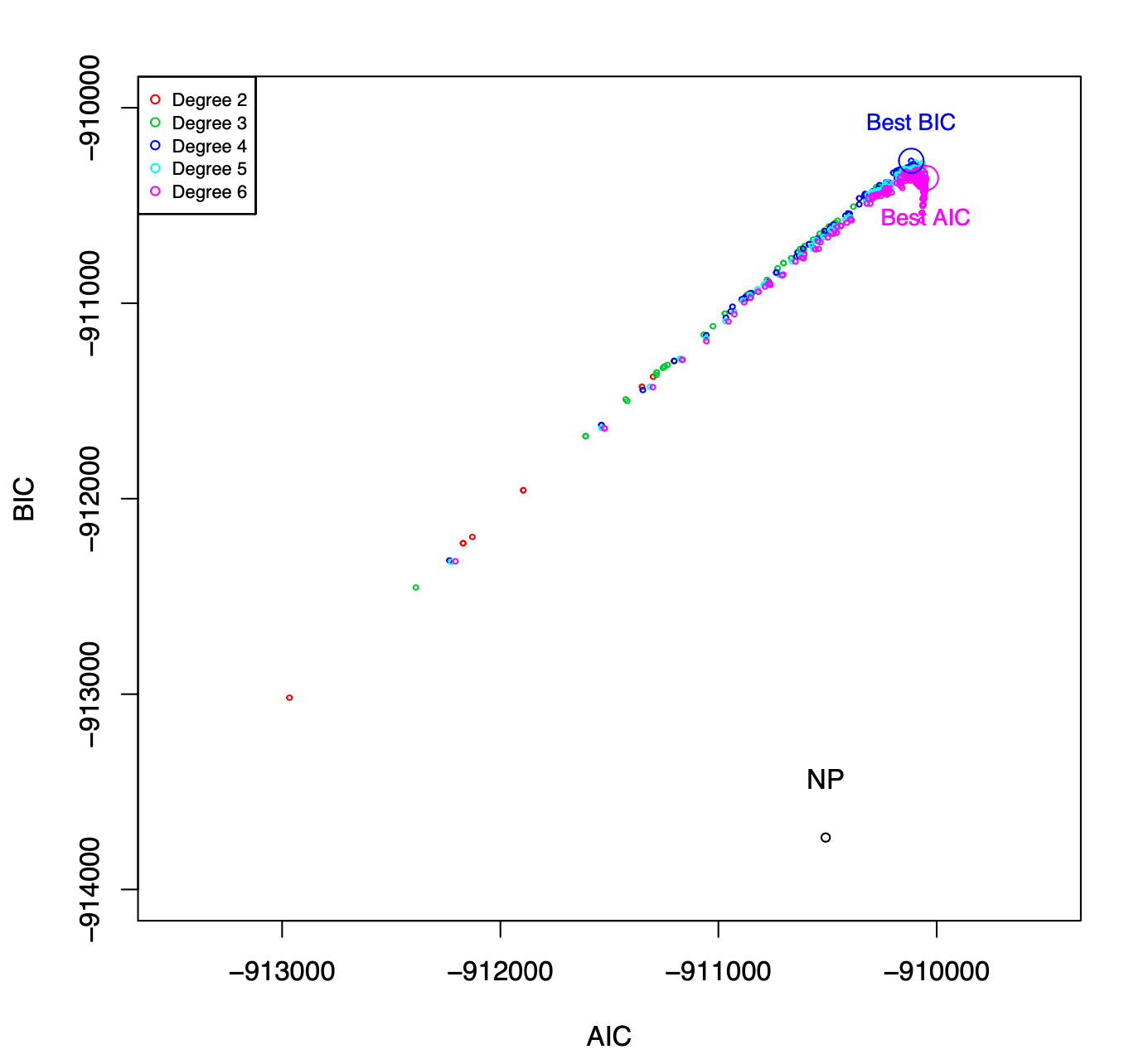}
\end{center}
\caption{AIC and BIC Values for the Parametric and Nonparametric Choo and
Siow Models}
\label{fig:AICandBIC}
\end{figure}

Figure~\ref{fig:AICandBIC} plots the values of the Akaike Information
Criterion (on the horizontal axis) and of the Bayesian Information Criterion
(on the vertical axis) for the 625 models, and for the nonparametric model
NP. The location of NP shows that even for our sample of a couple hundred
thousand observations, it is severely overparameterized: no fewer than 490
of our 625 models have a better AIC, and all of them have a better BIC.

Our best AIC model is still large: it has 60 coefficients, of which 49 are
significant at 5\%. With such a large sample, we could probably have
included even higher-degree terms and improved the AIC slightly. While the
AIC criterion subtracts twice the number of parameters from the
log-likelihood, the BIC criterion penalizes it by half of the logarithm of
the number of observations. With our 224,068 observations, this amounts to $%
6.2$ rather than $2$ times the number of parameters. As a result, the
BIC-selected model only has 30 coefficients, of which 28 differ
significantly from 0 at the 5\% level. For model selection (as opposed to
forecasting), BIC is more appropriate than AIC and we will work with its 30
selected basis functions from now on: all terms $x^m y^n$ and $x^m y^n D$
for $1\leq m\leq 2$ and $1\leq n\leq 4$.

\subsubsection{The Homoskedastic Choo and Siow Model}

Table~\ref{tab:CShomo} gives the estimated coefficients and their
bootstrapped standard errors and Students for the BIC-preferred modelin this
class.

\begin{table}[th]
{\small \ \centering
\begin{tabular}{r|rrr}
\hline
& Estimates & Standard Errors & Students \\ \hline
$1$ & -11.163 & 0.023 & -490.9 \\
$D$ & 1.147 & 0.066 & 17.3 \\
$X$ & -14.759 & 0.336 & -44.0 \\
$XD$ & 5.204 & 0.134 & 38.7 \\
$X^2$ & -13.211 & 0.208 & -63.4 \\
$X^2 D$ & 5.656 & 0.104 & 54.2 \\
$Y$ & -1.220 & 0.066 & -18.4 \\
$YD$ & 4.757 & 0.127 & 37.5 \\
$Y^2$ & -2.064 & 0.041 & -50.7 \\
$Y^2 D$ & 5.950 & 0.118 & 50.5 \\
$Y^3$ & 1.097 & 0.054 & 20.4 \\
$Y^3 D$ & 1.659 & 0.029 & 57.4 \\
$Y^4$ & -0.563 & 0.033 & -17.0 \\
$Y^4 D$ & -0.637 & 0.018 & -35.5 \\
$X Y$ & 26.379 & 0.403 & 65.4 \\
$X Y D$ & -16.697 & 0.336 & -49.7 \\
$X Y^2$ & -16.956 & 0.455 & -37.3 \\
$X Y^2 D$ & 10.238 & 0.298 & 34.3 \\
$X Y^3$ & 6.206 & 0.336 & 18.4 \\
$X Y^3 D$ & -3.936 & 0.227 & -17.3 \\
$X Y^4$ & -0.997 & 0.144 & -6.9 \\
$X Y^4 D$ & 0.881 & 0.092 & 9.6 \\
$X^2 Y$ & 12.940 & 0.276 & 46.9 \\
$X^2 Y D$ & -11.549 & 0.226 & -51.1 \\
$X^2 Y^2$ & -5.636 & 0.303 & -18.6 \\
$X^2 Y^2 D$ & 4.938 & 0.229 & 21.5 \\
$X^2 Y^3$ & 1.131 & 0.196 & 5.8 \\
$X^2 Y^3 D$ & -1.053 & 0.137 & -7.7 \\
$X^2 Y^4$ & 0.085 & 0.060 & 1.4 \\
$X^2 Y^4 D$ & -0.072 & 0.050 & -1.4 \\ \hline
\end{tabular}
}
\caption{Estimates for the Homoskedastic Logit Model }
\label{tab:CShomo}
\end{table}

\paragraph{Estimates}

Table~\ref{tab:CShomo} in Appendix~\ref{Appx:estims} collects our estimates
for the coefficients of the BIC-preferred model with iid standard type I EV
errors. Since the distributions $\mathbb{P}_x$ and $\mathbb{Q}_y$ are
parameter-free in this model, the table shows the estimated coefficients for
the 30 basis functions in its first column. We evaluated their standard
errors (third column) with a bootstrap procedure based on 999 draws from the
estimated variance-covariance matrix of the observed matching patterns $%
\bm{\hat{\mu}}$.

The bootstrap also allows us to compute a $p$-value for the entropy test
described in Theorem~\ref{thm:comoments}. The value of the entropy test
statistic in the sample has a bootstrapped $p$-value is $0.856$. Recall that
this tests the hypothesis that the true surplus function is a linear
combination of our 30 basis functions, conditional on the distributional
assumptions being true. The $p$-value tells us that this ``spanning
hypothesis'' would only be rejected at the 15\% level. This confirms that
the 30-bases model is a very good approximation to the data-generating
process. The Choo and Siow model aims at explaining marriage patterns by
age, from age 16 to age 75. In the early 1970s, close to 80\% of marriages
occurred before either partner was 30 years old, so that the number of data
points to fit is rather small. Even using BIC to reward parsimony, with more
than 200,000 observations we end up with a rich model and an excellent fit.

As a consequence, the distributional parameters we introduce can only
improve the fit marginally. We did find, however, that allowing for gender-
and age-dependent heteroskedasticity yielded a notable improvement in the
fit. Interestingly, it also changes the profile of surplus-sharing within
couples: the share that goes to the husband increases much more steeply than
in the original (homoskedastic) Choo and Siow specification. We also fitted
several Generalized Extreme Values models. The most promising ones seem to
be those of the FC-MNL family \citep{DavisSchiraldiRand2014}, which
incorporate the type of local correlation patterns that are missing from the
multinomial logit framework. While they do not outperform the basic Choo and
Siow specification in our application, they are easy to implement and seem
to us to have much potential in matching models.

\paragraph{Heteroskedastic Logit Models}

We explored several ways of adding heteroskedasticity to the benchmark
model. It is clear from~\ref{eq:sepPhi} that the parameters can only be
identified up to a scale normalization: multiplying both $\bm{\Phi}$ and the
error terms $\bm{\varepsilon}$ and $\bm{\eta}$ by the same positive number
has no effect on the equilibrium matching. The \cite{choo-siow:06} model
normalizes the scale (twice) by using standard type I EV errors for both $%
\bm{\varepsilon}$ and $\bm{\eta}$. When adding heteroskedasticity to $%
\bm{\varepsilon}$ and $\bm{\eta}$, we need to maintain one normalization.

Our simplest heteroskedastic model still uses a standard type I EV $%
\bm{\varepsilon}$ (our scale normalization) and adds only one parameter $%
\tau $, with
\begin{equation*}
\tau^2=\frac{V\eta}{V\varepsilon}.
\end{equation*}
This model allows for heteroskedasticity across genders, but not across
types. Somewhat surprisingly, the profiled loglikelihood of the model is
very flat with respect to $\tau$. While we did obtain an estimate of $0.927$
that is slightly lower than one, the improvement in the loglikelihood is so
small that the values of both AIC and BIC deteriorate.

Going further, we allow for type- and gender-dependent heteroskedasticity%
\footnote{\cite{csw:17} attempted to estimate a similar model, with
education as the type.}. To do this, we multiply the terms $\bm{\varepsilon}%
_{i\cdot}$ (resp.\ $\bm{\eta}_{j\cdot}$) by scale factors $\sigma_x$ (resp.\
$\tau_y$). We experimented with specifications of the form
\begin{align*}
\sigma_x &=\exp(\sigma_1 x+\ldots +\sigma_p x^p) \\
\tau_y &=\exp(\tau_0+\tau_1 y+\ldots+\tau_q y^q)
\end{align*}
Note that we do not allow for a constant term $\sigma_0$; this gives us the
requisite scale normalization.

Of all such specifications for $0\leq p\leq 4$ and $1\leq q\leq 4$, this
yields the largest improvement in the fit: a sizeable $+38.5$ points of
loglikelihood, and $+25.2$ points on BIC. The estimates of the parameters of
$\sigma_x$ and $\tau_y$ can be found in Table~\ref{tab:CSHXYstderrs}.

\begin{table}[ht]
{\small \centering
\begin{tabular}{r|rrr}
\hline
& Estimates & Standard Errors & Students \\ \hline
$\sigma_1$ & 0.793 & 0.051 & 15.4 \\
$\tau_0$ & -0.751 & 0.161 & -4.7 \\ \hline
\end{tabular}
}
\caption{Estimates for the Heteroskedastic Logit Model: Distributional
Parameters }
\label{tab:CSHXYstderrs}
\end{table}

\paragraph{Two-level, Two-nest Nested Logit}

We estimated a two-level nested logit model in which we separate the
singlehood option from all others. This model has two nests: one
corresponding to singlehood, and one to the 25 possible ages of the partner.
It introduces two additional parameters, $\gamma_m$ on the men side and $%
\gamma_w$ for women. The familiar equation from \cite{choo-siow:06}:
\begin{equation*}
2\log \mu_{xy}=\log\mu_{x0}+\log\mu_{0y}+\Phi_{xy}
\end{equation*}
becomes
\begin{equation*}
\gamma_m \log \frac{\mu_{xy}}{\sum_{t\in \mathcal{Y}}\mu_{xt}} +\gamma_w
\log \frac{\mu_{xy}}{\sum_{z\in \mathcal{X}}\mu_{zy}} =\log\frac{\mu_{x0}}{%
\sum_{t\in \mathcal{Y}}\mu_{xt}}+\log\frac{\mu_{0y}}{\sum_{z\in \mathcal{X}%
}\mu_{zy}} +\Phi_{xy}.
\end{equation*}
The values of $(1-\gamma_m)$ and $(1-\gamma_w)$ can be interpreted as
``within-nest correlations''; they equal zero in the \cite{choo-siow:06}
model.

We chose this specific nested logit model because we showed in \cite%
{galsalmatchingiia:19} that it satisfies a ``weak IIA'' property--and we
conjectured that it is the only separable model that does. When we tried to
estimate this two-nest specification, we consistently found a corner maximum
at $\gamma_m=1$. The other parameter $\gamma_w$ has a weak maximum at $0.91$%
, and the loglikelihood barely improves.

\paragraph{FC-MNL}

\cite{DavisSchiraldiRand2014} show that for any admissible values of $\sigma$
and $\tau$, there exist values of the $b$ matrix that rationalize a given
set of elasticities of substitution. We followed their suggestion of using $%
\sigma=0.5$ and $\tau=1.1$; and we chose the very parsimonious specification
of the $b$ matrix described in Section~\ref{subsec:mnlestims}. The maximum
likelihood estimates for the distributional parameters\footnote{%
Given the small gain in the loglikelihood, the standard errors are large.}
are in Table~\ref{tab:FCMNLestimates}.

\begin{table}[ht]
{\small \ \centering
\begin{tabular}{r|rrr}
\hline
& Estimates &  &  \\ \hline
$b_m(16)$ & 0.011 &  &  \\
$b_m(40)$ & 0.000 &  &  \\
$b_w(16)$ & 0.060 &  &  \\
$b_w(40)$ & 0.000 &  &  \\ \hline
\end{tabular}
}
\caption{Estimates for the FC-MNL Model: Distributional Parameters }
\label{tab:FCMNLestimates}
\end{table}

\end{document}